\documentclass[12pt]{article}

\usepackage{epsfig}
\usepackage{color,graphicx}
\usepackage{cite}
\usepackage{amssymb,amsmath}

\usepackage{a4wide}
\usepackage{amssymb}
\usepackage{amsmath}
\usepackage{graphicx}
\usepackage{mathdots}
\usepackage{youngtab}
\usepackage{caption}
\usepackage{subcaption}
\usepackage{enumerate}

\newcommand{\be}{\begin{equation}}
\newcommand{\ee}{\end{equation}}
\newcommand{\bea}{\begin{eqnarray}}
\newcommand{\eea}{\end{eqnarray}}

\numberwithin{equation}{section}

\begin{document}

\begin{center}  

\vskip 2cm 

\centerline{\Large {\bf 5d fixed points from brane webs and O7-planes}}
\vskip 1cm

\renewcommand{\thefootnote}{\fnsymbol{footnote}}

   \centerline{
    {\large \bf Oren Bergman}\footnote{bergman@physics.technion.ac.il}
    {\bf and}
    {\large \bf Gabi Zafrir}\footnote{gabizaf@techunix.technion.ac.il}}

\vspace{1cm}
\centerline{{\it Department of Physics, Technion, Israel Institute of Technology}} \centerline{{\it Haifa, 32000, Israel}}
\vspace{1cm}

\end{center}

\vskip 0.3 cm

\setcounter{footnote}{0}
\renewcommand{\thefootnote}{\arabic{footnote}}   
   
\begin{abstract}
We explore the properties of five-dimensional supersymmetric gauge theories living on 5-brane webs in orientifold 7-plane backgrounds.
These include $USp(2N)$ and $SO(N)$ gauge theories with fundamental matter, as well as $SU(N)$ gauge theories with symmetric and antisymmetric matter.
We find a number of new 5d fixed point theories that feature enhanced global symmetries.
We also exhibit a number of new 5d dualities.
\end{abstract}
      
\newpage

\tableofcontents

\section{Introduction}

Although 5d gauge theories are perturbatively non-renormalizable, in many ${\cal N}=1$ supersymmetric cases they are UV complete
 \cite{SEI,SM,SMI}.
Namely, there exist interacting 5d ${\cal N}=1$ superconformal theories with relevant deformations corresponding to an inverse Yang-Mills coupling
of a 5d ${\cal N}=1$ supersymmetric gauge theory.
The superconformal theory is in some sense the infinite coupling limit of the gauge theory.
The existence of such fixed point theories is also suggested by the existence of a (unique) 5d superconformal algebra $F(4)$
(which has a bosonic subgroup $SO(5,2)\times SU(2)_R$).
However these UV fixed point theories do not admit a Lagrangian description.
%

A useful way to describe such theories is using 5-brane webs in Type IIB string theory \cite{HA,HAK}.
This realizes a 5d SCFT as an intersection of 5-branes at a point. The moduli and mass parameters of the SCFT are described as motions 
of the internal and external 5-branes, respectively. In particular, the 5d SCFT may posses a deformation leading to a low-energy gauge theory
described by a simple 5-brane configuration containing stacks of D5-branes.
This allows us to study various aspects of 5d superconformal theories and the associated gauge theories.

First, it provides a new way to classify 5d gauge theories that have UV fixed points,
since the gauge groups and matter content are constrained by the brane construction.
In some cases this extends the perturbative classification of \cite{SMI}.

5-brane webs also allow us to uncover dual gauge theories in five dimensions.
Since in 5d masses are real, the IR theories obtained by positive and negative mass deformations may be different.
In particular these may be different gauge theories related by a continuation of the Yang-Mills coupling past infinity.
This can be seen in the 5-brane web construction by reversing the deformation leading to the original gauge theory,
and using the $SL(2,\mathbb{Z})$ symmetry of Type IIB string theory to transform the web into a configuration
with D5-brane stacks.

Finally, 5-brane webs can also motivate and assist in demonstrating non-perturbatively enhanced global symmetries.
5d gauge theories possess $U(1)$ global symmetries associated to the instanton number
currents, $j_I = *\mbox{Tr}(F\wedge F)$,  in each non-abelian gauge group factor.
In some cases the instanton operators associated to these currents provide additional
conserved currents, and lead to a larger global symmetry than is apparent in the gauge theory Lagrangian.
Such is the case, for example, for the $SU(2)$ theory with $N_F\leq 7$,
where the classical $SO(2N_F)\times U(1)_I$ symmetry is enhanced by instantons to $E_{N_F+1}$ \cite{SEI}.
%
%
Since conserved currents belong to BPS multiplets of operators,
enhancement of global symmetries is exhibited by the 5d superconformal index,
which can in turn be computed for the IR gauge theory using localization \cite{KKL,BGZ,Zaf,HKKP}.
The crucial ingredient 
is of course the instanton contribution, which can be obtained from known expressions for instanton 
partition functions in four dimensions.
However in some cases 
corrections are required \cite{BGZ,BMPTY,HKT,HKKP}. 
In the 5-brane web description these corrections are identified with extraneous states, corresponding to 
strings external to the web whose contributions have to be removed by hand from the partition function.


Most 5-brane web constructions so far have been for theories with $SU(N)$ gauge groups and matter fields
in fundamental or bi-fundamental representations\footnote{One exception being \cite{KB} which contains a short discussion on webs for $USp(2N)$ and $SO(N)$ gauge groups using $O5$ planes.}.
Our main purpose in this paper is to extend the study of 5d ${\cal N}=1$ gauge theories and fixed point 
theories via 5-brane webs to theories with $USp(2N)$ and $SO(N)$ gauge groups, as well as to $SU(N)$
theories with rank-two antisymmetric and symmetric matter.
We will do this by including orientifold 7-planes.
We will exhibit 5d fixed point theories, some of them new, with relevant deformations given by gauge theories
of the above form. We will also find new 5d dualities, and examples of enhanced global symmetry
involving the above types of theories.
        


The outline for the rest of the paper is as follows. 
Section 2 is devoted to $USp(2N)$ theories, constructed using 5-brane webs in an $O7^-$ plane background.
In section 3 we construct $SO(M)$ theories using an $O7^+$ plane.
In sections 4 and 5 we construct $SU(N)$ theories with rank 2 antisymmetric and symmetric matter,
by adding a fractional NS5-brane to the $O7^-$ and $O7^+$ plane, respectively.
We conclude in section 6.
We also include an appendix containing a discussion of corrections to instanton partition functions in the relevant cases of $USp(2N)$ and $SO(M)$.


\section{$O7^-$ and $USp$ gauge theories}

\begin{figure}
\center
\includegraphics[width=0.33\textwidth]{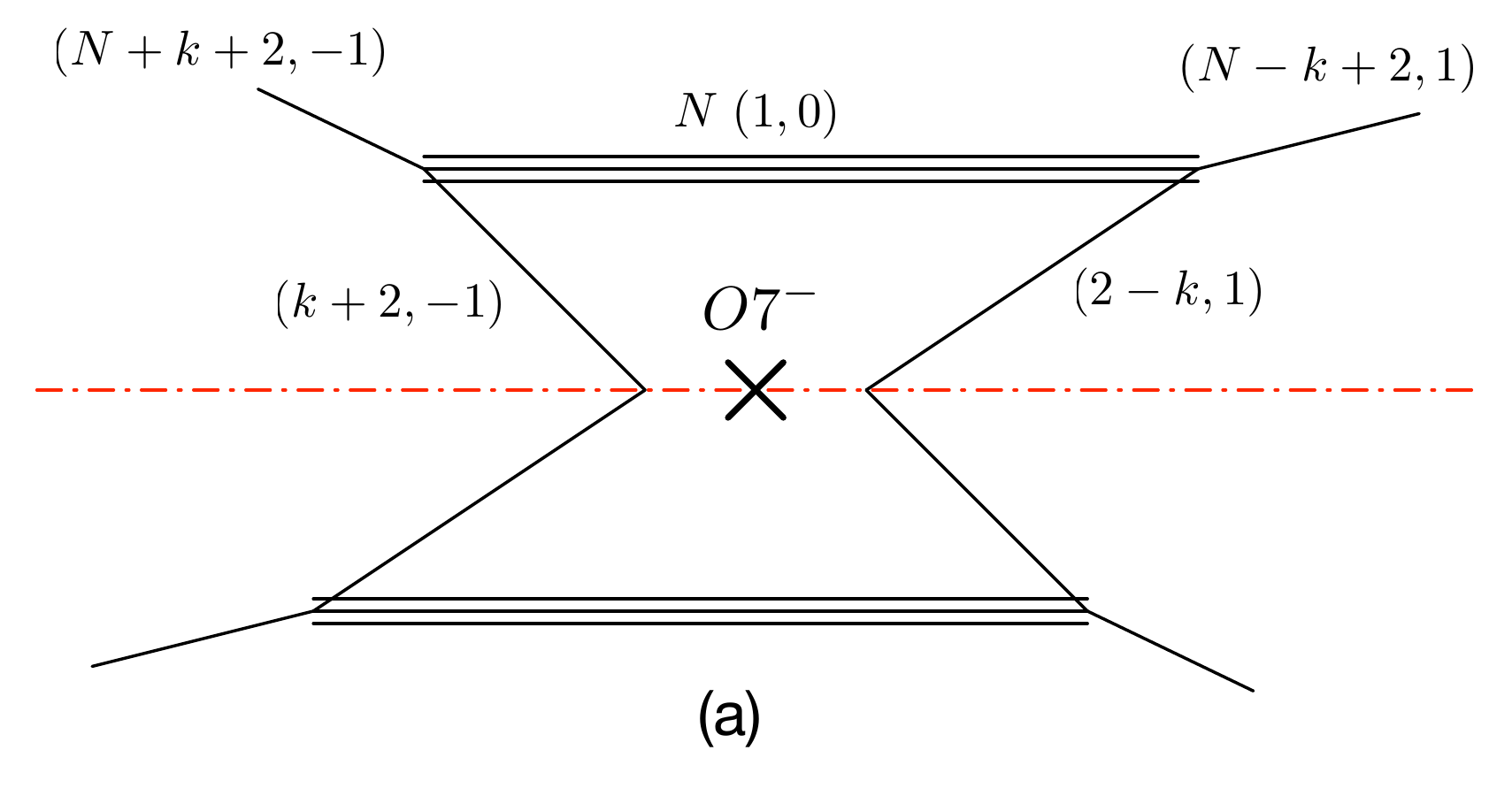} 
\hspace{0.5cm}
\includegraphics[width=0.3\textwidth]{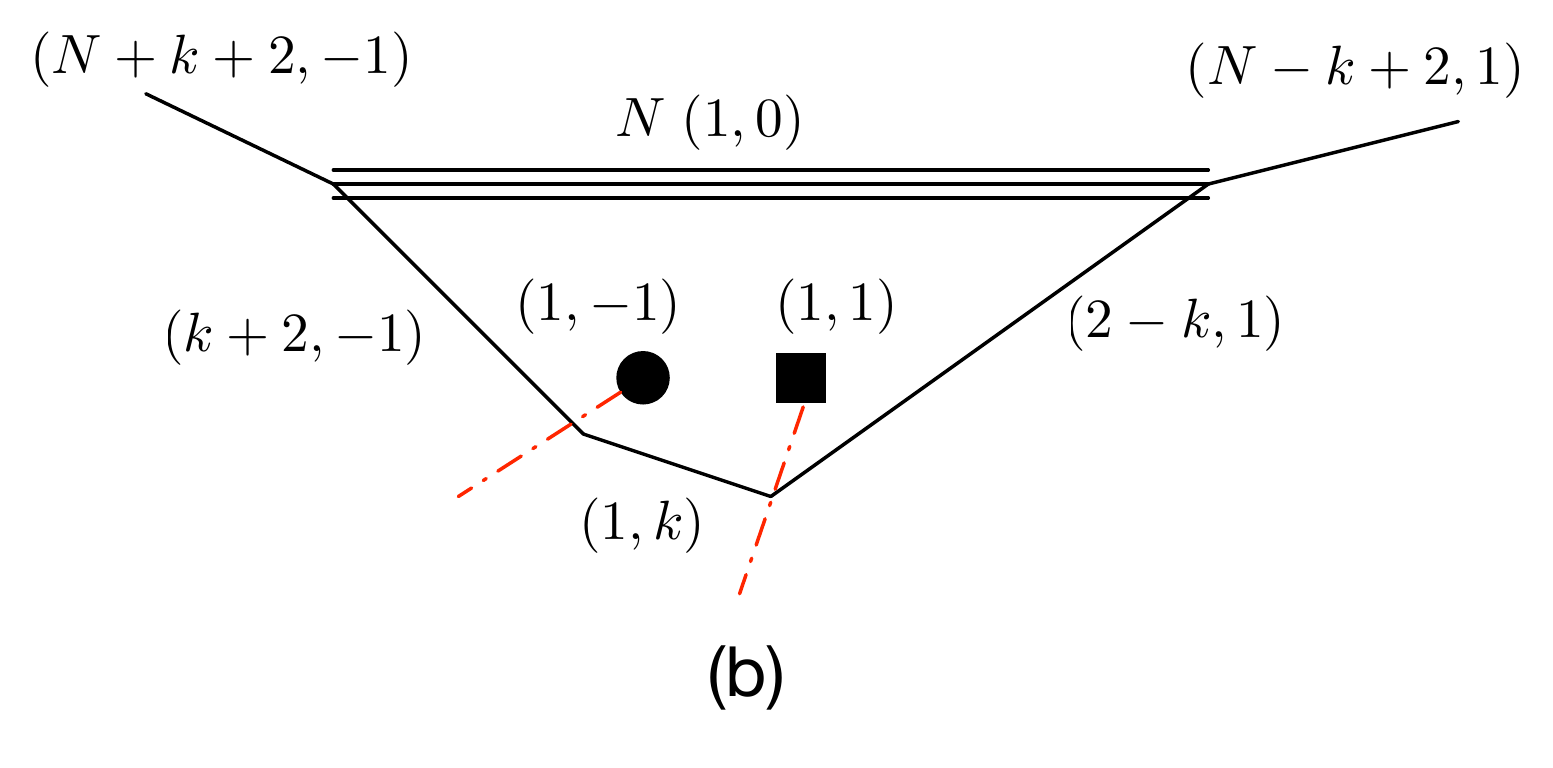} 
\hspace{0.5cm}
\includegraphics[width=0.25\textwidth]{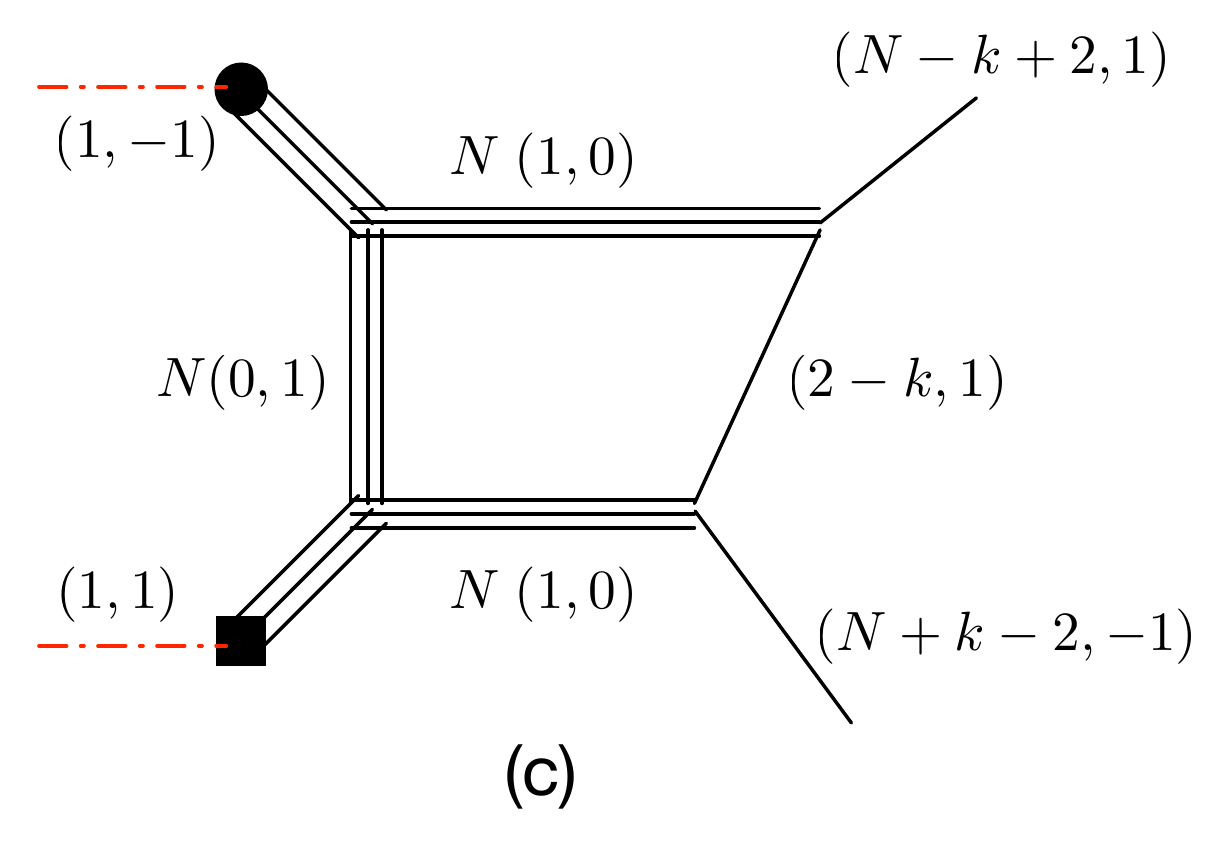} 
\caption{5-brane webs for $USp(2N)$} 
\label{USp2N}
\end{figure}

A classical 5-brane web configuration for a pure $USp(2N)$ theory (without an antisymmetric matter field) is shown in Fig.~\ref{USp2N}a.
This includes an orientifold 7-plane of type $O7^-$, which is parallel to the various 7-branes on which the external 
5-branes end.\footnote{That this web does not include an antisymmetric hypermultiplet is seen from 
the inability to separate the D5-branes along the orientifold plane.
On the other hand, for $N$ infinite D5-branes such a mode exists, and corresponds to the Higgs branch associated to
an antisymmetric hypermultiplet.}
This is essentially a 5d generalization of the 4d constructions in \cite{LL,EGKT,GK}.

In the figure we show the covering space, including two copies of the reduced web related by a reflection across the origin.
We have also included two copies of the cut associated with the monodromy of the $O7^-$ plane, $M_{O7^-} = -T^{-4}$, where
\be
T = \left(
\begin{array}{cc}
1 & 1 \\
0 & 1
\end{array}
\right) \,.
\ee
The physical space is the upper half plane, with the left and right halves of the cut identified.
The discontinuity of $(p,q)$5-brane charges across the cut corresponds to a clockwise action of the monodromy.
The bare 5d Yang-Mills coupling $g_0^{-2}$ is given by the separation along the cut.

We are still left with a choice of an integer $k$, corresponding to the D5-brane charge of one external 5-brane.
Naively the different choices are all related by $T\in SL(2,\mathbb{Z})$, which shifts $k \rightarrow k+1$,
and therefore all describe the same $USp(2N)$ gauge theory.
On the other hand, we know that the $USp(2N)$ theory admits a discrete theta parameter associated with $\pi_4(USp(2N)) = \mathbb{Z}_2$.
Apparently, the $O7^-$ plane is not invariant under $T$, but only under $T^2$.
The theta parameter is then related to the parity of $k$.
We will say more about this below.

%

%
%


At the quantum level, the $O7^-$ plane is resolved into a pair of mutually non-local $(p,q)$ 7-branes, whose combined monodromy
is given by that of the $O7^-$ plane \cite{Sen}.\footnote{The monodromy associated with a $(p,q)$7-brane is given by
\be
M_{p,q} = \left(
\begin{array}{rr}
1-pq & p^2 \\
-q^2 & 1+ pq
\end{array}
\right) \,.
\ee
}
There is some ambiguity in the choice of the $(p,q)$ charges of the two 7-branes.
In particular, the element $T\in SL(2,\mathbb{Z})$ transforms $(p,q)\rightarrow (p+q,q)$, but clearly
leaves the total monodromy $M_{O7^-} = -T^{-4}$ invariant.
One common choice for the $(p,q)$ charges of the 7-branes is $\{(2,1),(0,-1)\}$.
Another, related by $T$, is $\{(1,1),(1,-1)\}$.
We will now argue that these two resolutions correspond to physically distinct $O7^-$ planes,
and more generally that there are two variants of the $O7^-$ plane related by $T$.

%
Taking the 7-brane charges to be $\{(1,1),(1,-1)\}$, we get the brane web shown in Fig.~\ref{USp2N}b (in the reduced space).
Note the change in the 5-brane charges across the two cuts.
We can get a simpler presentation by moving the 7-branes outside.
Accounting for brane creation and for the effect of moving the cuts, we end up with the 5-brane web shown in Fig.~\ref{USp2N}c.
In this presentation the fact that the gauge group is $USp(2N)$, rather than say $SU(2N)$, is not immediately obvious.
It follows from the constraint imposed on the $N$-junctions by the s-rule.
We can now understand the connection between $k$ and the theta parameter.
Consider the $N=1$ case, namely the $USp(2)$, or $SU(2)$, theory. 
We identify the $k=1$ web with the $\theta = \pi$ theory, and the $k=2$ web with the $\theta = 0$ theory.
Furthermore, shifting $k\rightarrow k+2$ leaves $\theta$ invariant.

If we instead resolve the orientifold plane into 7-branes with charges $\{(2,1),(0,-1)\}$ the resulting web would be different.
Acting with $T$ brings it to the form of the web in Fig.~\ref{USp2N}c, but with $k\rightarrow k+1$.
This describes the $USp(2N)$ theory with the other value of $\theta$.
We conclude from this that there are two physically distinct variants of the $O7^-$ plane, one of which is
resolved to a 7-brane pair with charges $\{(1,1),(1,-1)\}$ up to an action of $T^{2n}$, and the other
to a pair with charges $\{(2,1),(0,-1)\}$ up to an action of $T^{2n}$.
It would be interesting to see this directly from the point of view of the orientifold.
This is presumably also related to the transition between the two values of $\theta$ in the Type I' brane construction
given in \cite{Bergman:2013ala}.

\subsection{Flavors}

\begin{figure}[h]
\center
\includegraphics[width=0.35\textwidth]{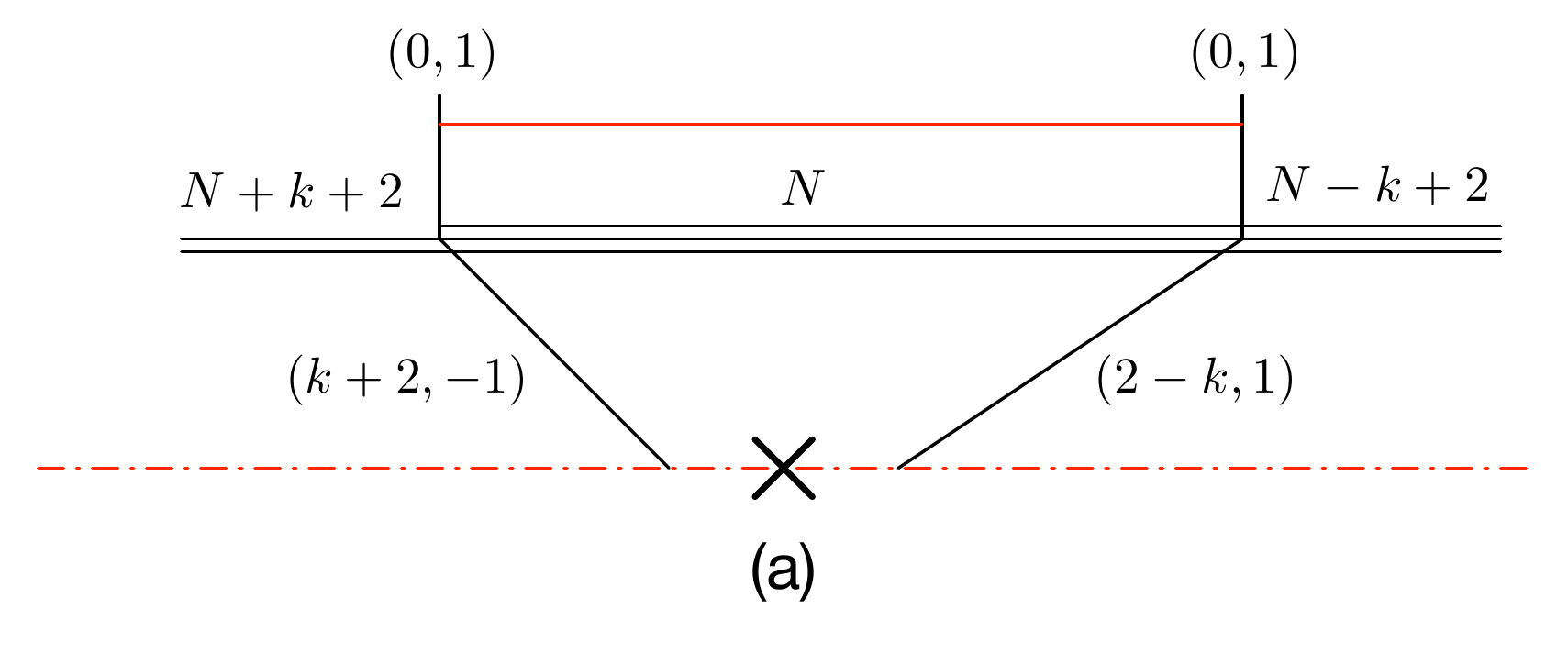} 
\hspace{1cm}
\includegraphics[width=0.35\textwidth]{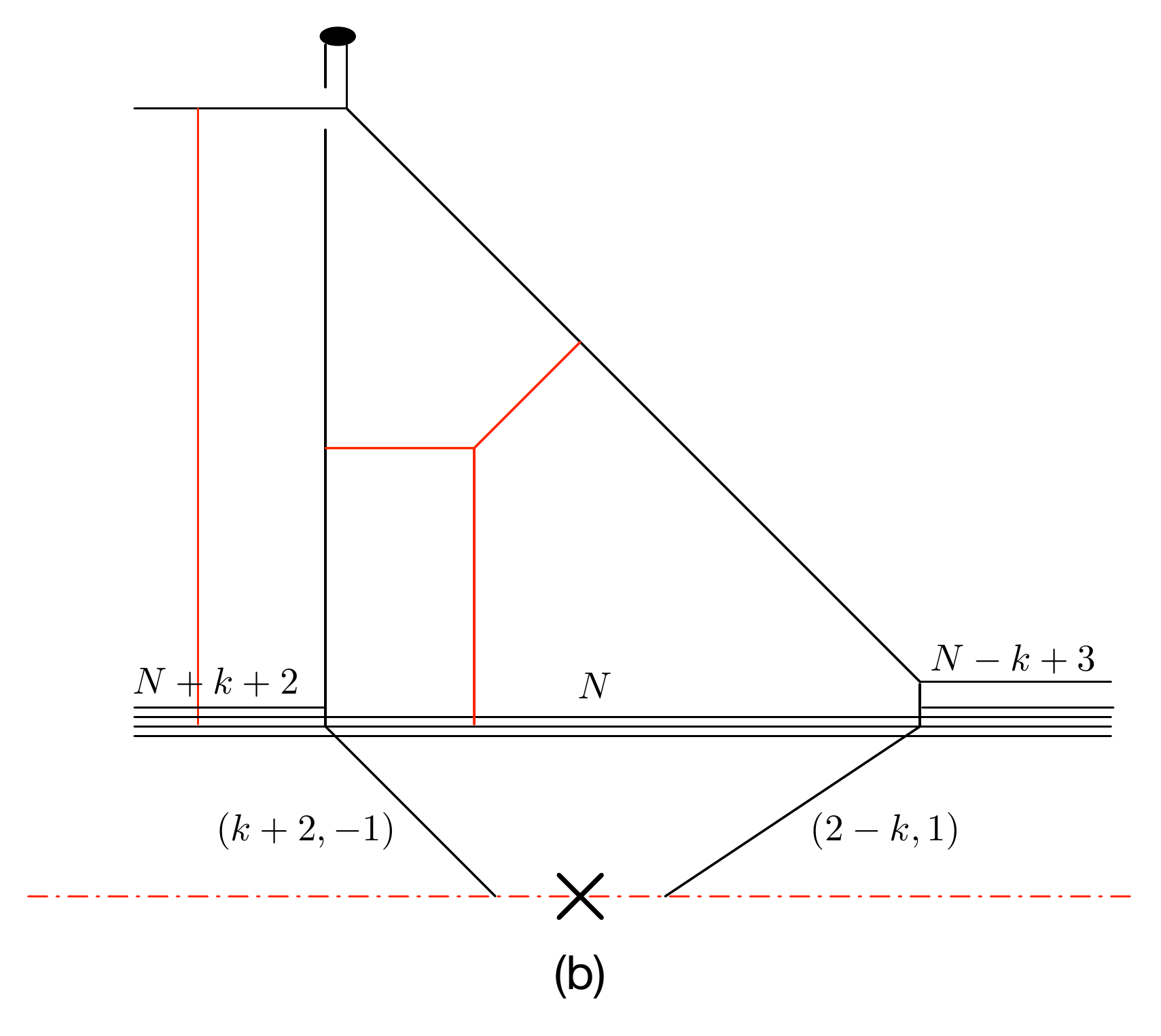}
\caption{$USp(2N)$ with $N_F=2N+4$ and $N_F=2N+5$.}
\label{USpflavors}
\end{figure}

Matter in the fundamental representation (flavor) can be added by attaching external D5-branes.
Requiring that external 5-branes do not intersect (which would lead to additional massless degrees of freedom)
leads to the condition that $N_F \leq 2N + 4$ (Fig.~\ref{USpflavors}a), in agreement with the classification of \cite{SMI}.
We claim however that also with $N_F = 2N +5$ one remains in the realm of consistent 5d theories.
In particular, the $N=1$ case is the rank one $E_8$ theory.\footnote{The higher rank $E_8$ theories correspond to
$USp(2N)$ with $N_F = 7$ and an antisymmetric hypermultiplet. For $N=1$ the antisymmetric field is a singlet and decouples.}
The 5-brane web for $N_F = 2N+5$ is shown in Fig.~\ref{USpflavors}b.
The dangerous intersection is avoided as a consequence of the s-rule.
This is similar to the situation for $SU(N)$ with $N_F = 2N+1$ \cite{BZ},
which is also one more flavor than allowed by \cite{SMI}.

As before, the $O7^-$-plane is resolved quantum mechanically into a pair of 7-branes,
and one can obtain an alternative 5-brane web realization of the theories by Hanany-Witten transitions.
In particular for $N=1$ we get the familiar 5-brane webs for $SU(2)$ with $N_F = 1,2,3$ and 4 flavors,
as well as those describing $N_F = 5,6$ and 7 flavors (the $E_6, E_7$ and $E_8$ theories) \cite{BB}.

The 5-brane webs shown in Fig.~\ref{USpflavors}  suggest that the global symmetry is enhanced at the fixed points in these cases.
For $N_F=2N+4$ the classical global symmetry of the IR gauge theory is $SO(4N+8)_F\times U(1)_I$.
The pair of parallel external legs suggests that the $U(1)_I$ factor is enhanced to $SU(2)$, 
like in the case of $SU(N)_0$ with $N_F=2N$ \cite{BGZ}.
For $N_F=2N+5$ on the other hand, the parallel external legs, D5-branes in this case, suggest that the classical
global symmetry $SO(4N+10)_F\times U(1)_I$ is enhanced to $SO(4N+12)$.

This can be seen explicitly in the 5d superconformal index, 
the key ingredient of which is the contribution of instanton states corresponding to the 5d lift of the multi-instanton partition function \cite{KKL}.
In the above two cases the instanton computation exhibits the same kind of pathologies 
associated with extraneous ``decoupled" states that were encountered in other cases in \cite{BGZ,BZ}, 
such as for $SU(N)$ with $N_F = 2N$ and $N_F= 2N+1$.
The relevant states in the present cases are shown in red in Fig.~\ref{USpflavors}.
For $N_F=2N+4$ it is a D-string between the external parallel NS5-branes,
and for $N_F = 2N +5$ there is a fundamental string between the flavor D5-branes (or their images)
and the extra external D5-brane, and a 3-pronged string attached to the color D5-branes (or their images).
Note that all of these carry two units of instanton charge.\footnote{The instanton charge of the fundamental string state can be seen
by the dependence of its mass on the bare YM coupling, {\em i.e.}, the horizontal separation.}
This follows from the fact that the D-string worldvolume gauge symmetry is $O(2)$, 
which is the ADHM dual group for two instantons of $USp(2N)$.
The minimally instanton-charged object corresponds to 
a fractional D-string that intersects the O7-plane, and cannot move away from it. 
The string states in the $N_F=2N+5$ web are also charged under other symmetries.
The fundamental string state is in the vector representation of $SO(4N+10)_F$, and the 3-pronged string
is in the fundamental representation of $USp(2N)$.

The contribution of these states to the 2-instanton term then has the form
\be
\Delta Z_{2} = \mbox{} - \frac{x^2}{(1-x y)(1-\frac{x}{y})} \, \left[ \mbox{charge factor}   \right] \,,
\label{subtraction}
\ee
where $x$ and $y$ are the fugacities associated with the Cartans of $SO(4)\subset SO(5,2)$.
This must be subtracted from the 2-instanton partition function to obtain a consistent 
result (see Appendix for details).
In the case of $N_F = 2N +4$ we expect the subtraction to plethystically exponentiate to a correction factor
for the full multi-instanton partition function\footnote{In the case of $SU(2)+6F$, such a correction was also noticed in \cite{HKKP}, and for $USp(4)+8F$ in \cite{GC}. }
\be
\mathcal{Z}_c = PE\left[\frac{x^2 q^2}{(1-x y)(1-\frac{x}{y})}\right] \mathcal{Z} \,.
\label{multi-instanton}
\ee
We verify this up to the 4-instanton level in the Appendix.
In the $N_F=2N+5$ case, the presence of the gauge-charged state complicates the counting,
and we can only perform the subtraction at the basic 2-instanton level.

Combining the corrected instanton partition functions with the perturbative contributions from the gauge and matter supermultiplets we find 
that the superconformal indices are given by:
\be
I^{N_F=2N+4} = 1 + x^2 (1+\chi_{\bf Ad}^{SO(4N+8)} + q^2 + \frac{1}{q^2}) + O(x^3) \,,
\ee
and 
\be
I^{N_F=2N+5} =  1 + x^2 (1+\chi_{\bf Ad}^{SO(4N+10)} + (q^2 + \frac{1}{q^2})\chi_{\bf 4N+10}^{SO(4N+10)}) + O(x^3) \,.
\ee
The $x^2$ terms correspond to the contributions of conserved current multiplets.
We see that the 2-instanton states provide additional charged currents.
For $N_F=2N+4$ these lead to an enhancement of $U(1)_I$ to $SU(2)$, and for $N_F=2N+5$, of $SO(4N+10)_F\times U(1)_I$ to $SO(4N+12)$.

\subsection{Duality} 
\label{USpduality}

Since we do not have a simple (perturbative) description for the S-dual of the O7-plane, we cannot identify
the S-dual gauge theory directly in these cases.
However for $N=1$ we already know the answer in some cases. The $USp(2)=SU(2)$ theory with $N_F\leq 7$ is a self-dual gauge theory
description of the $E_{N_F+1}$ fixed point theory \cite{MPTY}.
We can use this to determine the duals in higher rank cases.

Let us consider 
the specific class of theories with gauge group $USp(2N)$ and $N_F=2N+2$ fundamental hypermultiplets.
This is an interesting class of examples, since it is related by dimensional reduction, as in \cite{BZ}, to a duality in four dimensions.

\begin{figure}
\center
\includegraphics[width=0.3\textwidth]{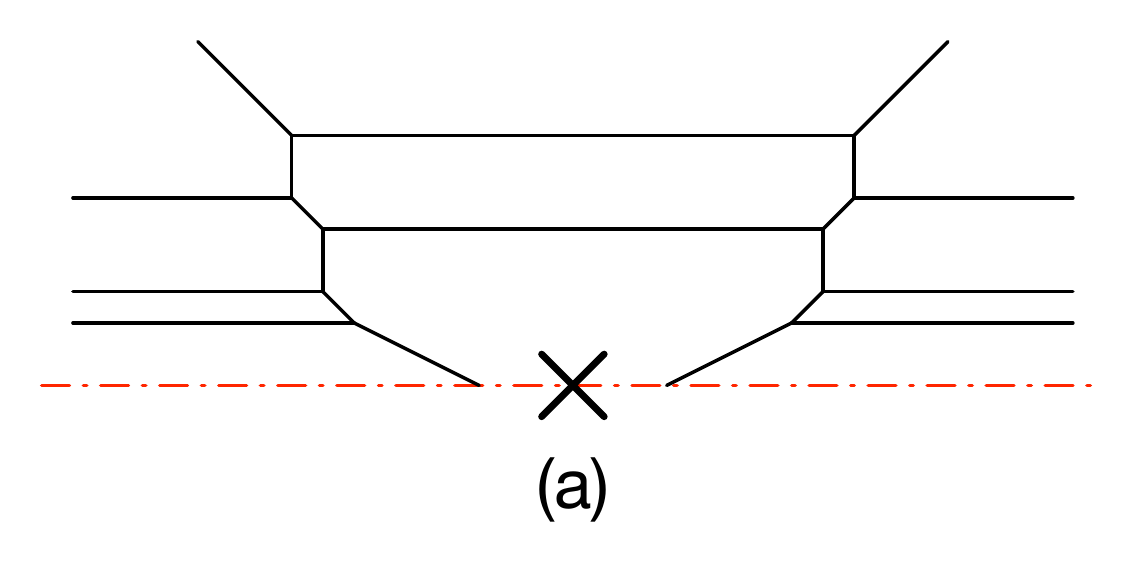} 
\hspace{1cm}
\includegraphics[width=0.2\textwidth]{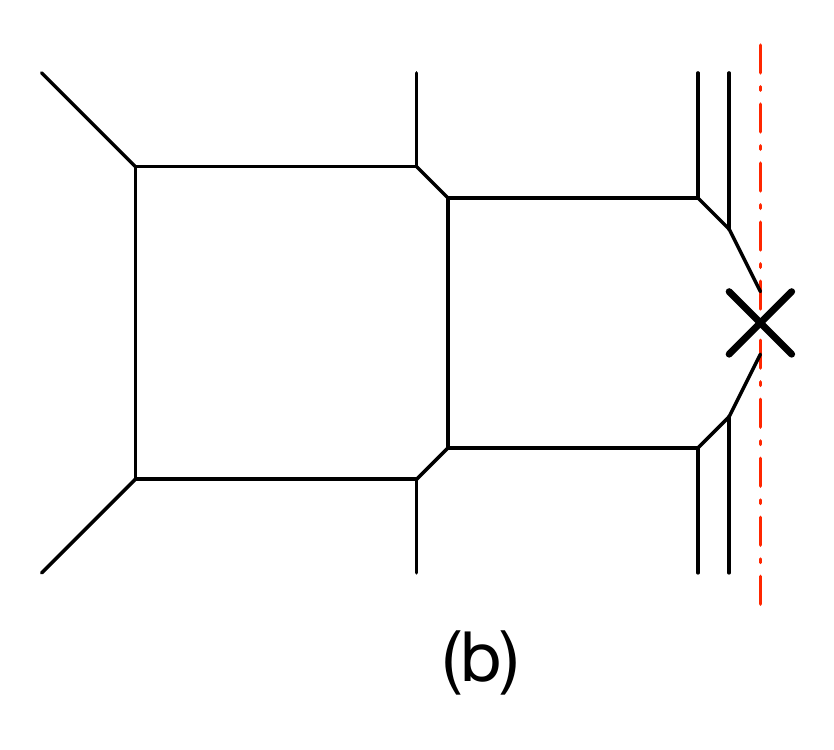} 
\caption{The orientifold 5-brane web for $USp(4)+6$ and its S-dual ($k=0$).}
\label{USp(4)+6}
\end{figure}

\begin{figure}
\center
\includegraphics[width=0.2\textwidth]{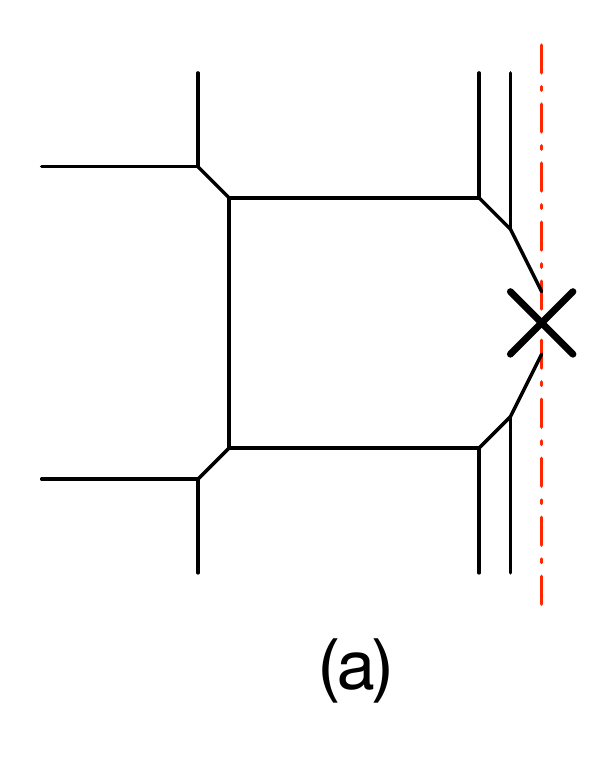} 
\hspace{1cm}
\includegraphics[width=0.3\textwidth]{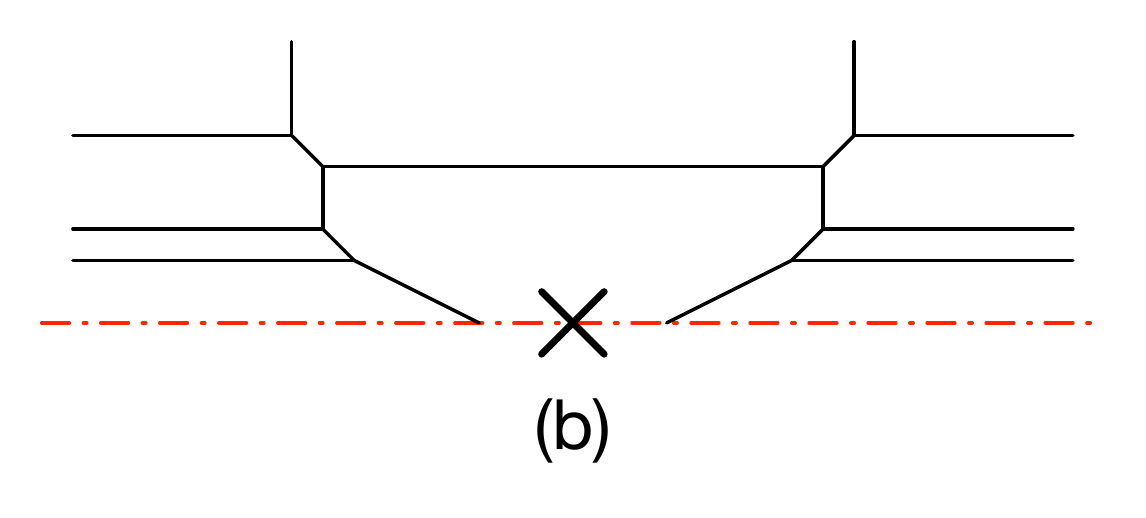} 
\caption{The orientifold 5-brane web for $SU(2)+6$ and its S-dual, which also describes $SU(2)+6$ (we have taken $k=0$).}
\label{SU(2)+6}
\end{figure}

%
The simplest interesting case is $USp(4)$ with $N_F=6$.
The orientifold 5-brane web for this theory is shown in Fig.~\ref{USp(4)+6}a.
The S-dual web, Fig.~\ref{USp(4)+6}b, corresponds to the same UV fixed point, but describes a different IR gauge theory.
The dual theory appears to be a linear quiver with gauge group $SU(2)\times SU(2)$, but the matter content is not obvious.
We can identify it more precisely by ``ungauging" the first $SU(2)$ factor, leading to the web in Fig.~\ref{SU(2)+6}a.
The S-dual of this web, Fig.~\ref{SU(2)+6}b, corresponds to $SU(2)$ with $N_F=6$. Since this theory is self-dual,
the original web also describes $SU(2)$ with $N_F=6$, albeit with only an 
$SU(2)\subset SO(12)_F$
exhibited manifestly.
The dual of $USp(4)$ with $N_F=6$, Fig.~\ref{USp(4)+6}b, therefore corresponds to a gauging of
$SU(2)\subset SO(12)_F$ in the $SU(2), N_F=6$ theory.
The resulting gauge theory is the linear quiver theory $SU(2)_\pi \times SU(2) + 4$.
The remaining matter global symmetry is $SO(8)_F\times SU(2)_{BF}$, where $SU(2)_{BF}$ is associated
to the bi-fundamental hypermultiplet.\footnote{The non-trivial $\theta$ parameter of the unflavored $SU(2)$ factor
can be demonstrated by mass-deforming the web along the lines described in \cite{BGZ,BZ}.}
We can therefore describe the 5d SCFT corresponding to the webs of Fig.~\ref{USp(4)+6} as gauging an $SU(2)\subset E_7$
in the rank one $E_7$ theory and flowing to the UV.
The dual gauge theory descriptions of the resulting rank two SCFT correspond to different embeddings of $SU(2)$
in $E_7$, in one case leaving $SO(12)_F$ and in the other $SO(8)_F\times SU(2)_{BF}$.

This is actually the 5d lift of the second 4d ${\cal N}=2$ duality of Argyres and Seiberg \cite{AS},
which relates the strong coupling limit of the 4d $USp(4)+6$ superconformal gauge theory to 
the weak gauging of an $SU(2)\subset E_7$ in the Minahan-Nemeschansky $E_7$ theory \cite{Minahan:1996fg}.

\begin{figure}
\center
\includegraphics[width=0.3\textwidth]{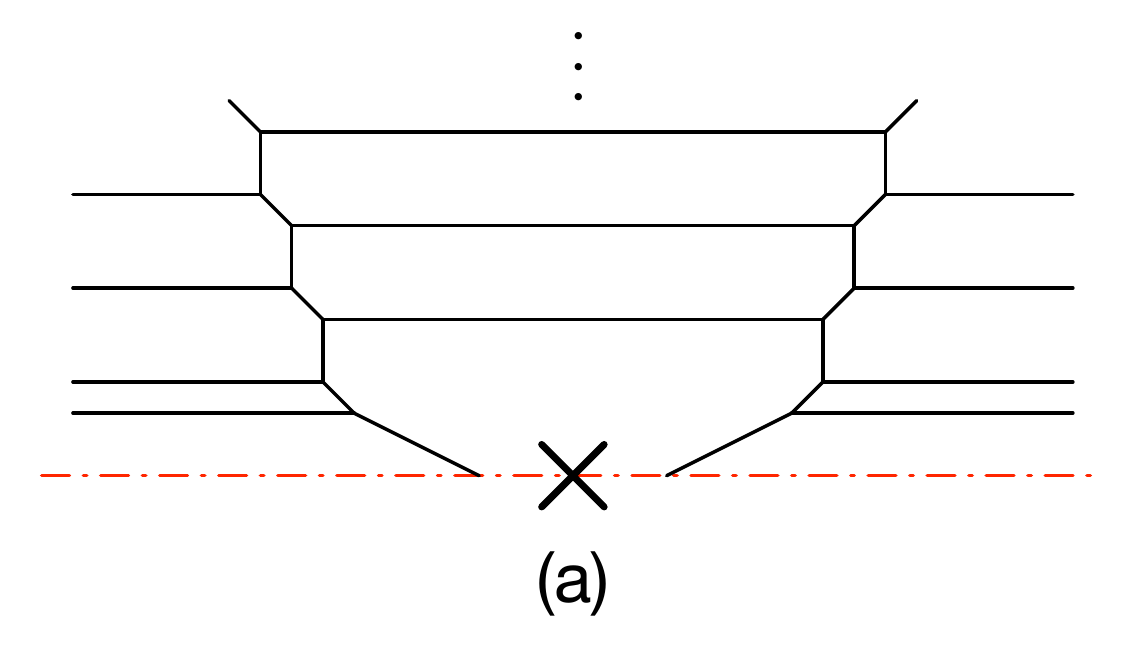} 
\hspace{1cm}
\includegraphics[width=0.3\textwidth]{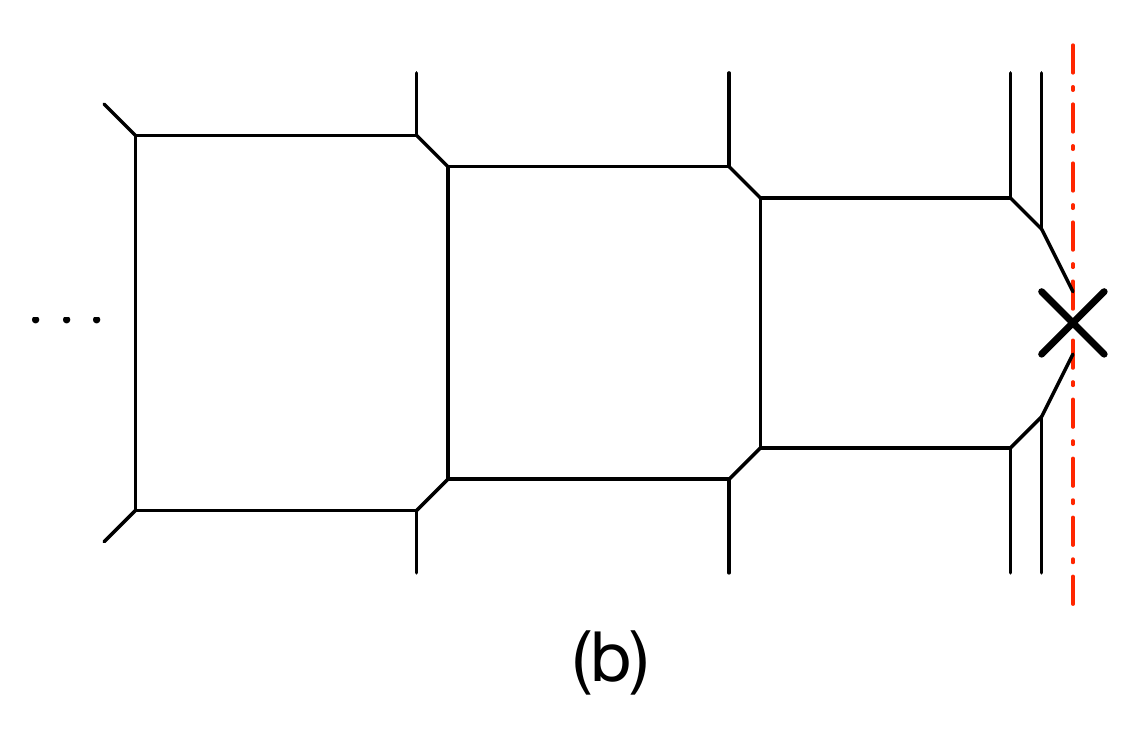} 
\caption{The orientifold 5-brane web for $USp(2N)+2N+2$ and its S-dual ($k=0$).}
\label{USp(2N)+2N+2}
\end{figure}

Generalizing to $USp(2N)$ with $N_F = 2N+2$ is straightforward
(see Fig.~\ref{USp(2N)+2N+2}).
The S-dual web corresponds to 
the linear quiver theory $SU(2)_\pi \times SU(2)_0^{N-2} \times SU(2) +4$.
The shared UV fixed point is a 5d rank $N$ SCFT with a $(2N+3)$-dimensional space of mass parameters,
and corresponds to gauging an $SU(2)$ inside a particular 5d rank $N-1$ SCFT.
The latter has dual IR descriptions as the quiver theory $2+SU(2)_0^{N-2}\times SU(2)+4$, or as
a $USp(2N-2)$ theory with $N_F= 2N+2$.

\begin{figure}
\center
\includegraphics[width=0.3\textwidth]{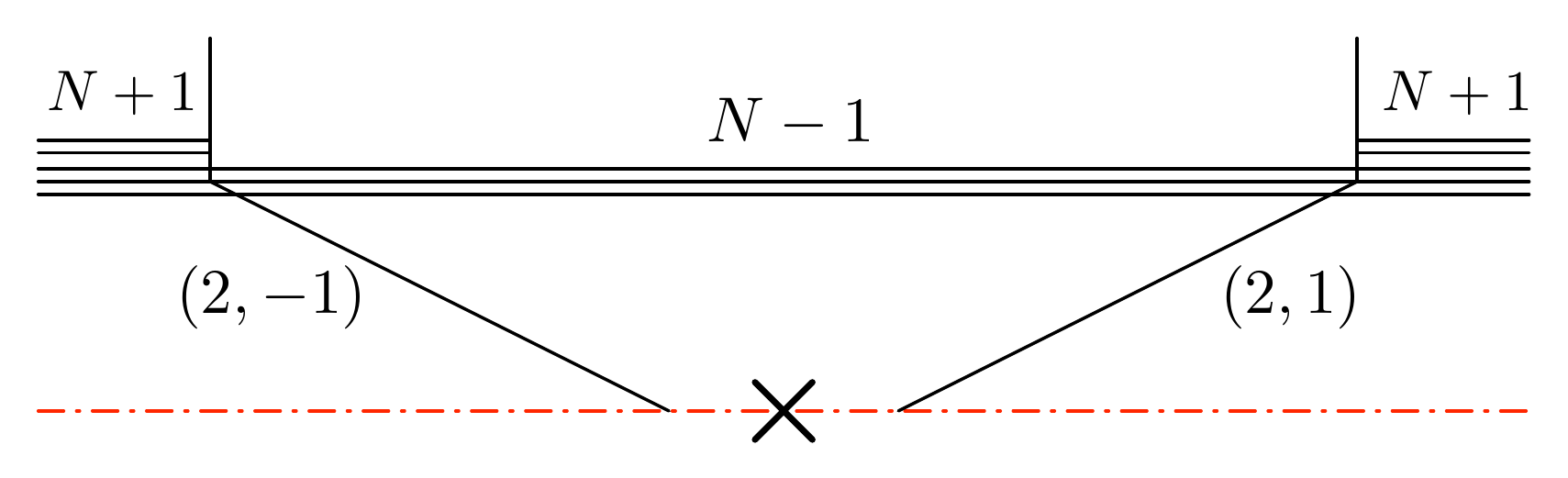} 
\hspace{0.5cm}
\includegraphics[width=0.3\textwidth]{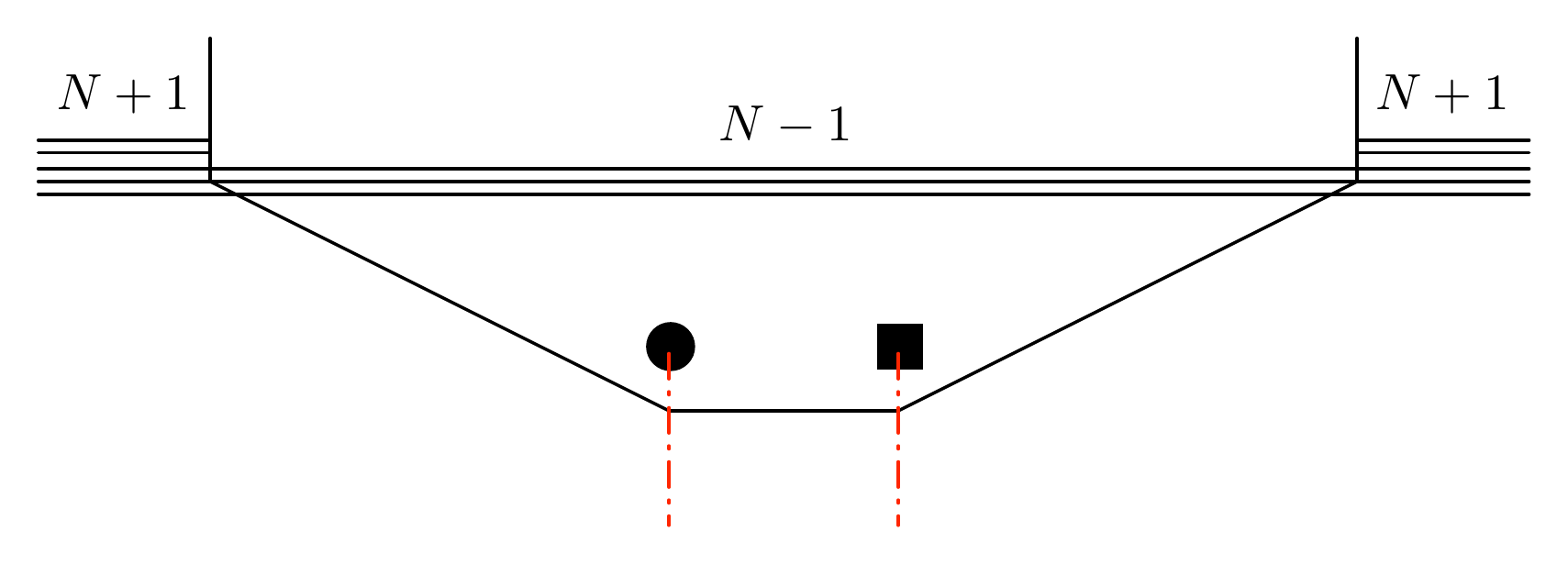}
\hspace{0.5cm}
\includegraphics[width=0.25\textwidth]{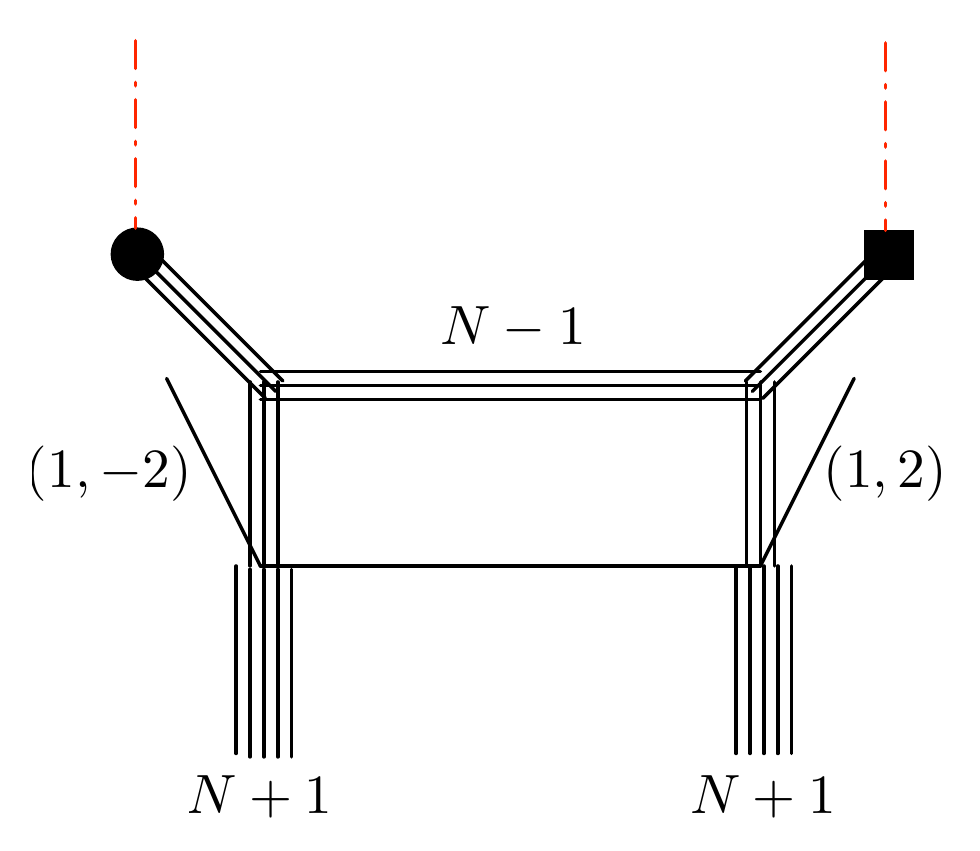}
\hspace{0.5cm}
\includegraphics[width=0.35\textwidth]{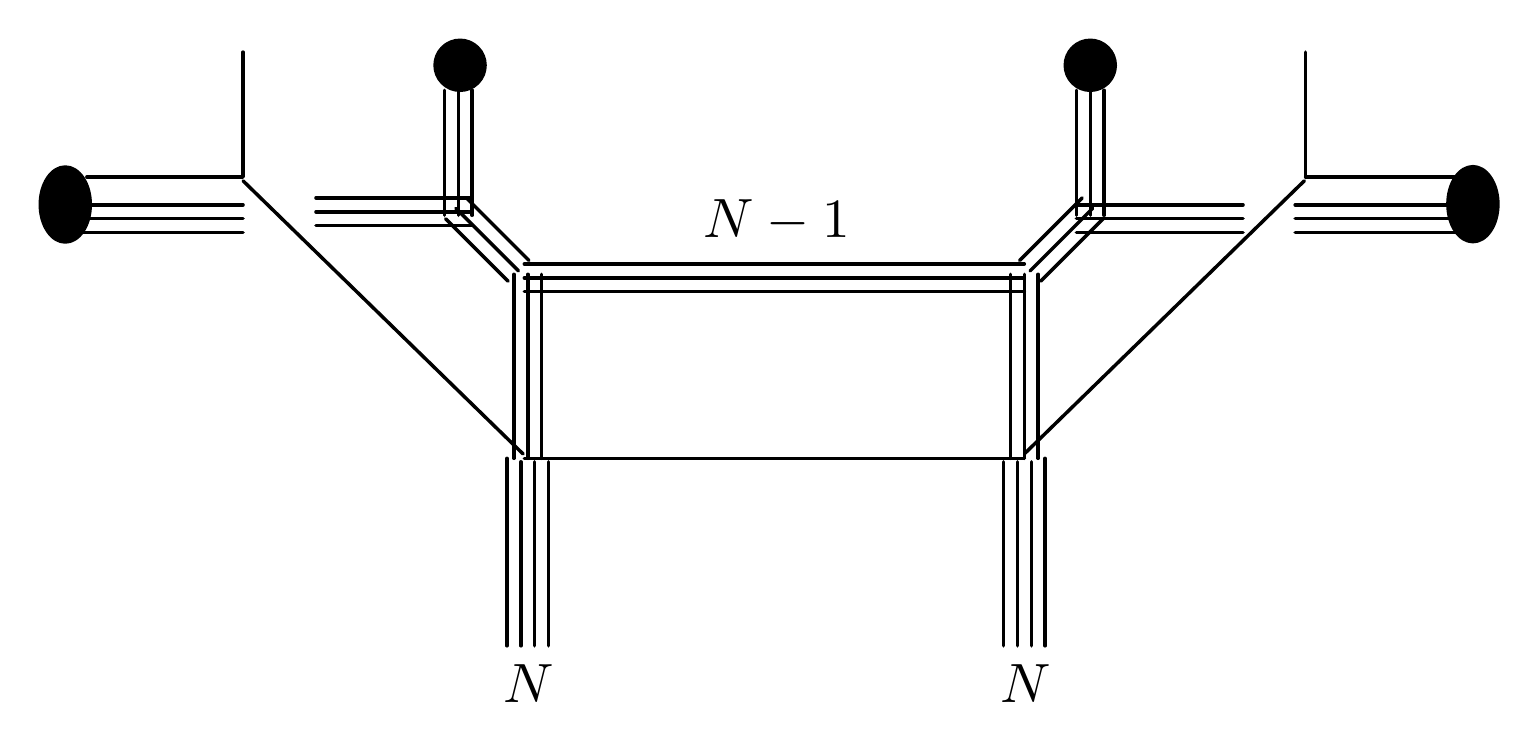}
\hspace{0.5cm}
\includegraphics[width=0.23\textwidth]{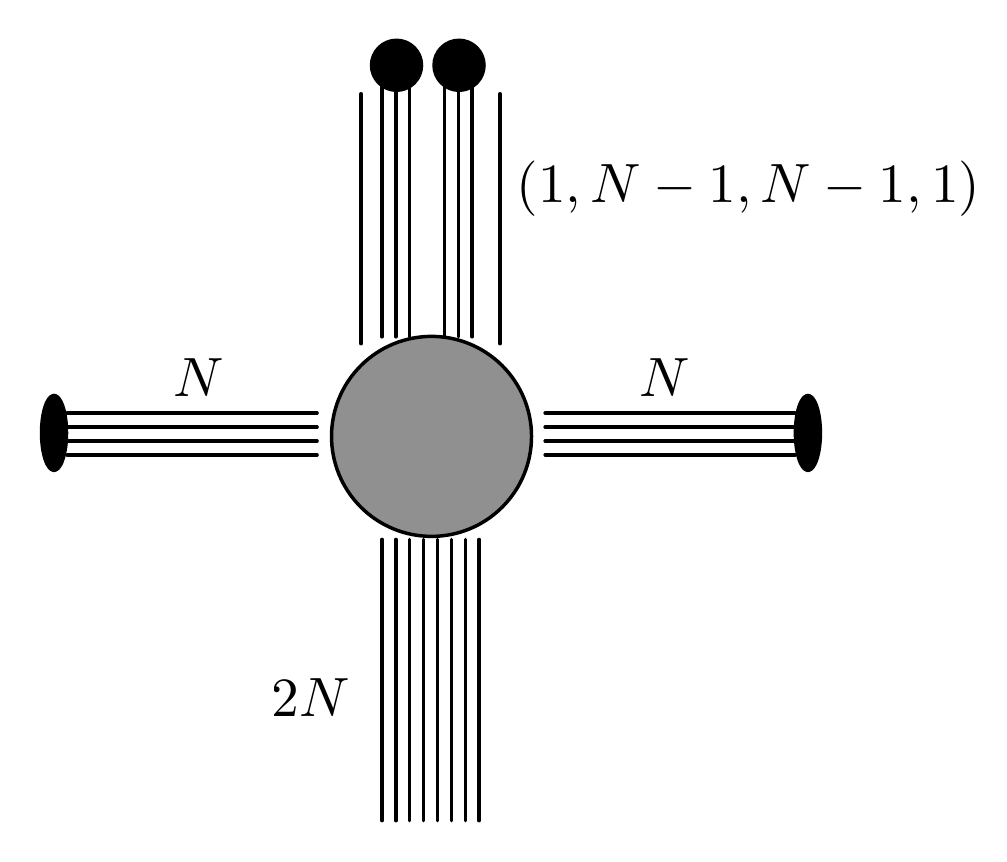}
\hspace{0.5cm}
\includegraphics[width=0.23\textwidth]{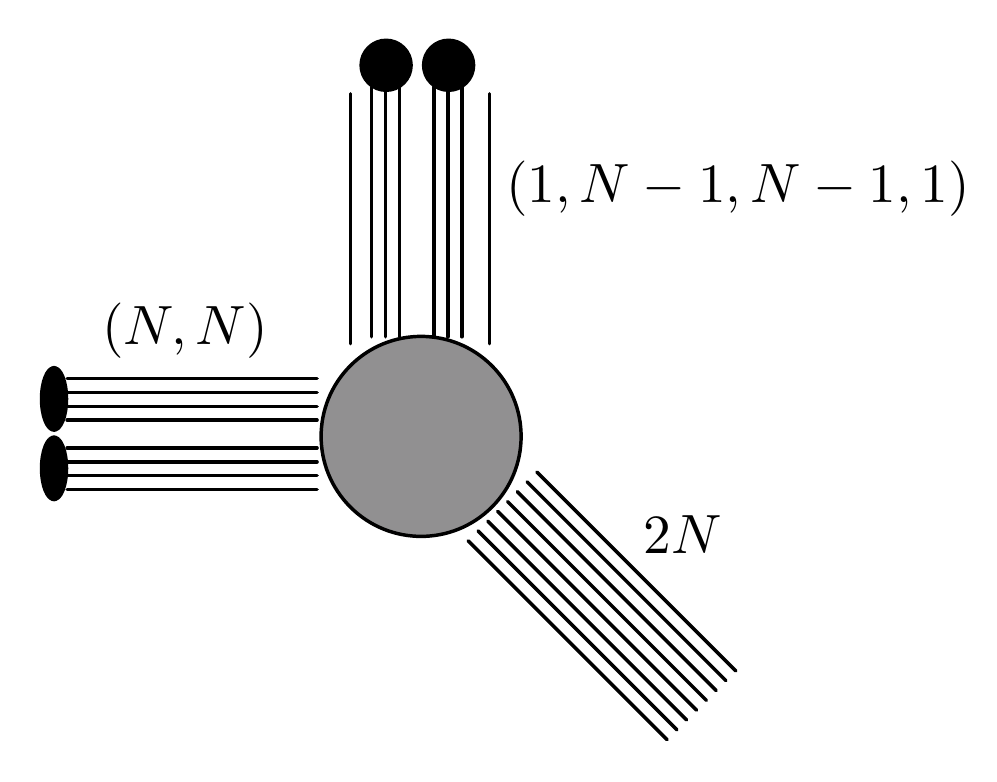}
\caption{Manipulating the 5-brane web for $USp(2N-2) + 2N+2$ via various HW transitions.}
\label{USp(2N-2)+2N+2}
\end{figure}

The reduction to 4d then gives a duality between the 4d ${\cal N}=2$ superconformal gauge theory with gauge group $USp(2N)$ and
$N_F=2N+2$,
and an $SU(2)$ gauging of a particular isolated 4d ${\cal N}=2$ SCFT given by the dimensional reduction of the above 5d rank $N-1$ SCFT.
We can identify the 4d SCFT according to the classification of \cite{Gaiotto:2009we} in terms of a 3-punctured sphere with a specific set
of punctures, by manipulating the 5-brane web to a standard triple 5-brane junction form, as shown in Fig.~\ref{USp(2N-2)+2N+2}.
This represents the theory as a limit of the $T_{2N}$ theory, in which two of the maximal punctures (corresponding to the fully
symmetrrized $2N$-box Young tableau) are replaced by non-maximal punctures;
an $(N,N)$ puncture (corresponding to a Young diagram with two columns of $N$ boxes), and an $(N-1,N-1,1,1)$ puncture.
This 4d duality first appeared in \cite{Tachi}.

\section{$O7^+$ and $SO$ gauge theories}

\begin{figure}[h]
\center
\includegraphics[width=0.35\textwidth]{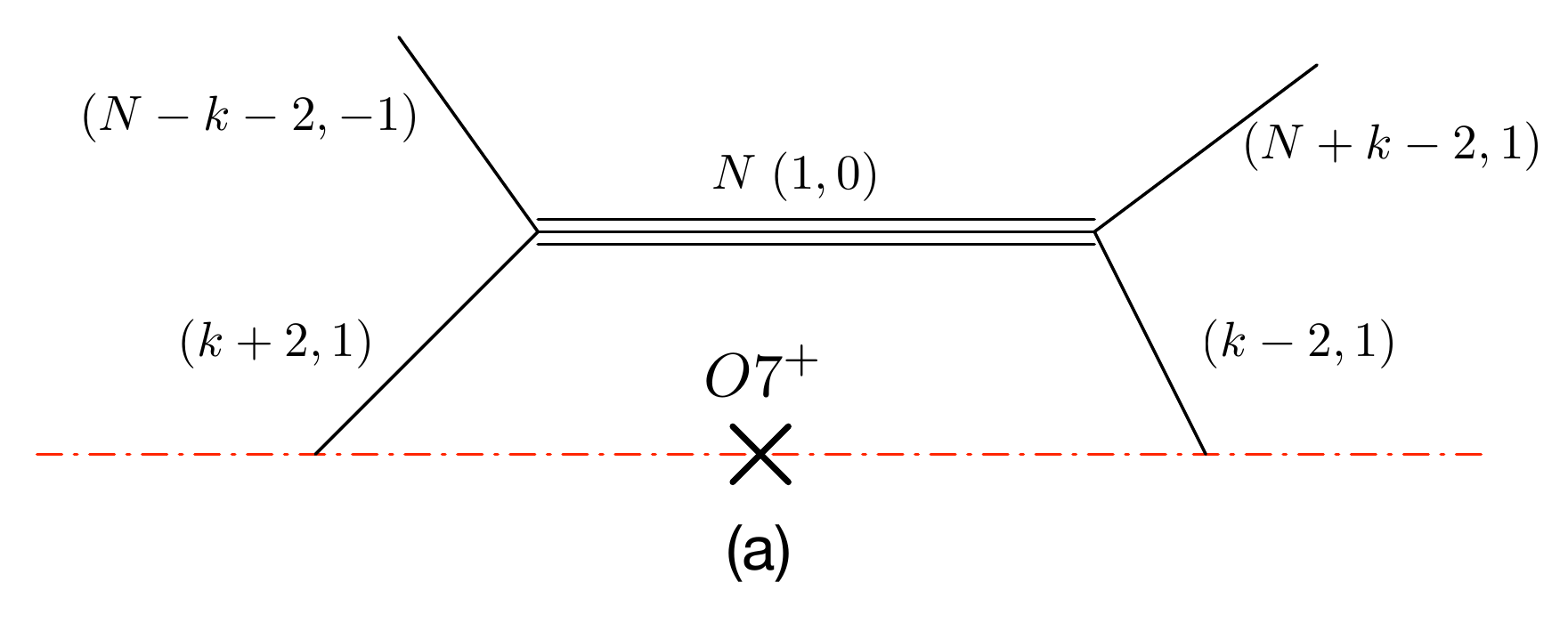} 
\hspace{1cm}
\includegraphics[width=0.35\textwidth]{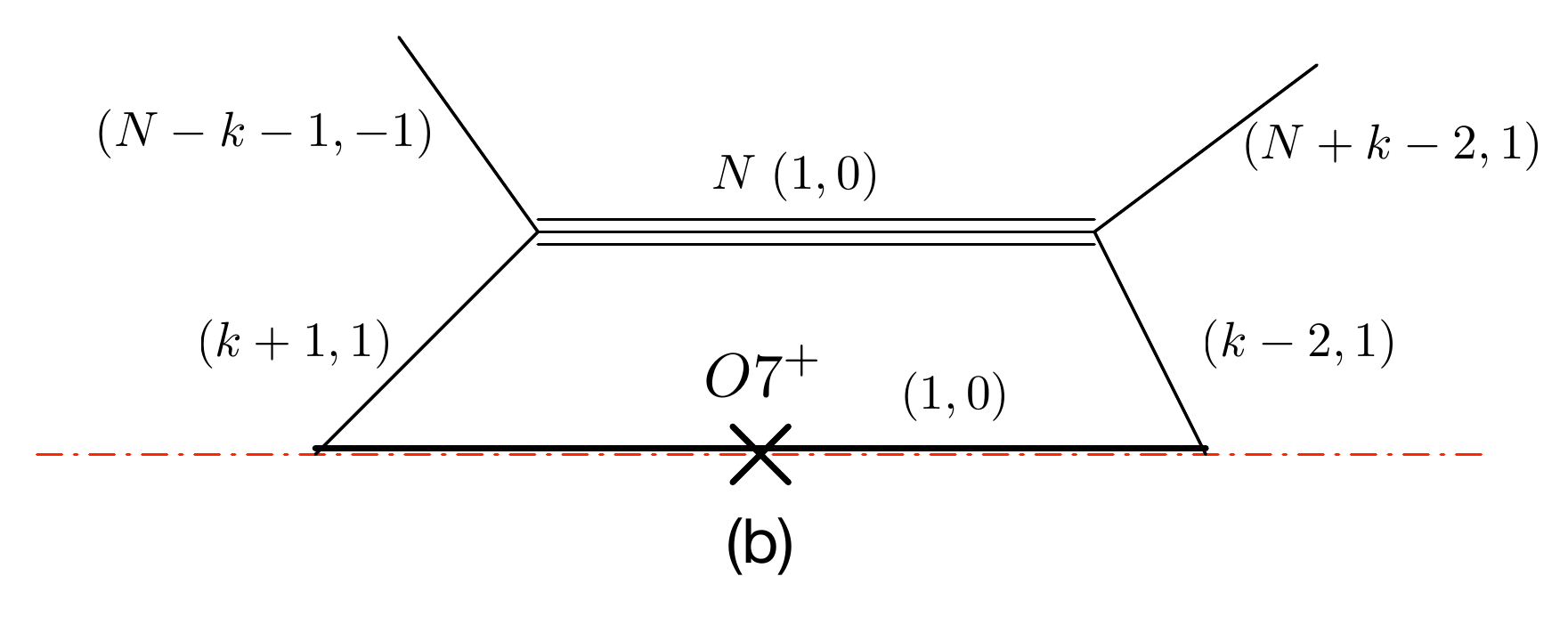} 
\caption{The orientifold 5-brane webs for $SO(2N)$ and $SO(2N+1)$.}
\label{SO}
\end{figure}

Replacing the $O7^-$ plane with an $O7^+$ plane, we obtain a 5-brane web for an $SO(2N)$ gauge theory.
The monodromy matrix of the $O7^+$ plane is given by $M_{O7^+} = - T^{4}$.
The 5-brane web for the pure $SO(2N)$ theory is shown in Fig.~\ref{SO}a.
This is the exact quantum configuration; unlike the $O7^-$-plane, the $O7^+$-plane is not resolved into 7-branes.
As before, $k\rightarrow k +1$ under $T$. 
Since there is no additional parameter in the $SO(2N)$ theory, the web must describe the same theory for all $k$.\footnote{For 
$SO(4) \sim SU(2)\times SU(2)$ and $SO(6) \sim SU(4)$ there are discrete choices ($\theta$ parameters in the former
and CS level in the latter), and it is possible that the different webs describe different theories.}
This implies that, unlike the $O7^-$ plane, the $O7^+$ plane must be invariant under $T$.
A slight generalization of this web including a fractional D5-brane gives the pure $SO(2N+1)$ theory (Fig.~\ref{SO}b).

\subsection{Flavors}

\begin{figure}[h]
\center
\includegraphics[width=0.3\textwidth]{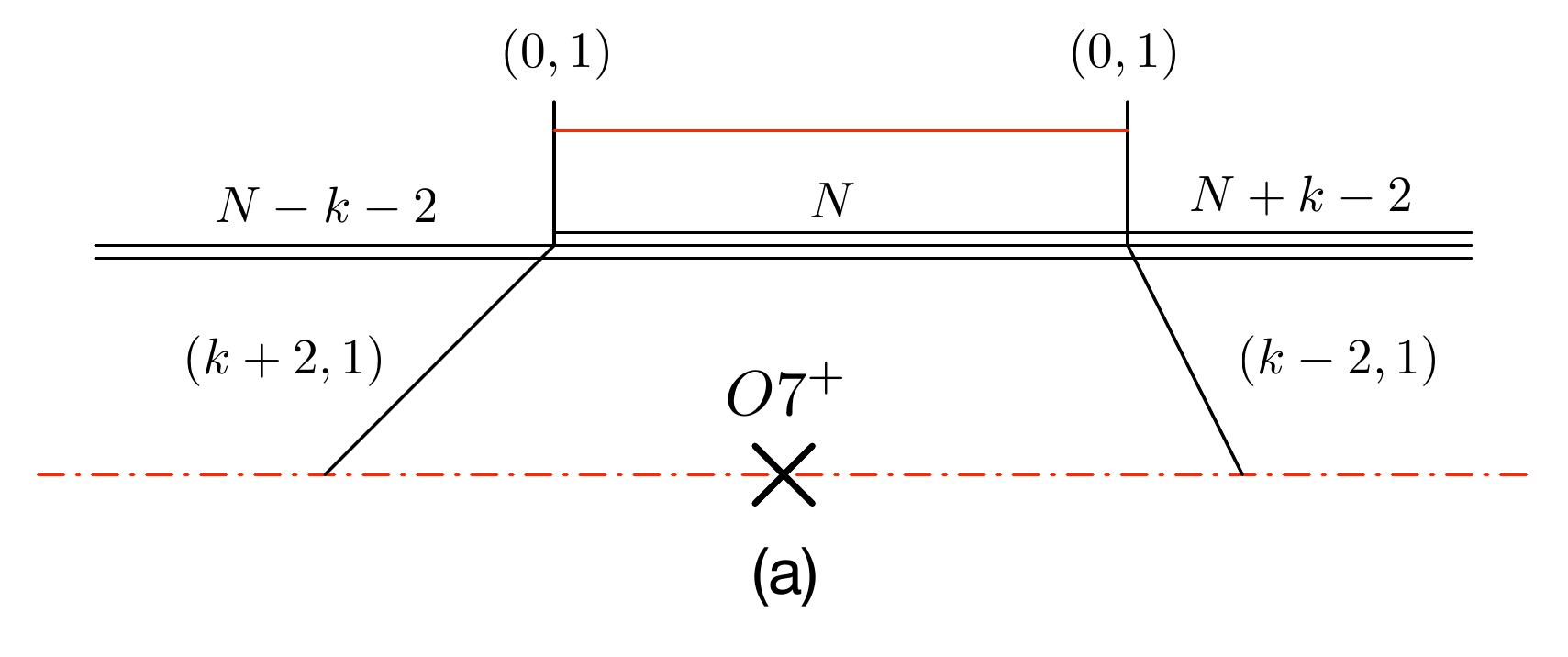} 
\hspace{1cm}
\includegraphics[width=0.3\textwidth]{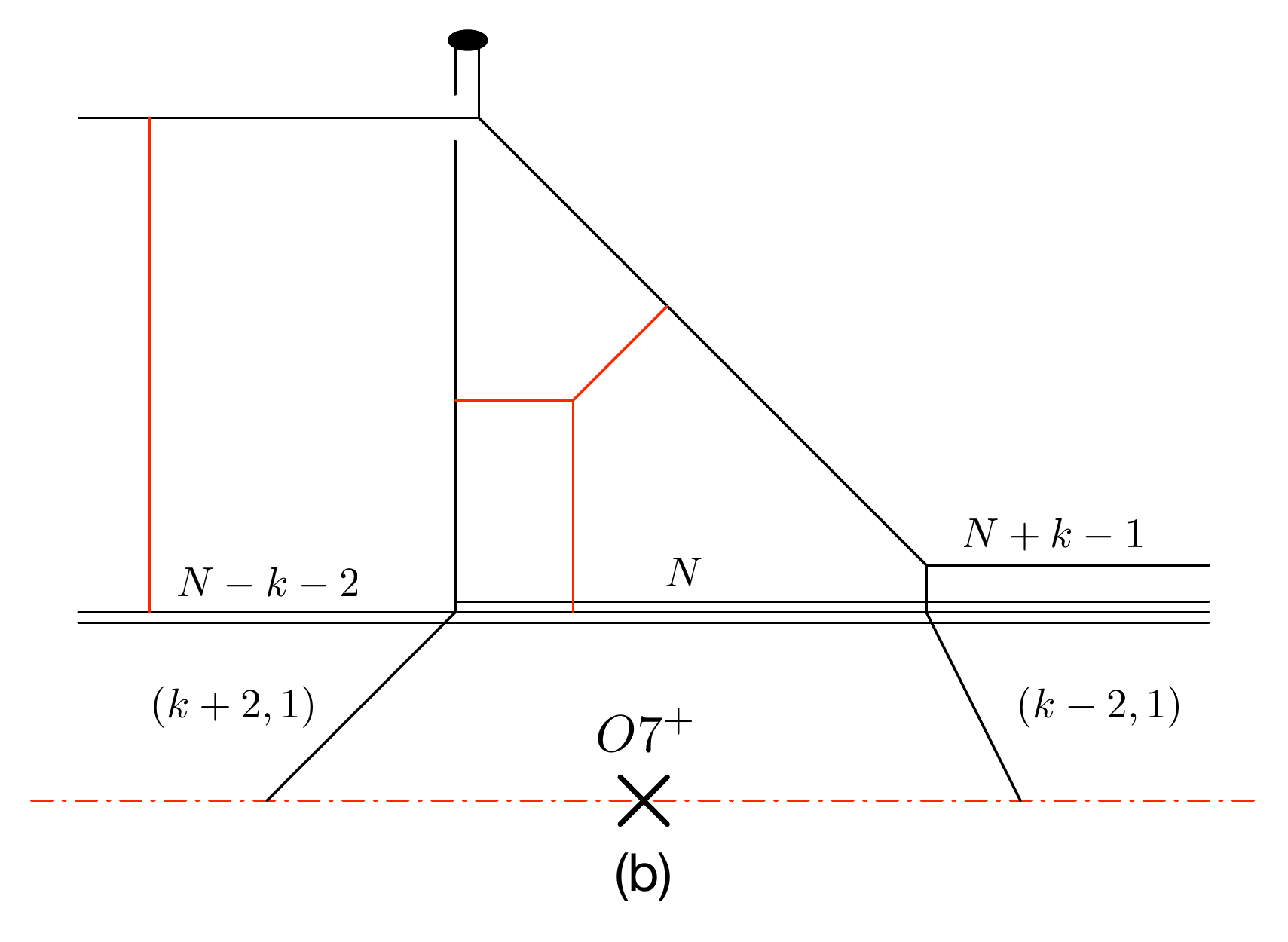} 
\caption{$SO(2N)$ with $N_V=2N-4$ and $N_V=2N-3$.}
\label{SO(2N)flavors}
\end{figure}

The condition in \cite{SMI} for an $SO(M)$ gauge theory to come from a 5d fixed point is $N_V\leq M-4$.
In the 5-brane webs this is again seen as the condition of no intersections. 
But, as in other cases, the 5-brane web construction appears to imply that one additional flavor is allowed.
The 5-brane webs for $SO(2N)$ with $N_V=2N-4$ and $N_V=2N-3$ are shown in Fig.~\ref{SO(2N)flavors}.
The cases of $SO(2N+1)$ with $N_V=2N-3$ and $N_V=2N-2$, respectively, are very similar.

As before, the 5-brane webs in these cases exhibit extraneous instanton-charged states (shown in red in Fig.~\ref{SO(2N)flavors})
that are not part of the 5d gauge theories, and whose contribution to the instanton partition function must be removed by hand.
Here they carry the minimal instanton charge, since there is no fractional BPS D-string.
This can also be understood from the fact that the D-string worldvolume gauge symmetry is $SU(2)$, 
which is the ADHM dual gauge group for one instanton of $SO(M)$.

For $SO(M)$ with $N_V=M-4$,
the proposed correction factor is identical to the one in (\ref{multi-instanton}), except with $q^2$ replaced by $q$.
This is verified in the Appendix to the 2-instanton level. 
In special cases one can also compare with other approaches.
For $SO(4)$ we compared with the instanton partition function of $SU(2)\times SU(2)$ (without matter fields) \cite{KKL}, 
finding complete agreement.
We have also compared the result for $SO(5)$ with $N_V=1$ with that of $USp(4)$ with $N_A=1$ \cite{KKL}, again finding complete agreement.
Going beyond the 2-instanton level is quite difficult. 
Taking into account the correction factor, the superconformal index takes the form, 
\be 
I^{N_V=M-4} = 1 + x^2(1+\chi_{\bf Ad}^{USp(2M-8)}+q+\frac{1}{q}) + O(x^3)
\ee
exhibiting the enhancement of $U(1)_I$ to $SU(2)$.

For $N_V=M-3$, the subtraction can only be carried out at the one-instanton level, and has the form of (\ref{subtraction}),
with the fundamental string state contributing charges in the fundamental representation of the $USp(2M-6)$ flavor symmetry,
and the 3-pronged string contributing charges in the vector representation of the $SO(M)$ gauge symmetry.
The analysis for higher instanton level is made complicated by this gauge charge.
See the Appendix for details.
For $SO(4)$ with $N_V=1$, we compared with the 1-instanton partition function for 
$SU(2)\times SU(2)$ with a bi-fundamental \cite{BGZ},
and find complete agreement.
Including the corrected 1-instanton contribution, the superconformal index is given by

\be
I^{N_V=M-3} = 1 + x^2 \left(1+\chi_{\bf Ad}^{USp(2M-6)} + (q + \frac{1}{q})\chi_{\bf 2M-6}^{USp(2M-6)}\right) + O(x^3)
\ee
The $x^2$ terms in the fundamental representation of $USp(4N-6)_F$ point to an enhancement 
of $USp(4N-6)_F\times U(1)_I$ to $USp(4N-4)$.
But this also requires the existence of flavor-singlet currents carrying two units of instanton charge, 
which we are currently unable to demonstrate.

Both of these enhancements agree with the results of \cite{Zaf1} using simplified instanton analysis.

\subsection{Duality}


As before, we can ask whether the continuation past infinite coupling yields a dual gauge theory.
In the case of the $O7^+$ plane we cannot describe the S-dual webs in terms of gauge theories, since we do
not have a simple case in which there is a known dual gauge theory description.
However we may still be able to provide an alternative formulation of the theory in terms of gauging 
a subgroup of the global symmetry in a lower rank SCFT, which in turn has an IR gauge theory description.

\begin{figure}[h]
\center
\includegraphics[width=0.3\textwidth]{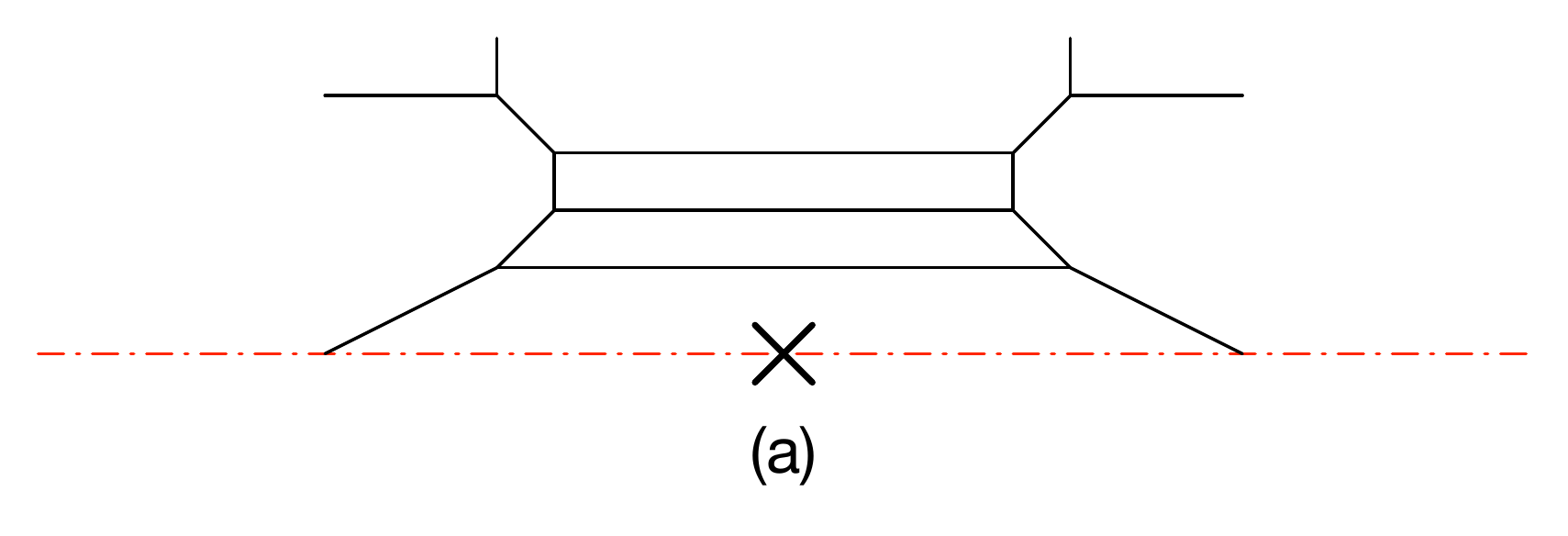} 
\hspace{1cm}
\includegraphics[width=0.3\textwidth]{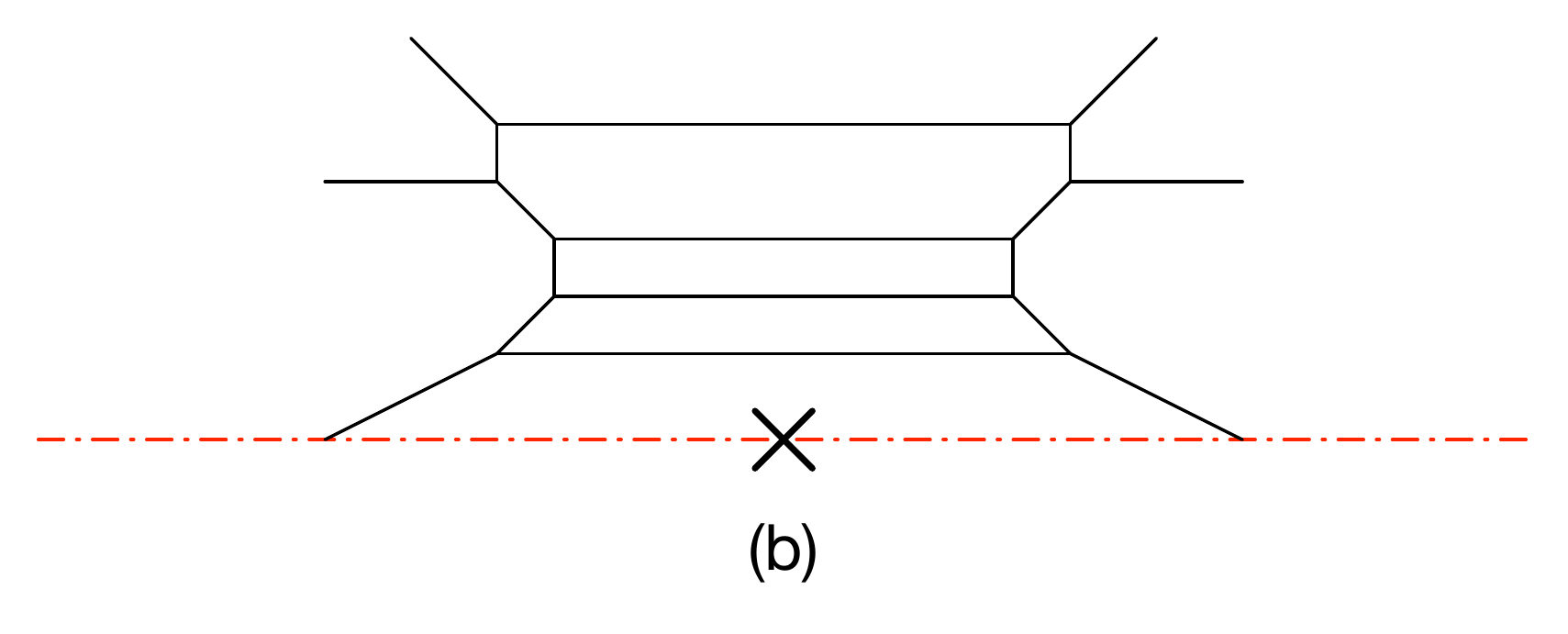} 
\caption{Gauging an (instantonic) $SU(2)$ in the $SO(6)+2$ SCFT leads to the $SO(8)+2$ SCFT.}
\label{SO(2N)duality}
\end{figure}

Let us take the SCFT corresponding to $SO(6)$ with $N_V=2$ as our starting point.
The global symmetry of the UV fixed point is $SU(2)\times USp(4)_F$, with the $SU(2)$ realized non-perturbatively in the gauge theory.
In the 5-brane web, Fig.~\ref{SO(2N)duality}a, the $SU(2)$ is associated with the pair of parallel external NS5-branes.
Gauging it is described by the 5-brane web of Fig.~\ref{SO(2N)duality}b. This corresponds to a SCFT with an IR
description as $SO(8)$ with $N_V=2$.
So although we do not have a dual gauge theory for the latter, we can describe it alternatively as an $SU(2)$ gauging of
a rank 3 SCFT which has an IR description as $SO(6)$ with $N_V=2$.
This can be generalized in a straightforward way to $SO(2N)$ with $N_V=2N-6$, leading to an alternative description
as the linear quiver with gauge group factors $SU(2)^{N-3}$, where the last factor gauges the $SU(2)$ part of the 
global symmetry of the aforementioned rank 3 SCFT.

\section{$SU(N)$ with an antisymmetric}

Incorporating matter in representations other than the fundamental or bi-fundamental in 5-brane webs is notoriously hard.
Even in the simplest example of $USp(2N)$ with an antisymmetric hypermultiplet (corresponding to the rank $N$ $E_1$
or $\tilde{E}_1$ theories), where the 5-brane web is known, the existence of the antisymmetric field is only 
understood indirectly, by analyzing the symmetry on the Higgs branch \cite{BG} (In fact there is a remaining puzzle in this case which we will review and resolve below).
Using the same strategy, we also proposed a 5-brane web for $SU(2N)$ with an antisymmetric 
hypermultiplet in \cite{BZ}.\footnote{Special cases of webs with two antisymmetric hypermultiplets appeared in \cite{BGZ}.}


Here we will give a more direct construction using the $O7^-$ plane, which makes the antisymmetric hypermultiplet manifest.
The construction involves a fractional NS5-brane that is stuck on the $O7^-$ plane. Analogous constructions exist in 4d \cite{LLL,GK}.

\subsection{$SU(2N)$}

\begin{figure}[h]
\center
\includegraphics[width=0.3\textwidth]{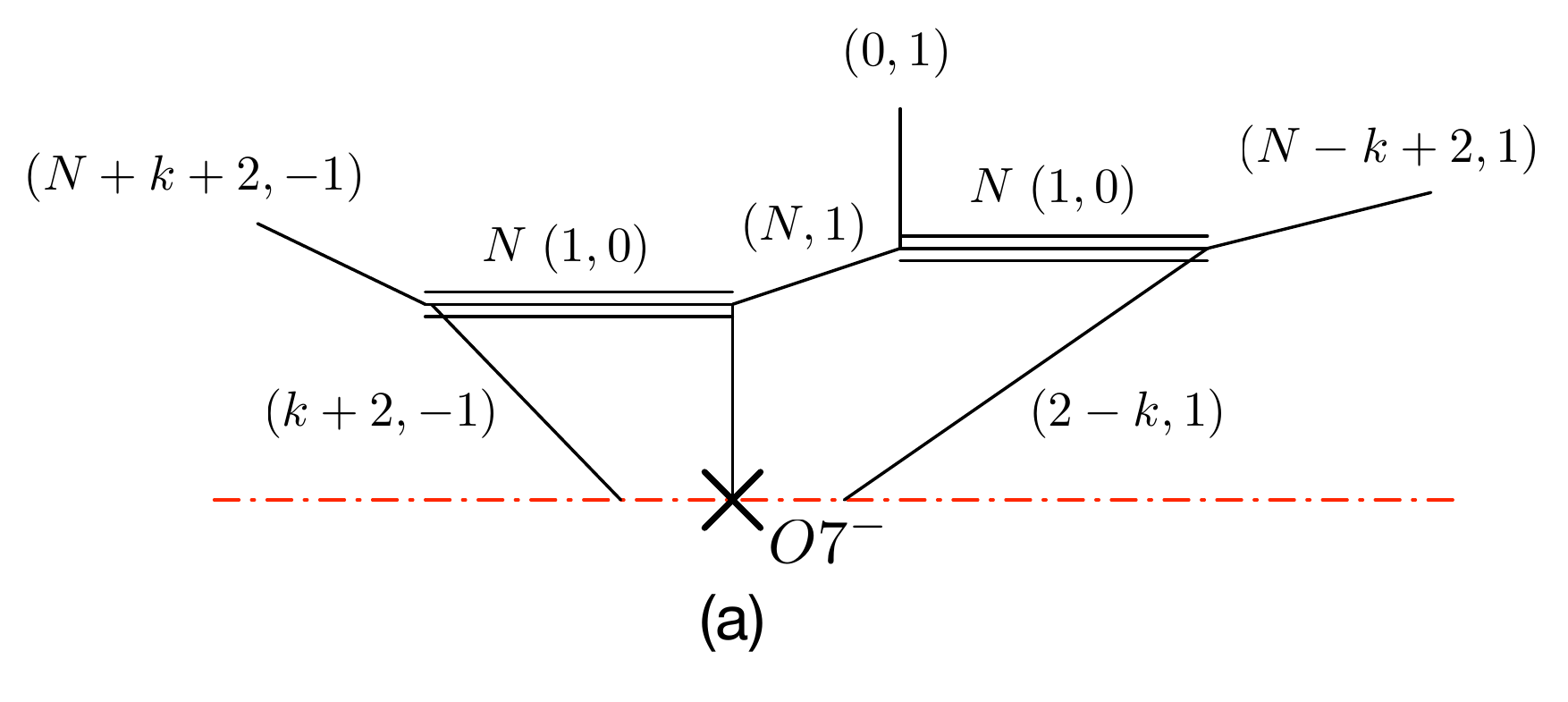} 
\hspace{0.5cm}
\includegraphics[width=0.3\textwidth]{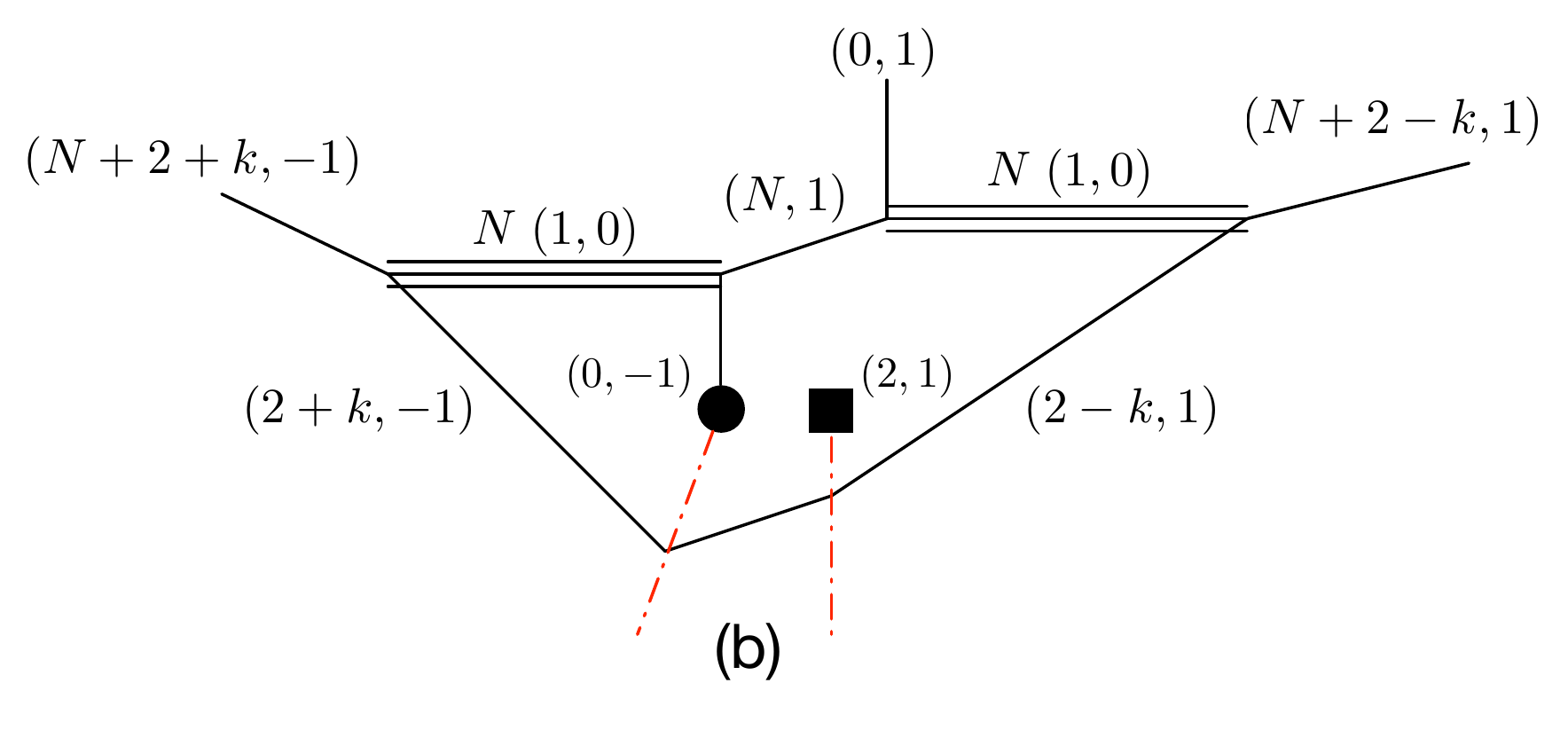} 
\hspace{0.5cm}
\includegraphics[width=0.3\textwidth]{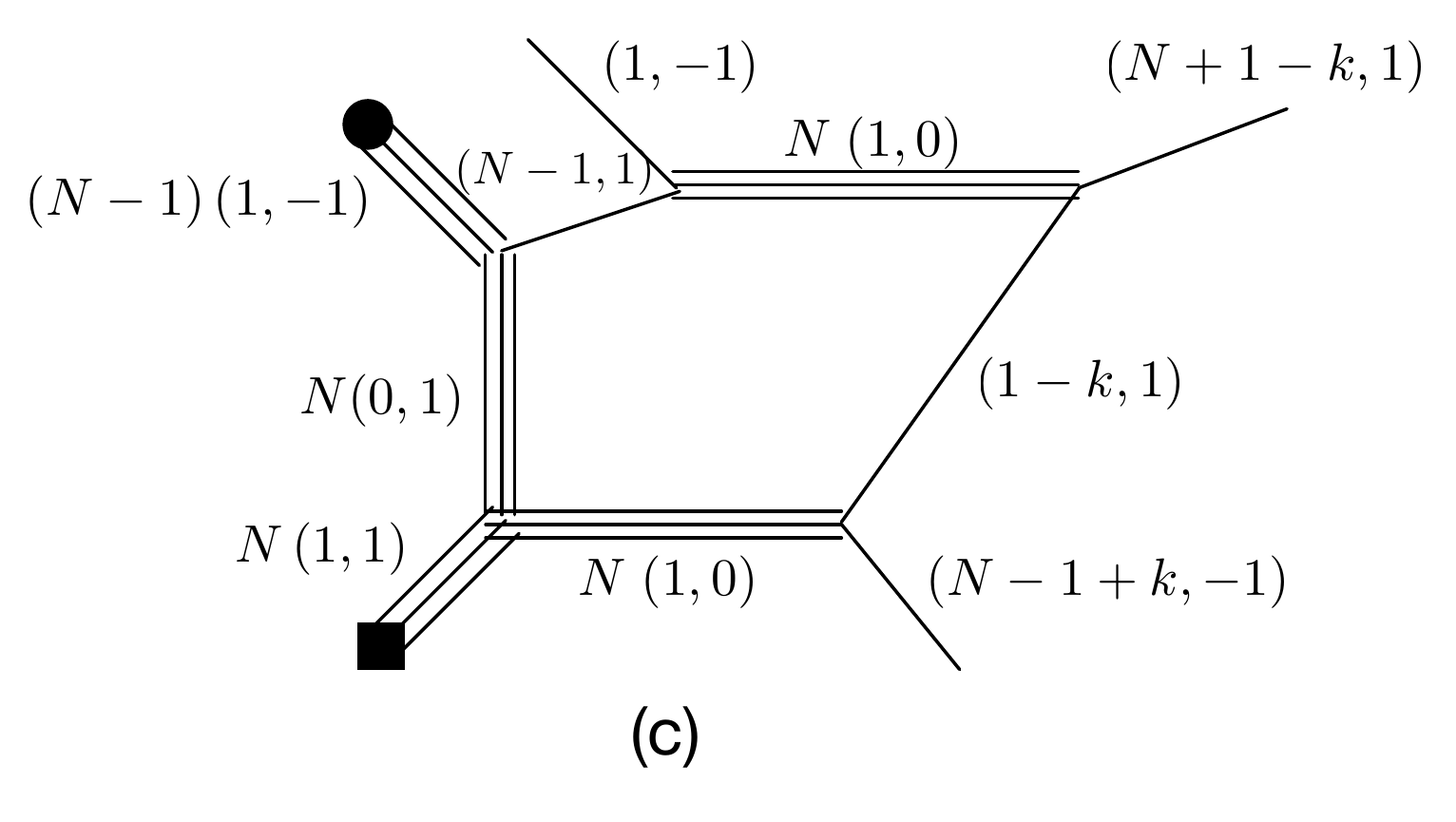} 
\caption{Orientifold 5-brane web for $SU(2N)_k$ with $N_A=1$, with the mass deformation corresponding to the antisymmetric
hypermultiplet.}
\label{SU(2N)+AS}
\end{figure}

The orientifold 5-brane web for $SU(2N)$ with $N_A=1$ is shown in Fig.~\ref{SU(2N)+AS}a.
Now different values of $k$ correspond to distinct theories since the NS5-brane is not invariant under $T\in SL(2,\mathbb{Z})$.
The integer $k$ is the CS level of the theory.
Note that for a vanishing CS level, $k=0$, the web is reflection symmetric, corresponding to charge-conjugation
symmetry (which the CS term breaks).
The hypermultiplet in the antisymmetric representation corresponds to an open string connecting the D5-branes
on either side of the NS5-brane. The position of the external NS5-brane in the plane transverse to the $O7^-$ plane
corresponds to the mass of the hypermultiplet, 
and its position along the 
$O7^-$ plane corresponds to its VEV, namely to the Higgs branch of the theory.
On the Higgs branch the web reduces to that of $USp(2N)$, as it should.
The orientifold web with the NS5-brane appears to exhibit $2N$ Coulomb moduli, since it has $2N$ faces.
However they are not independent. There is one constraint coming from the fact that the NS5-brane
cannot detach from the $O7^-$ plane. The number of independent parameters is then $2N-1$, the 
dimension of the Coulomb branch of $SU(2N)$.

To our knowledge, the resolution of the $O7^-$ plane with a fractional NS5-brane has not been previously studied,
but one can make a conjecture based on the resolution of the bare $O7^-$ plane.
The only possibility that makes sense is that the combination is resolved into a $(2,1)$ 7-brane and a $(0,-1)$ 7-brane with the NS5-brane
ending on it
(Fig.~\ref{SU(2N)+AS}b).
Using this, together with a couple of HW transitions and a $T^{-1}$ transformation,
we 
arrive at the web shown in Fig.~\ref{SU(2N)+AS}c.
This is precisely the web originally proposed in \cite{BZ}.

\begin{figure}[h]
\center
\includegraphics[width=0.25\textwidth]{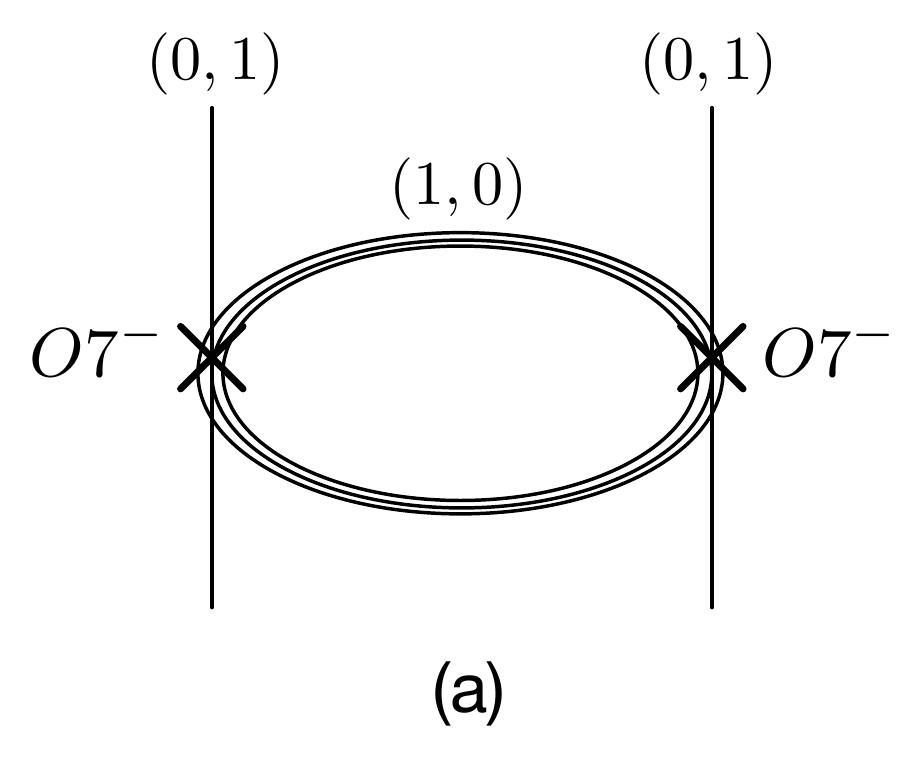} 
\hspace{0.5cm}
\includegraphics[width=0.3\textwidth]{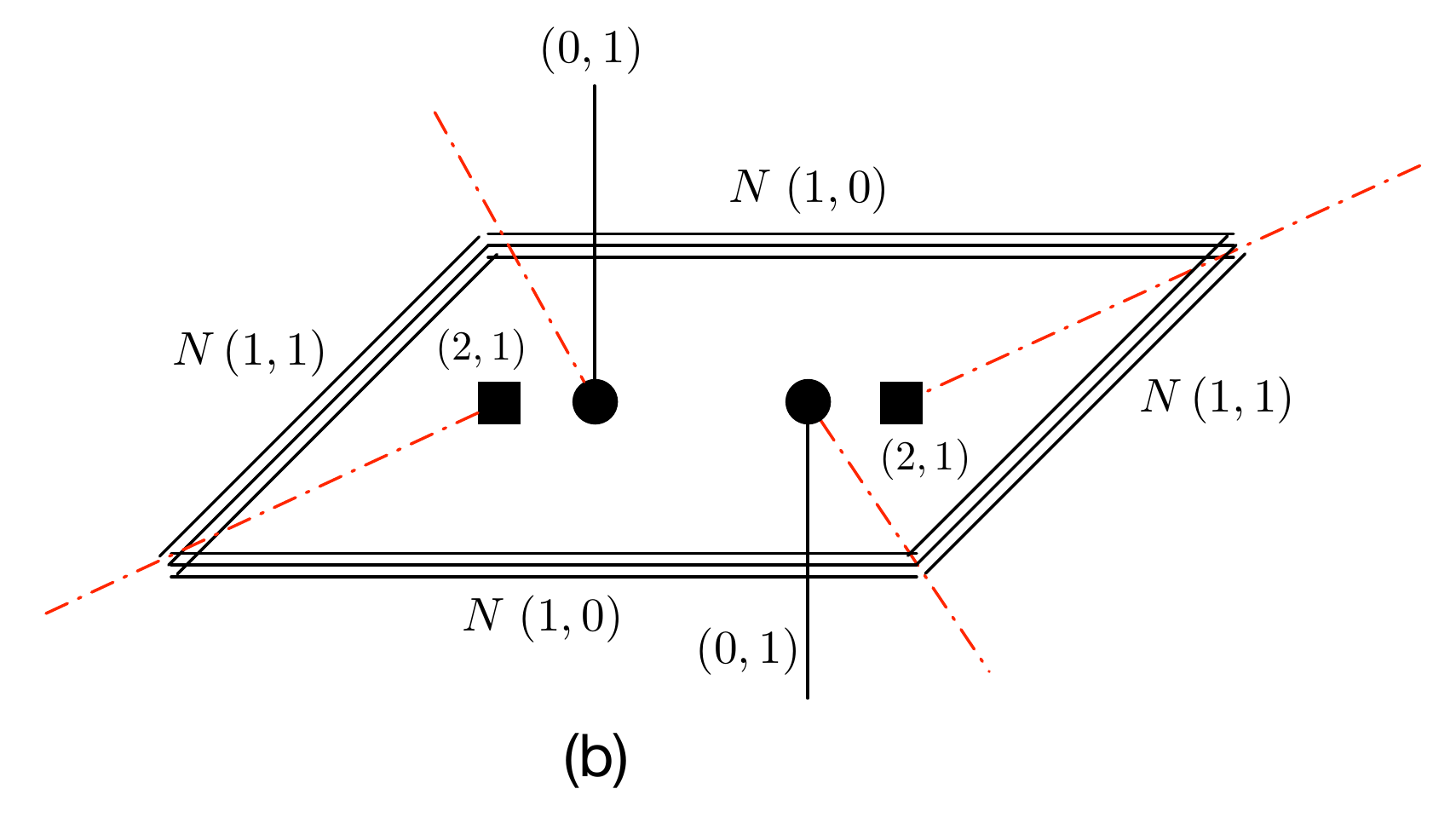} 
\hspace{0.5cm}
\includegraphics[width=0.3\textwidth]{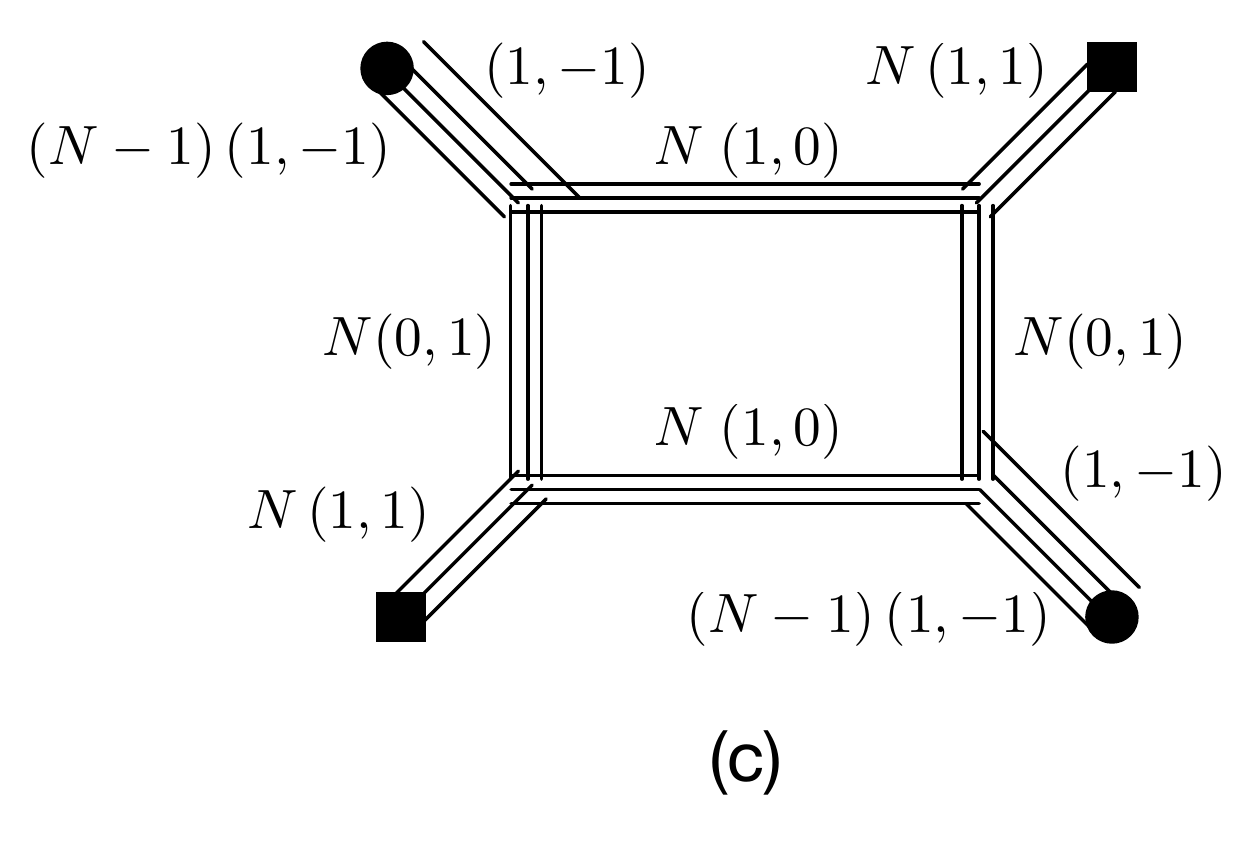} 
\caption{Orientifold brane configuration and 5-brane web for $SU(2N)_0$ with $N_A=2$.}
\label{SU(2N)+2AS}
\end{figure}

As an aside, we can now resolve two puzzles from \cite{BG}.
The first has to do with the 5-brane web resolution of the classical brane configurations for the theories
corresponding to the orbifolds without vector structure.
For example for the $\mathbb{Z}_2$ orbifold the theory is $SU(2N)_0$ with $N_A=2$.
The classical Type IIB brane configuration consists of $2N$ D5-branes on a circle, with two $O7^-$ planes 
at antipodal points, each supporting a fractional NS5-brane, Fig.~\ref{SU(2N)+2AS}a. 
We can now identify the resolved configuration, Fig.~\ref{SU(2N)+2AS}b, and using 
manipulations similar to the ones above, obtain a 5-brane web construction, Fig.~\ref{SU(2N)+2AS}c.

\begin{figure}[h]
\center
\includegraphics[width=0.25\textwidth]{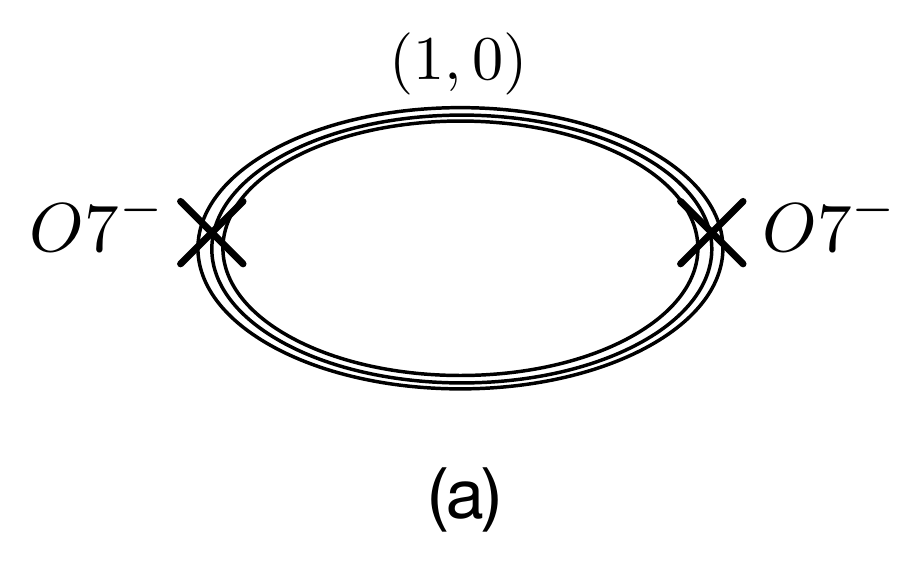} 
\hspace{0.5cm}
\includegraphics[width=0.25\textwidth]{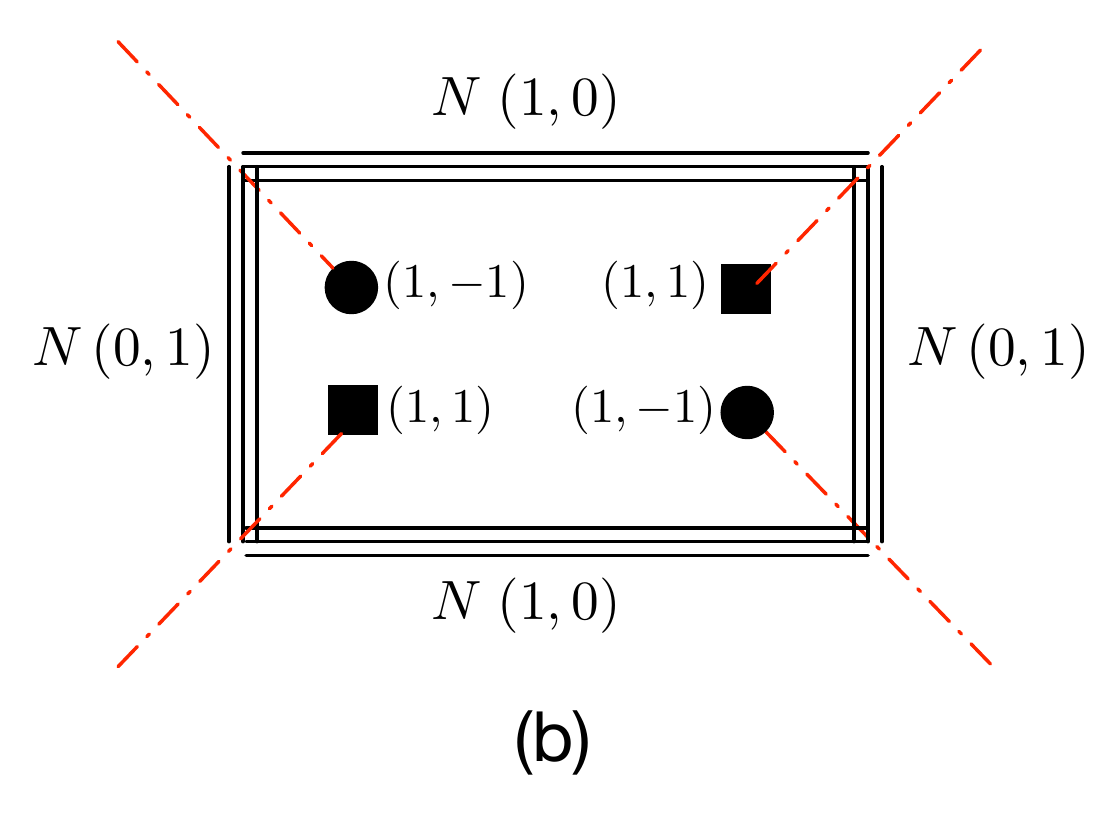} 
\hspace{0.5cm}
\includegraphics[width=0.3\textwidth]{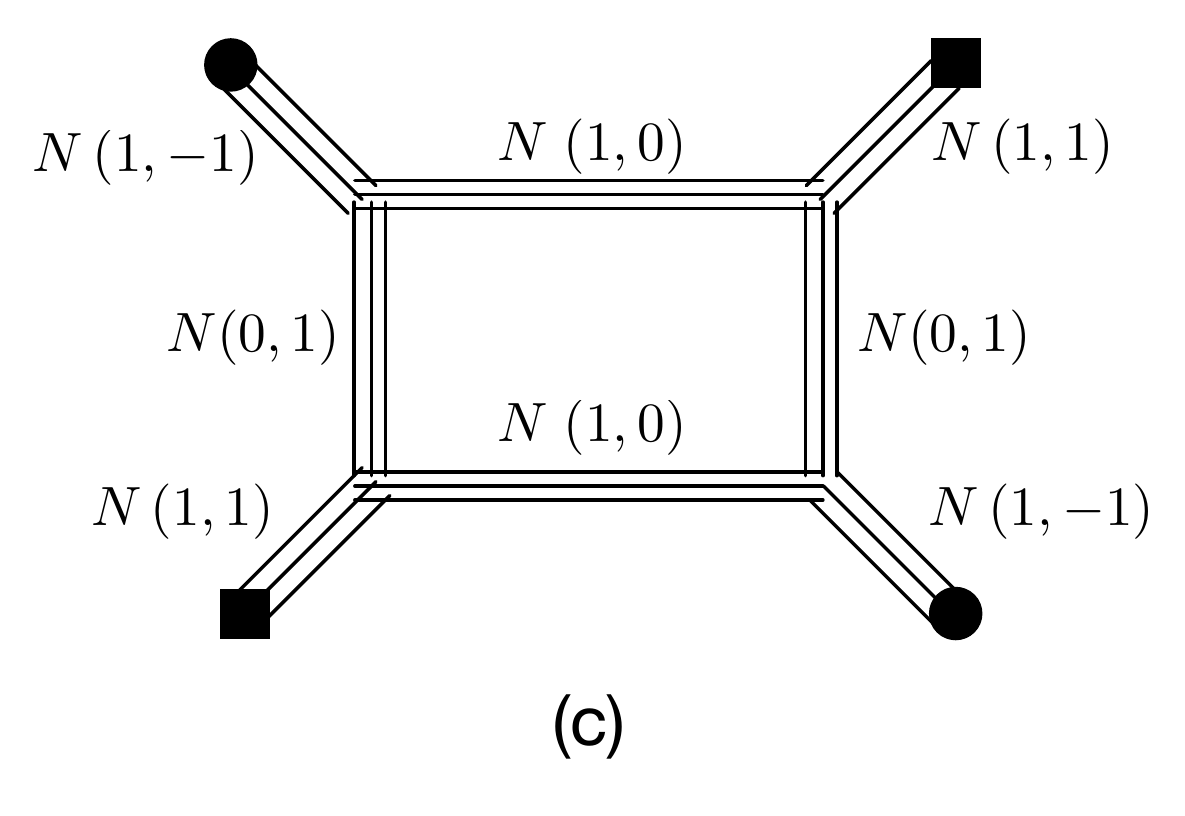} 
\caption{Orientifold brane configuration and 5-brane web for $USp(2N)$ with $N_A=1$.
This gives the theory with $\theta = 0$. The theory with $\theta = \pi$ is obtained by resolving the two $O7^-$ planes
differently, replacing $(1,1)+(1,-1)$ by $(2,1)+(0,-1)$ for one. Replacing both yields $\theta = 2\pi \sim 0$.}
\label{USp(2N)+AS}
\end{figure}

\begin{figure}[h]
\center
\includegraphics[width=0.25\textwidth]{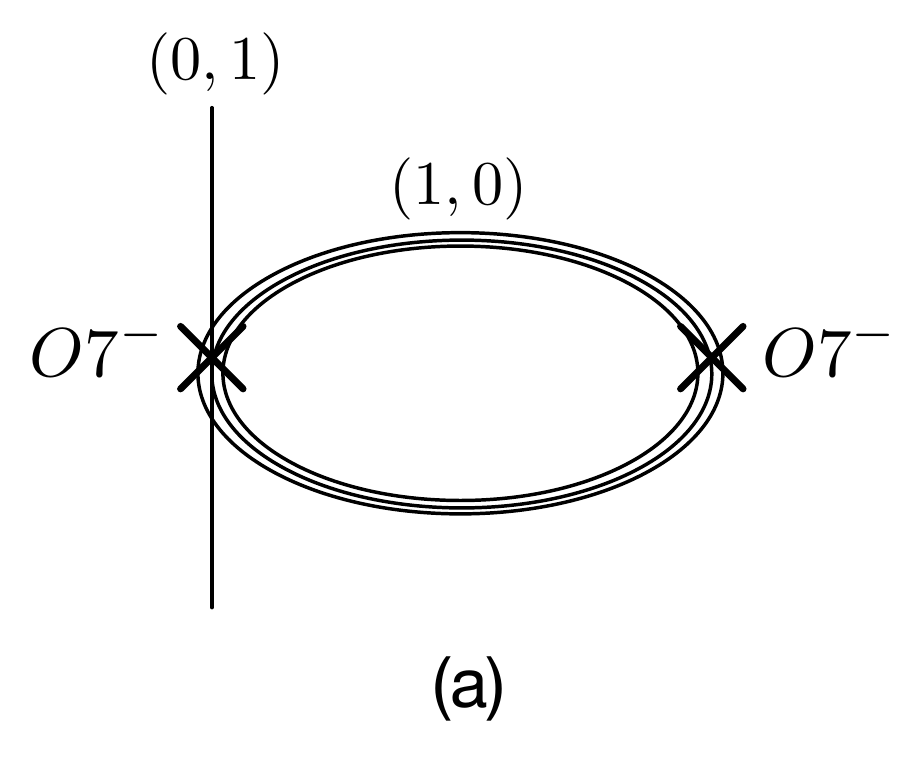} 
\hspace{0.5cm}
\includegraphics[width=0.3\textwidth]{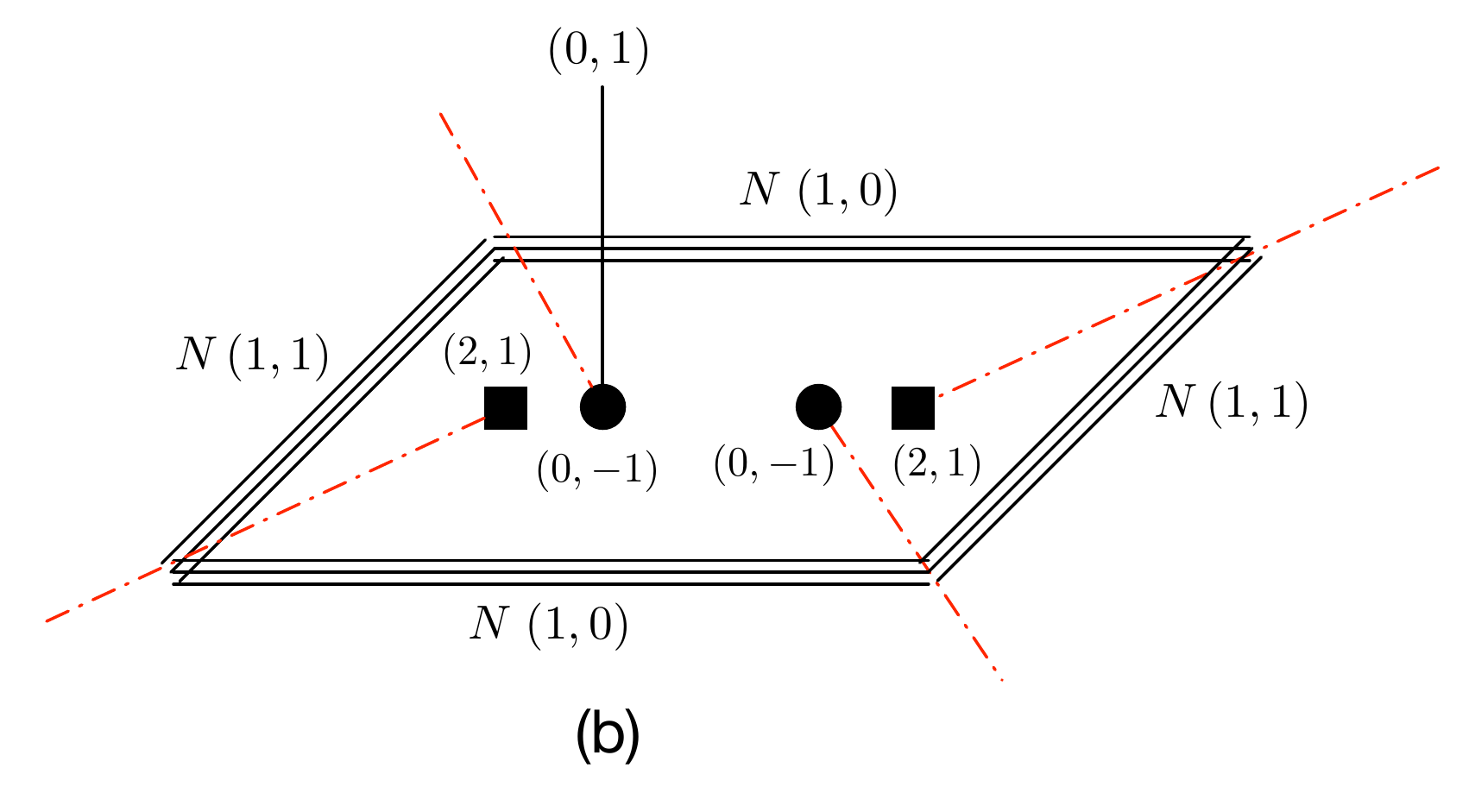} 
\hspace{0.5cm}
\includegraphics[width=0.3\textwidth]{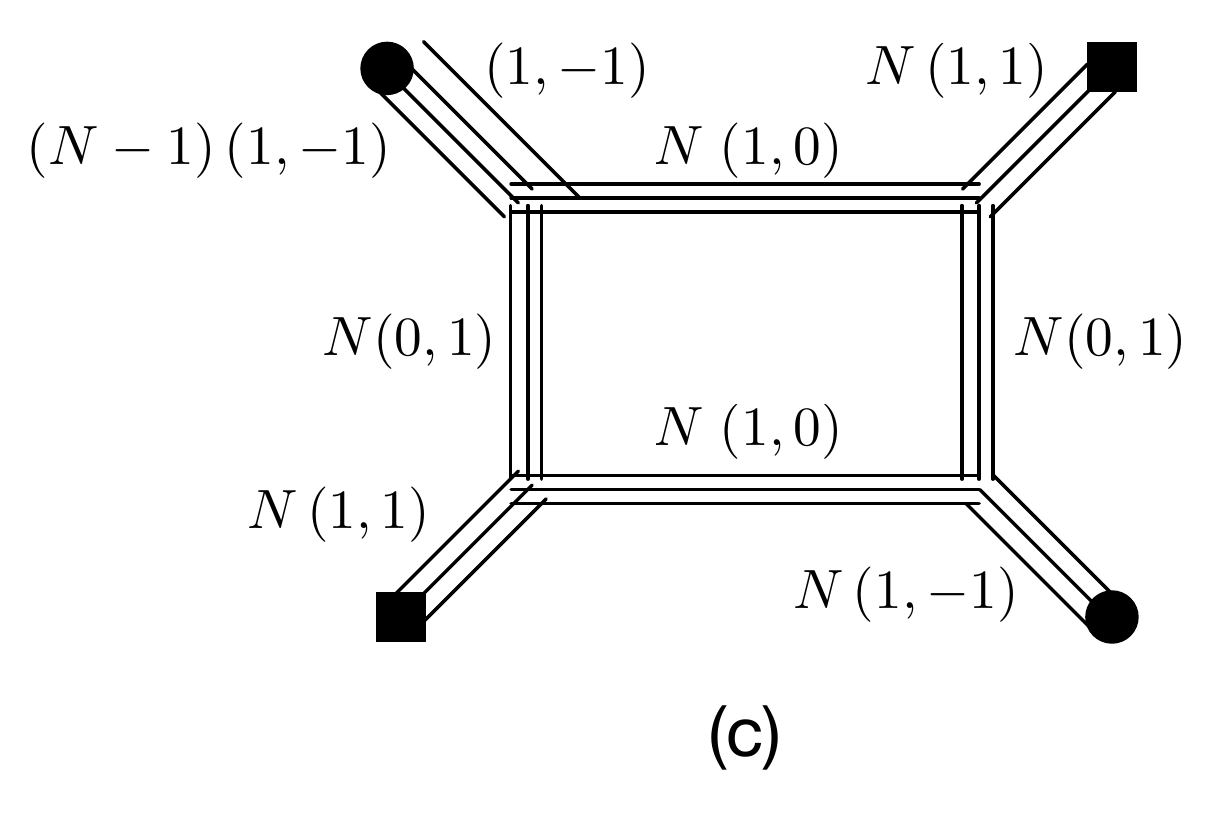} 
\caption{An alternative orientifold brane configuration and 5-brane web for $USp(2N)$ with $N_A=1$.
This again gives the theory with $\theta = 0$. Replacing the second pair of 7-branes with $(1,1)+(1,-1)$ gives $\theta = \pi$.}
\label{USp(2N)+ASalternative}
\end{figure}

The second puzzle has to do with the 5-brane web construction of the $USp(2N)$ theory with $N_A=1$ (Fig.~\ref{USp(2N)+AS}).
While this web exhibits the correct Coulomb and Higgs branches, it 
does not admit a deformation corresponding to giving a mass to the hypermultiplet.
It exhibits only one mass parameter, corresponding to the YM coupling.
We can now provide an alternative 5-brane web for the same theory, which does admit the hypermultiplet mass deformation.
If we attach a fractional NS5-brane to one of the $O7^-$ planes the theory remains the same; it is still $USp(2N)$ with $N_A=1$.
This is the ``$\mathbb{Z}_1$" orbifold. However the 5-brane web obtained from resolving this configuration is different,
and now admits an extra deformation corresponding to the mass of the antisymmetric hypermultiplet (Fig.~\ref{USp(2N)+ASalternative}).

\subsection{$SU(2N+1)$}

\begin{figure}[h]
\center
\includegraphics[width=0.3\textwidth]{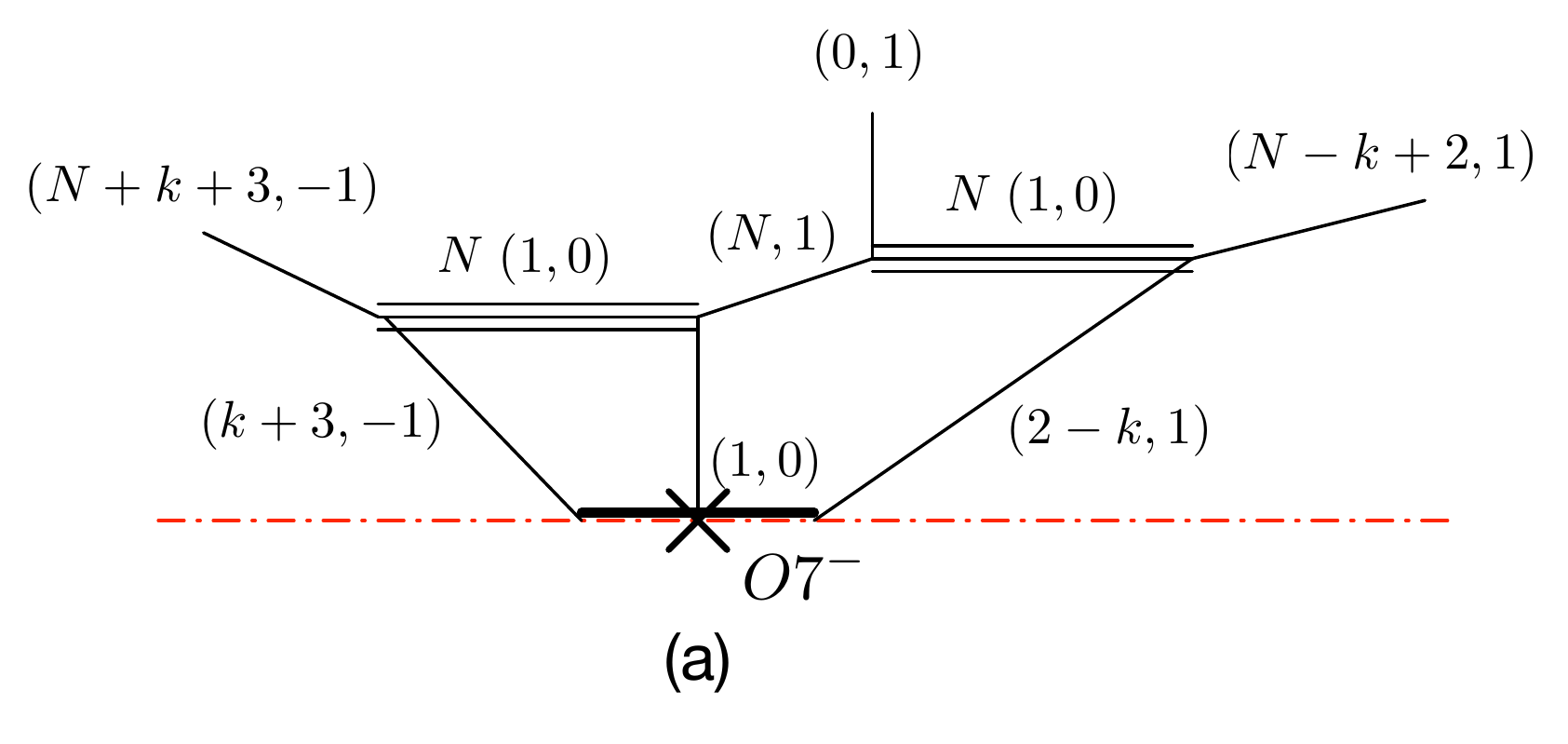} 
\hspace{0.5cm}
\includegraphics[width=0.3\textwidth]{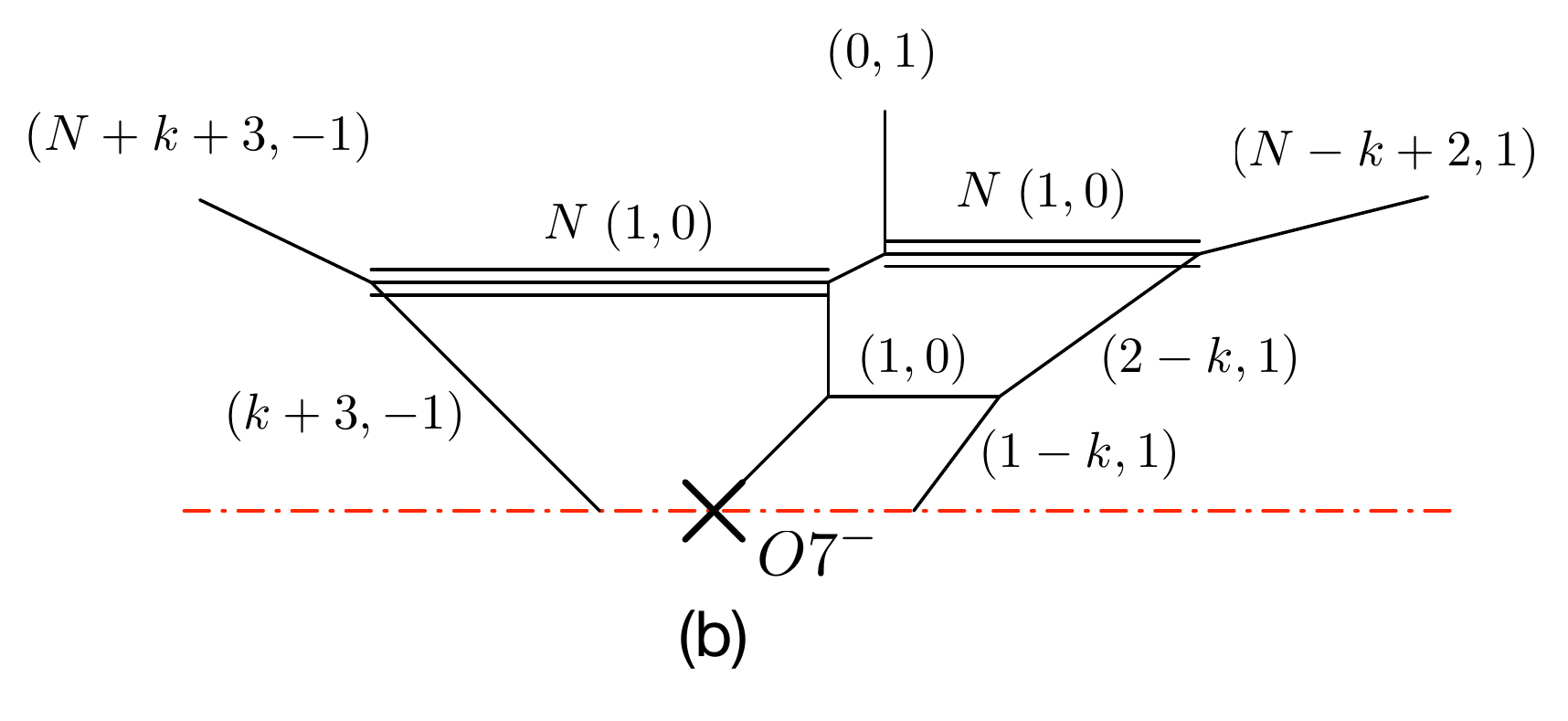} 
\hspace{0.5cm}
\includegraphics[width=0.3\textwidth]{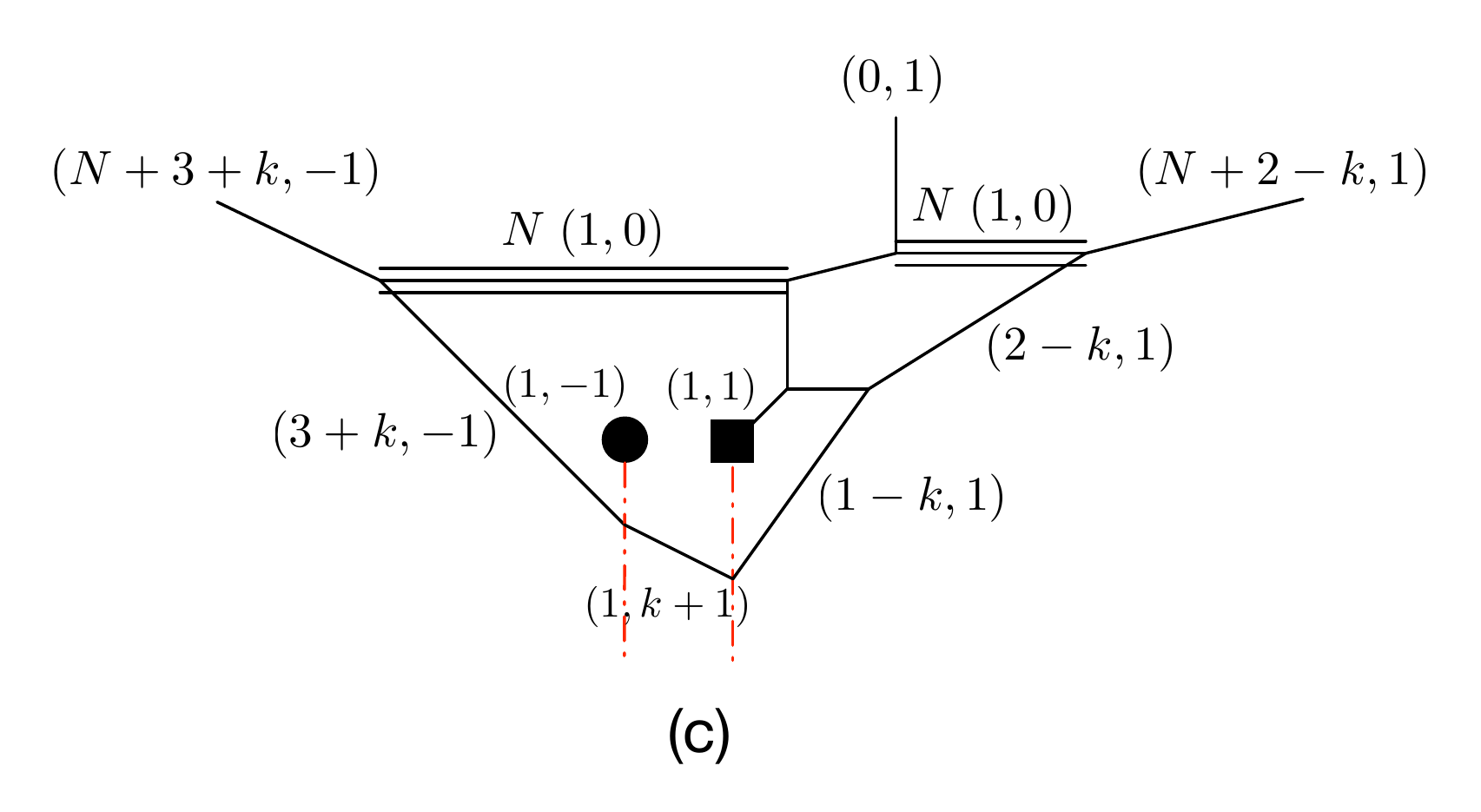} 
\caption{Orientifold 5-brane web for $SU(2N+1)_{k + \frac{1}{2}}$ with $N_A=1$.}
\label{SU(2N+1)+AS}
\end{figure}

The orientifold web for $SU(2N+1)$ with $N_A=1$ is obtained by adding a fractional D5-brane ending on the fractional NS5-brane, 
Fig.~\ref{SU(2N+1)+AS}a.
Note that this does not affect the NS5-brane, since the net number of D5-branes ending on it vanishes.
This configuration would not be possible in the absence of the fractional NS5-brane; the consistency conditions of \cite{GP} 
would require an even number of D5-branes.
The integer $k$ again determines the CS level of the theory, which must now be half-odd-integer due to the parity anomaly associated with the antisymmetric matter multiplet.\footnote{In general, there is a parity anomaly if the cubic Casimir of the matter representation is odd.
For the antisymmetric representation of $SU(M)$ the cubic Casimir is $M-4$.}

It is important here that the fractional D5-brane is ``broken" on the fractional NS5-brane.
In particular, the fractional NS5-brane can now detach from the $O7^-$ plane by combining with the fractional D5-brane, as shown in Fig.~\ref{SU(2N+1)+AS}b,
giving an extra modulus for a total of $2N$, in agreement with the dimension of the 
Coulomb branch for $SU(2N+1)$.
This also implies that there is no Higgs branch, since the NS5-brane cannot be separated from the fractional D5-brane.
This is also consistent with the gauge theory. 
For $SU(2N)$ the Higgs branch is spanned by the baryonic operator 
$\epsilon^{\alpha_1\cdots \alpha_{2N}} A_{\alpha_1\alpha_2}\cdots A_{\alpha_{2N-1}\alpha_{2N}}$.
For $SU(2N+1)$ there is no such gauge invariant operator.\footnote{There would be a Higgs branch if we added a hypermultiplet 
in the fundamental representation. The corresponding operator is
$\epsilon^{\alpha_1\cdots \alpha_{2N+1}} A_{\alpha_1\alpha_2}\cdots A_{\alpha_{2N-1}\alpha_{2N}} \psi_{\alpha_{2N+1}}$.}

There is also an obvious proposal for the resolution of the $O7^-$ plane in this case, Fig.~\ref{SU(2N+1)+AS}c.

\subsection{Flavors}

\begin{figure}[h]
\center
\includegraphics[width=0.3\textwidth]{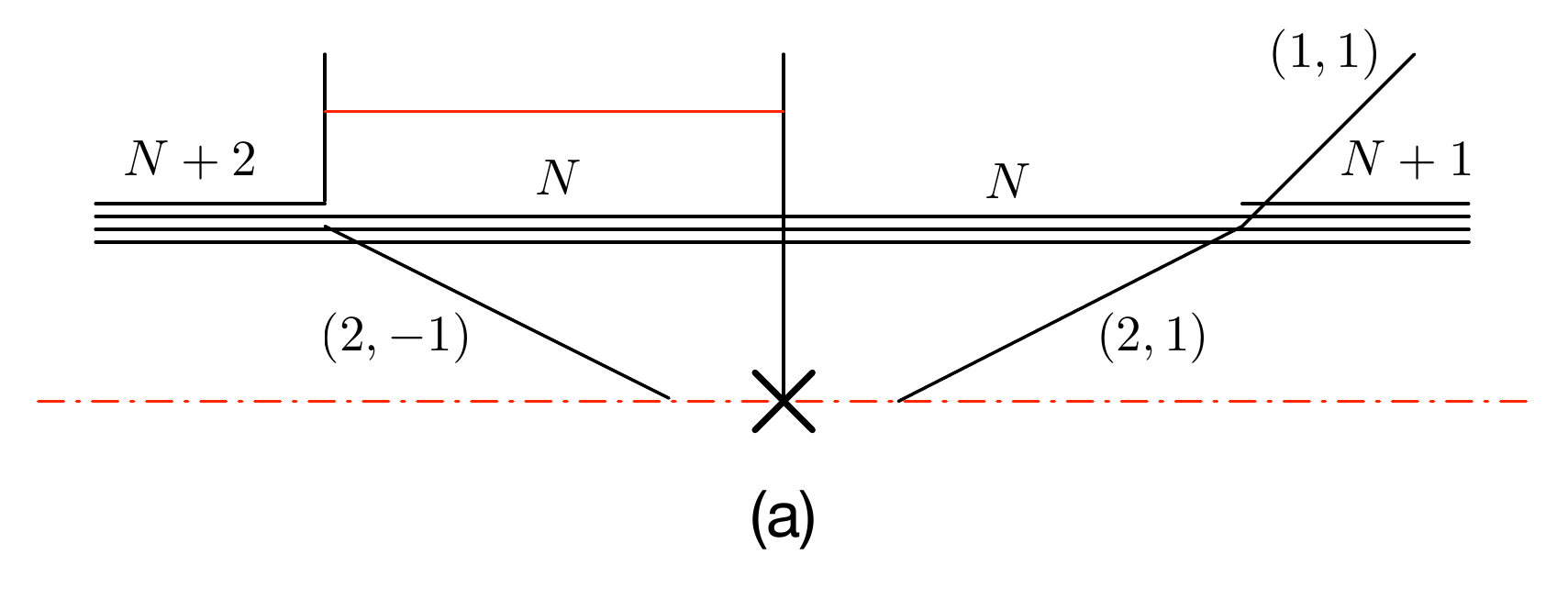} 
\hspace{0.5cm}
\includegraphics[width=0.3\textwidth]{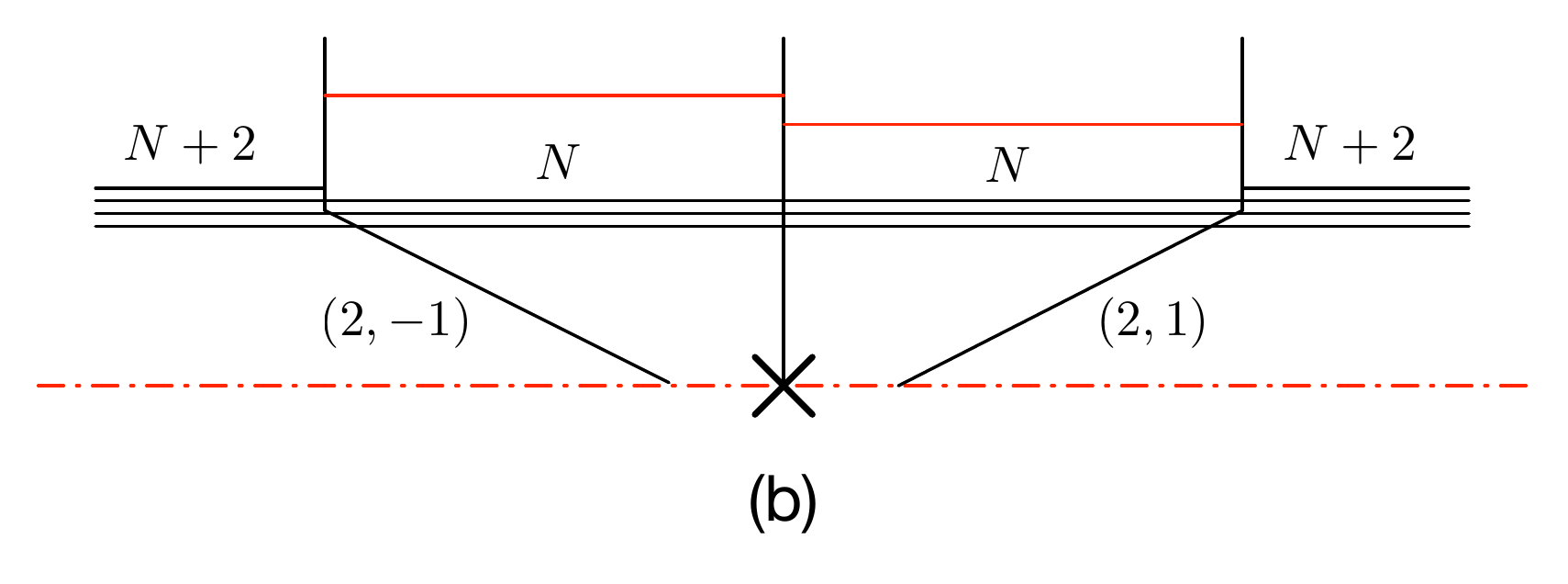} 
\hspace{0.5cm}
\includegraphics[width=0.3\textwidth]{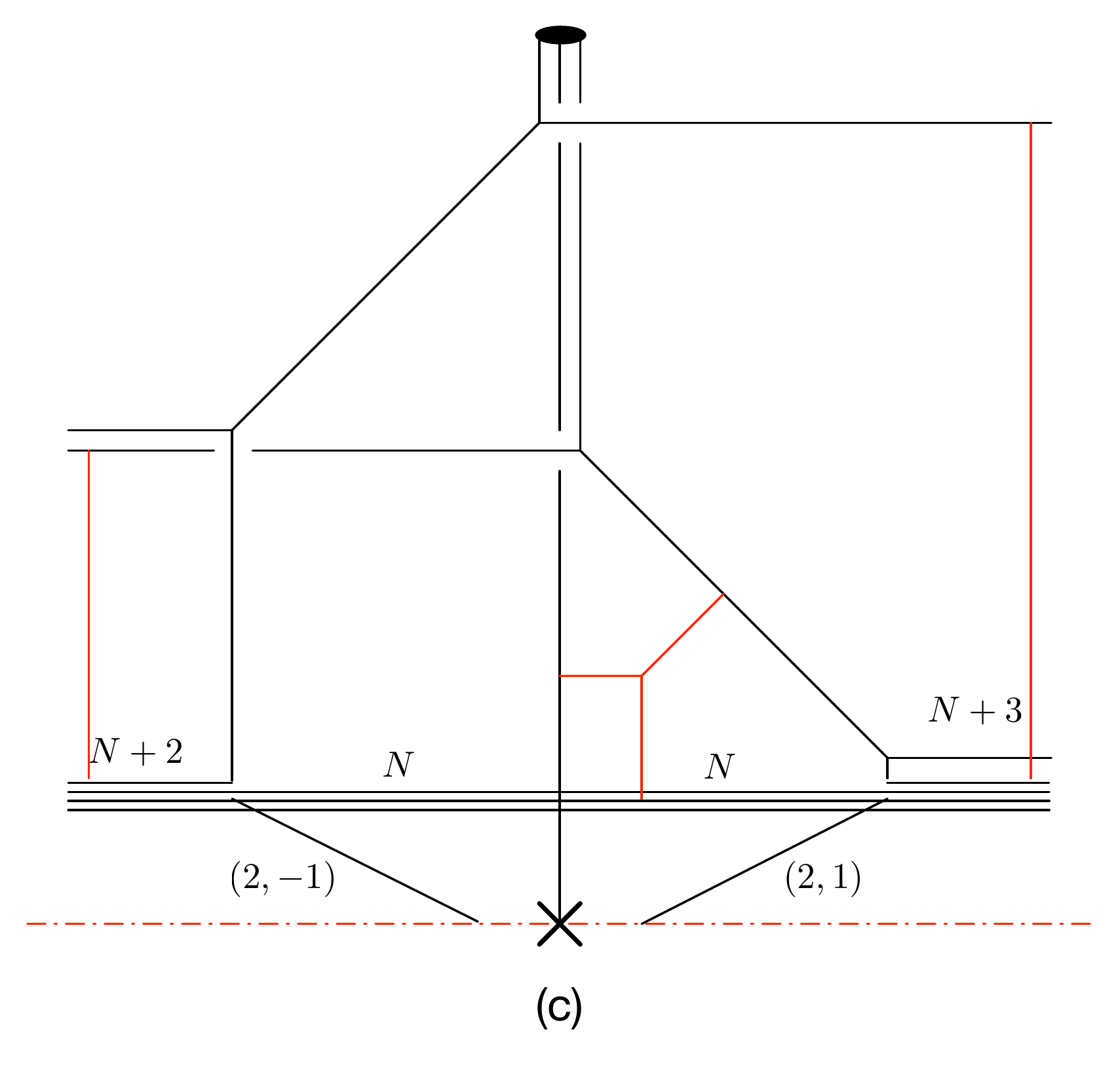} 
\caption{$SU(2N)$ with $N_A=1$ and (a) $N_F = 2N+3$, $\kappa_{CS}=\frac{1}{2}$, (b) $N_F=2N+4$, $\kappa_{CS}=0$, 
(c) $N_F=2N+5$, $\kappa_{CS}=\frac{1}{2}$.}
\label{SU(2N)+AS+flavor}
\end{figure}

We can again add matter in the fundamental representation by attaching external D5-branes.
The condition of \cite{SMI} for a fixed point to exist for $SU(M)$ with $N_A=1$ is $N_F + 2|\kappa_{CS}|\leq 8-M$.
In this case we find a very different bound. 
The orientifold 5-brane web construction allows $N_F + 2|k| \leq M+5$.
The CS level is either $k$ or $k \pm \frac{1}{2}$, depending on whether $N_F+M$ is even or odd, respectively.
In particular we can have an arbitrarily large rank.
The bound is saturated by the web with the avoided intersection shown (for $M=2N$ and $k=0$) in Fig.~\ref{SU(2N)+AS+flavor}c.
The two other interesting cases are $N_F + 2|k| = M+3$ and $N_F + 2|k| =M+4$, shown (again for $M=2N$ and $k=0$) 
in Figs.~\ref{SU(2N)+AS+flavor}a and \ref{SU(2N)+AS+flavor}b, respectively. 
These have parallel external NS5-branes.
In all three cases we exhibit some of the extraneous instanton-charged states that should be removed from the instanton partition function.
We expect these three theories to exhibit enhanced global symmetries in the UV, and in particular we expect the non-abelian
flavor symmetry to be enhanced in the maximally flavored theory. In fact, these theories were recently studied in \cite{Yon}, using simplified instanton analysis, whose finding is indeed consistent with the enhancement suggested by these webs.   

\subsection{Duality}

In formulating duality conjectures in this case we will follow the strategy we used for the $USp(2N)$ theories.
We begin with a theory whose dual we know, and which we can engineer as a 5-brane web,
and then gauge part of its global symmetry (and maybe add flavors) to obtain a new theory, whose dual we infer from the S-dual web.

\subsubsection{Example 1}

\begin{figure}
\center
\includegraphics[width=0.3\textwidth]{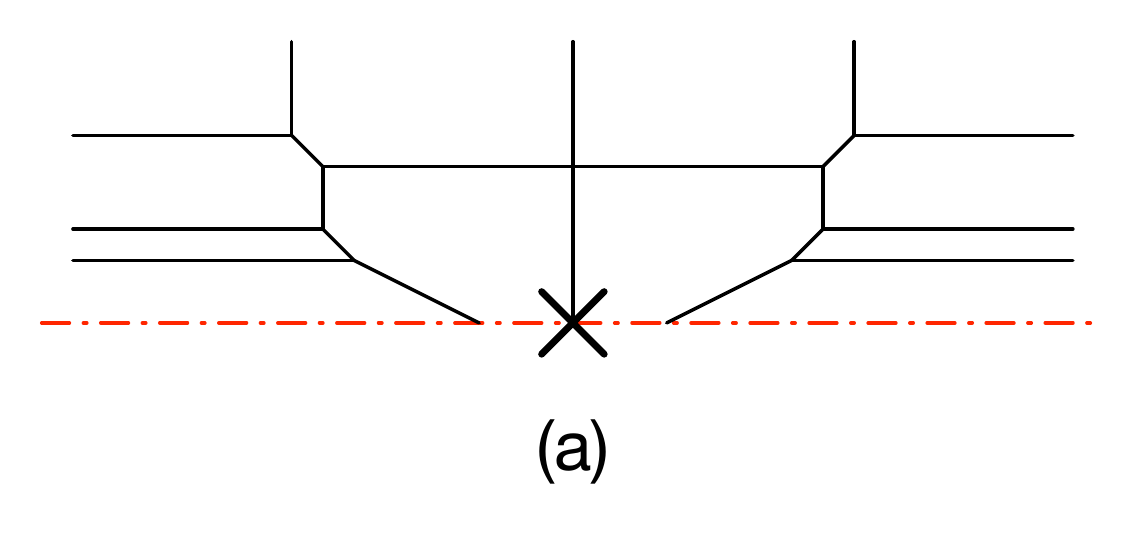} 
\hspace{1cm}
\includegraphics[width=0.18\textwidth]{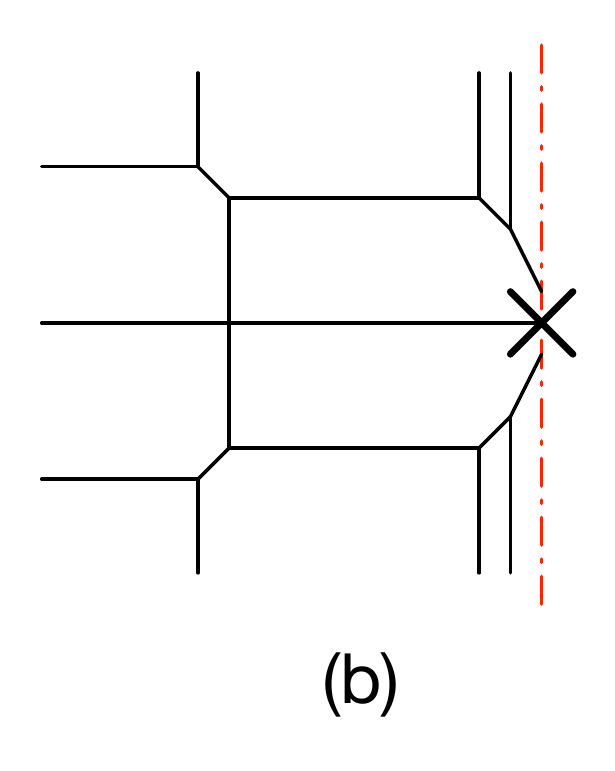} 
\caption{The orientifold 5-brane web for $SU(2)$ with $N_A=1$ and $N_F=6$ and its S-dual, which describes the same theory (we have taken $k=0$).}
\label{SU(2)+AS+6}
\end{figure}

\begin{figure}
\center
\includegraphics[width=0.2\textwidth]{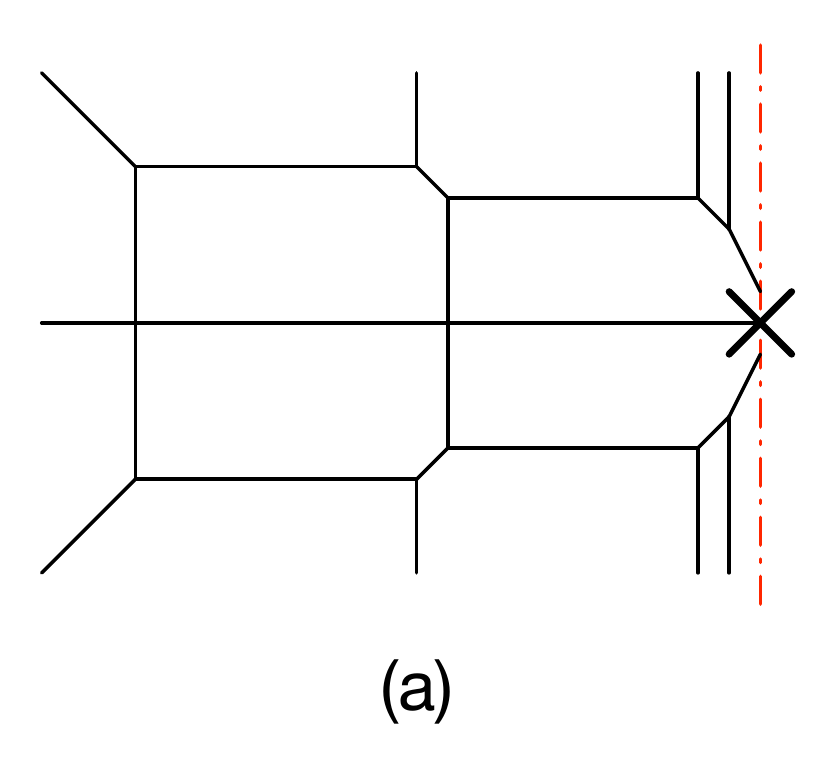}
\hspace{1cm}
\includegraphics[width=0.3\textwidth]{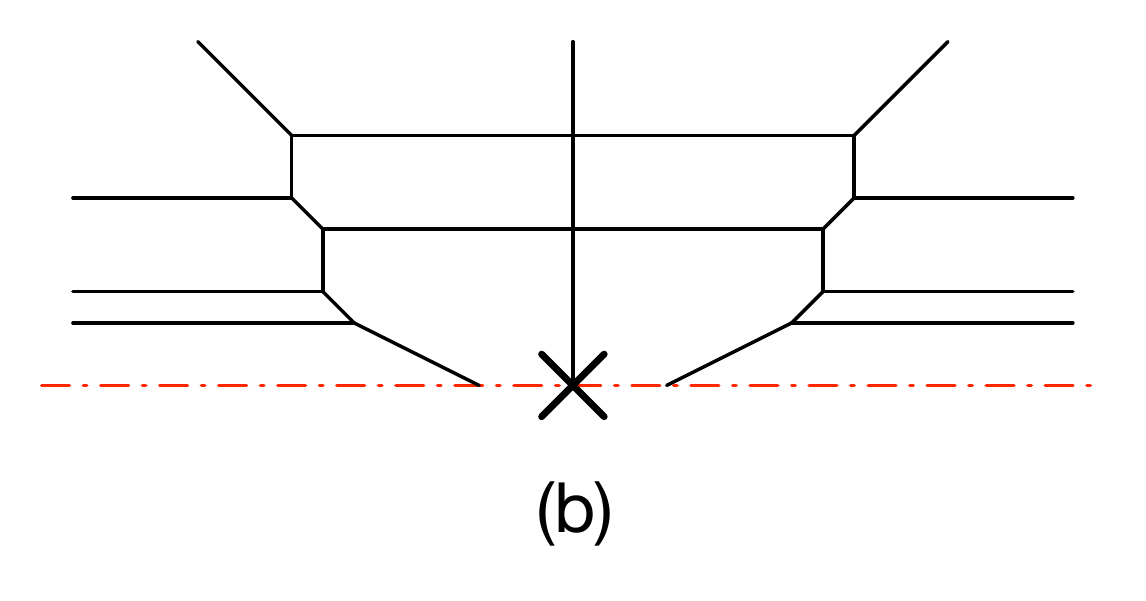} 
\caption{Orientifold 5-brane web for $2+SU(3)_0\times SU(2)+3$ and its S-dual, $SU(4)_0$ with $N_A=1$ and $N_F=6$.}
\label{SU(4)+AS+6}
\end{figure}

In our first example we begin with the self-dual $SU(2)+6$ theory.
This is an IR gauge theory description of the rank one $E_7$ theory.
The antisymmetric matter multiplet is neutral in this case.
The web for this theory and its S-dual in the present description are shown in Fig.~\ref{SU(2)+AS+6}.
The dual web makes manifest an 
$SU(3)$ subgroup of the global symmetry, 
acting on the three flavors corresponding to the three external D5-branes.
Now consider the 5-brane web shown in Fig.~\ref{SU(4)+AS+6}a. This corresponds to gauging 
this $SU(3)$
and adding one $SU(3)$ flavor. 
The resulting gauge theory is the linear quiver $2+SU(3)_0\times SU(2)+3$.
The extra $SU(3)$ flavor (other than the one we added) corresponds to the fractional D5-brane.
The fact that the CS level is zero follows from the symmetry of the web.
We can now read-off the dual gauge theory from the S-dual web, Fig.~\ref{SU(4)+AS+6}b, as
$SU(4)_0$ with $N_A=1$ and $N_F=6$.

\begin{figure}
\center
\includegraphics[width=0.3\textwidth]{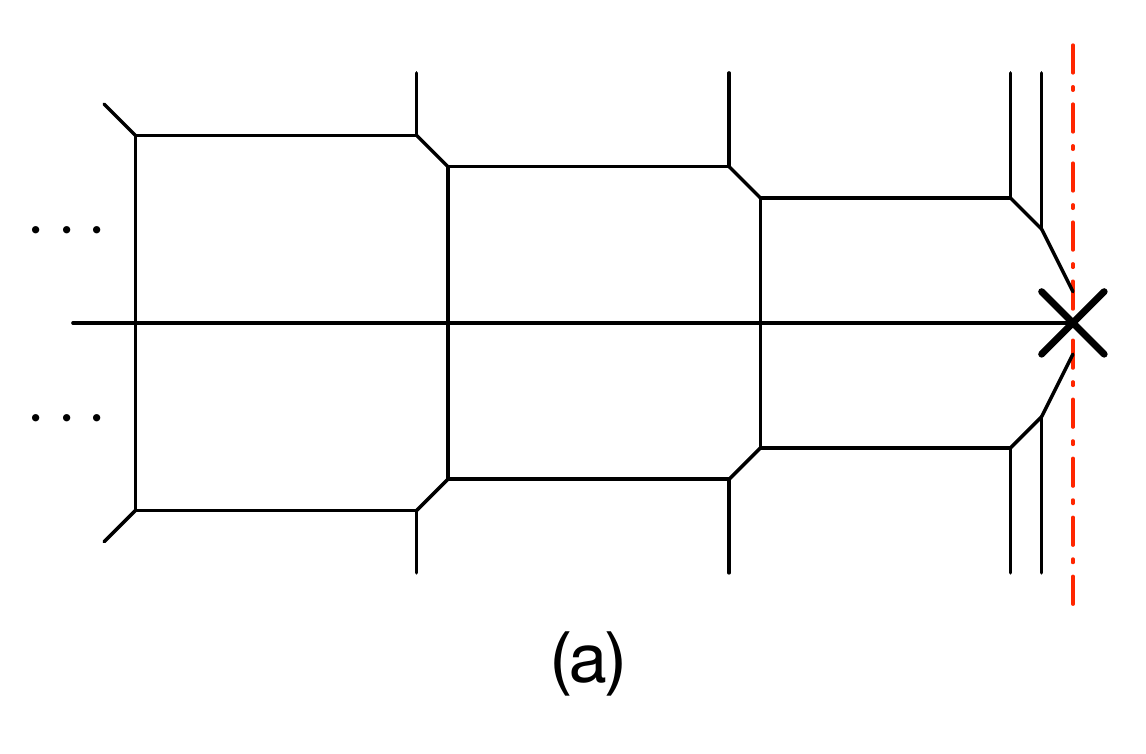} 
\hspace{1cm}
\includegraphics[width=0.3\textwidth]{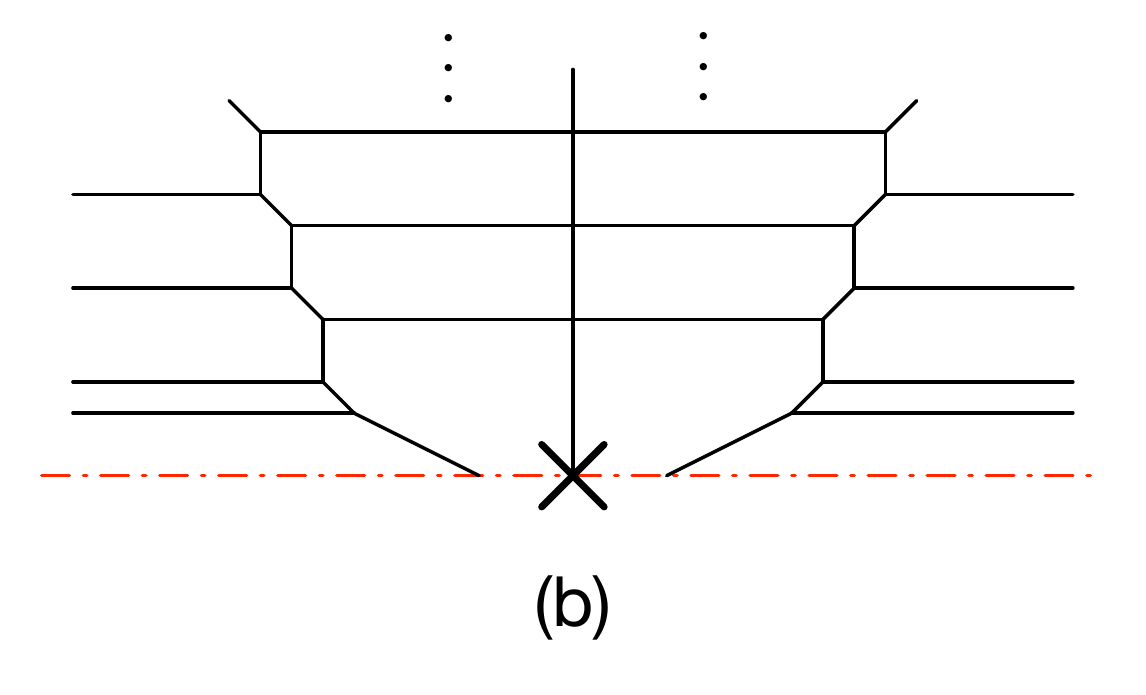} 
\caption{Orientifold 5-brane web for $1+SU(3)_0^{N-2} \times (SU(3)+1)\times SU(2)+3$ and its S-dual,
$SU(2N)_0$ with $N_A=1$ and $N_F=2N+2$.}
\label{SU(2N)+AS+2N+2}
\end{figure}

We can generalize this construction by adding two more $SU(3)$ flavors, gauging the global $SU(3)$ symmetry,
repeating $N-2$ times, and in the end adding a single flavor to the last $SU(3)$ factor, Fig.~\ref{SU(2N)+AS+2N+2}a.
The result is a quiver theory with the structure $1+SU(3)_0^{N-2}\times [SU(3)_0 +1]\times SU(2)+3$.
The S-dual web, Fig.~\ref{SU(2N)+AS+2N+2}b, describes $SU(2N)_0$ with $N_A=1$ and $N_F = 2N+2$.
As a 
consistency check, note that on the Higgs branch associated with the fractional 5-brane
this reduces at low energy to the duality of section \ref{USpduality},
between $USp(2N)$ with $N_F=2N+2$ and the quiver $SU(2)^N + 4$.
On the quiver side this corresponds to a VEV of an operator given by a product of the matter fields through
the quiver, starting with the fundamental of the first $SU(3)$ and ending with the flavor of the last $SU(3)$.

\subsubsection{Example 2}

\begin{figure}
\center
\includegraphics[width=0.3\textwidth]{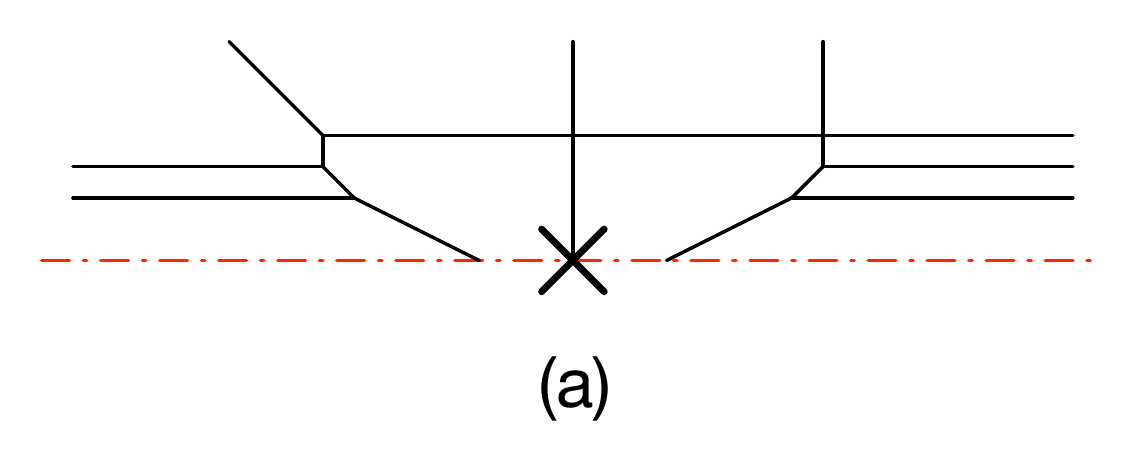} 
\hspace{1cm}
\includegraphics[width=0.15\textwidth]{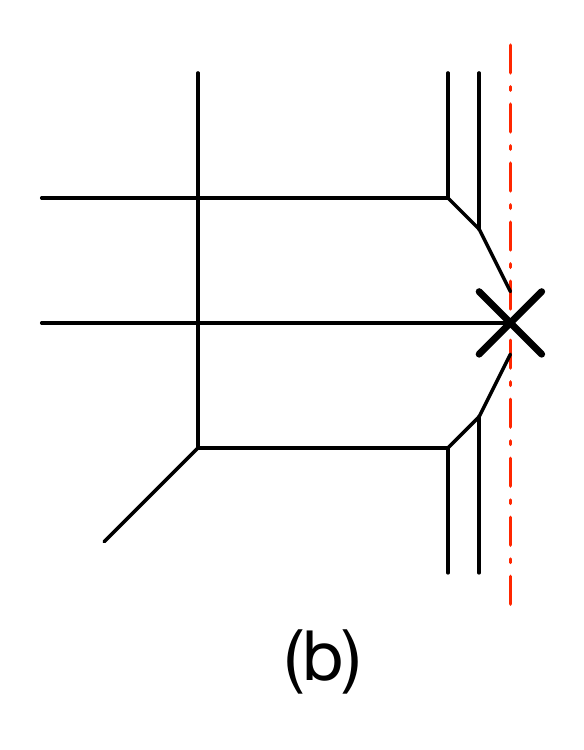} 
\caption{The orientifold 5-brane web for $SU(2)$ with $N_A=1$ and $N_F=5$ and its S-dual, which describes the same theory.}
\label{SU(2)+AS+5}
\end{figure}

\begin{figure}
\center
\includegraphics[width=0.2\textwidth]{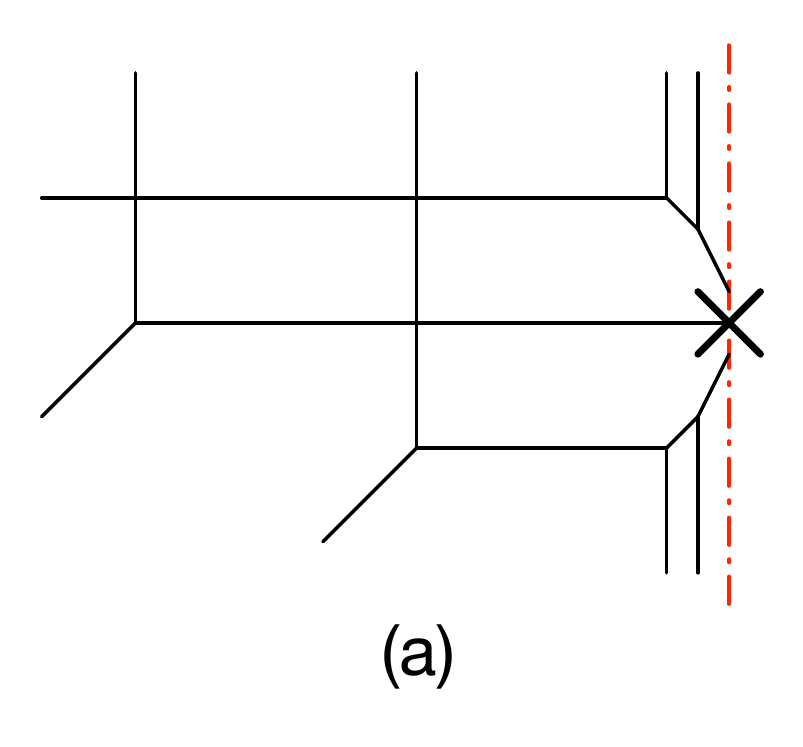} 
\hspace{1cm}
\includegraphics[width=0.2\textwidth]{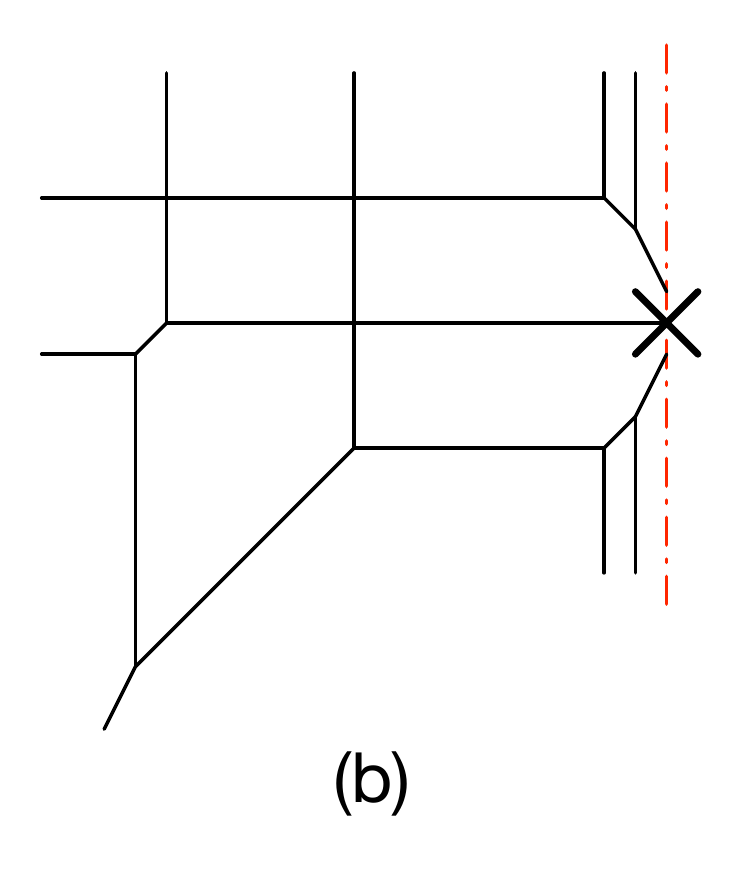} 
\hspace{1cm}
\includegraphics[width=0.3\textwidth]{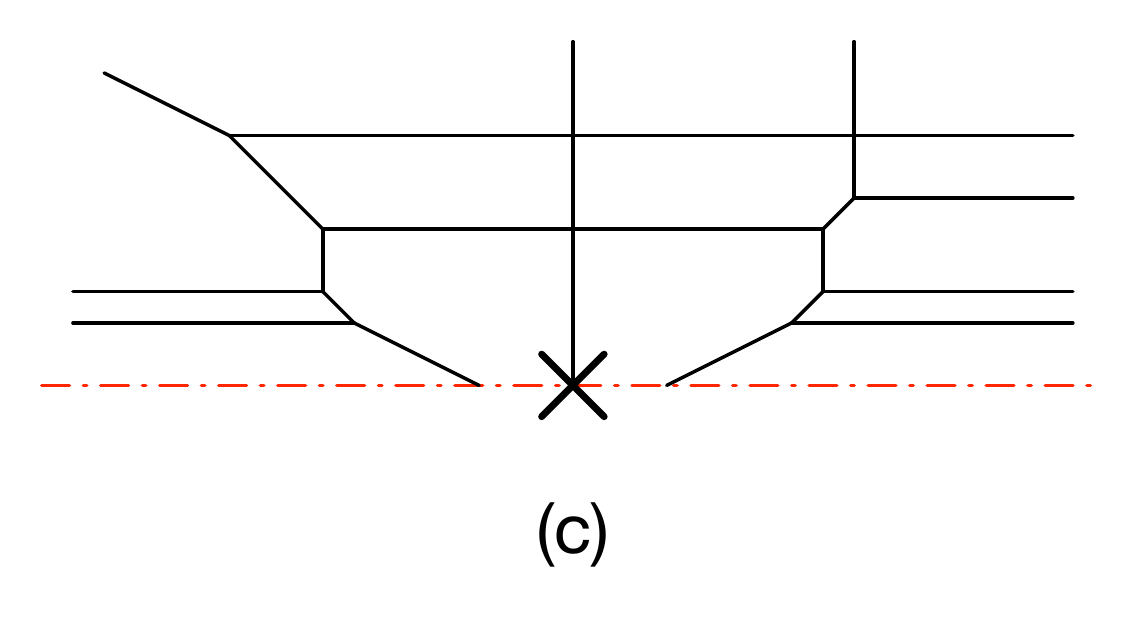} 
\caption{(a) Gauge an $SU(2)\subset SO(10)$ and add a flavor. (b) Then gauge a non-perturbative $SU(2)$.
(c) S-duality gives $SU(4)_{\pm 1}$ with $N_A=1$ and $N_F=6$.}
\label{SU(4)_1+AS+6}
\end{figure}

Our second example is slightly more intricate. We start with the self-dual $SU(2)+5$ theory (the IR gauge theory description 
of the rank one $E_6$ theory), described here by the
web and its S-dual shown in Fig.~\ref{SU(2)+AS+5}. 
First we gauge the $SU(2) \subset SO(10)_F$ 
associated with the two flavors given by the external D5-branes
on the left of the S-dual web, and add one flavor, Fig.~\ref{SU(4)_1+AS+6}a. This results in the quiver $2+SU(2)\times SU(2)+3$,
where the extra flavor on the left comes from the fractional D5-brane as before.
As the next step, we would like to gauge the $SU(2)$ global symmetry associated with the two external $(1,1)$5-branes.
This gives the web shown in Fig.~\ref{SU(4)_1+AS+6}b, with the S-dual shown in Fig.~\ref{SU(4)_1+AS+6}c. 
The latter describes $SU(4)_{\pm 1}$ with $N_A=1$ and $N_F=6$.
But what theory is this theory the dual of?
This is not immediately obvious since the $SU(2)$ that we gauged
is not a subgroup of the perturbative flavor symmetry of the quiver theory, which is $[SO(6)\times SO(4)\times U(1)^2] \times SU(2)$.

In fact the quiver theory corresponds to a known 5d fixed point, the so-called $R_{0,4}$ theory,
in which the factor $SO(6)\times SO(4)\times U(1)^2$
is enhanced to $SU(8)$ \cite{BZ}.\footnote{The 5-brane web description 
in \cite{BZ} is presumably related to the one here by resolving the orientifold plane.}
The $SU(2)$ that we are gauging is therefore contained in the $SU(8)$ factor.
%
%
The $R_{0,4}$ theory has another IR gauge theory description
as $SU(3)_{\frac{1}{2}} +7$ \cite{BZ}.
In this case the perturbative global symmetry is $SU(7)\times U(1)^2$, with one $U(1)$ getting enhanced to $SU(2)$, and
$SU(7)\times U(1)$ getting enhanced to $SU(8)$.
We can then use $SU(8)$ transformations to rotate the gauged $SU(2)$ into 
the non-Abelian flavor symmetry of either of the gauge theories.
This leads to three possible quiver theory duals of the $SU(4)_{\pm 1}$ theory with $N_A=1$ and $N_F=6$:
$SU(2)\times SU(2)\times SU(2)+3$, $2+SU(2)\times SU(2)\times SU(2)+1$, and 
$SU(2)\times SU(3)_{\frac{1}{2}}+5$.

We can generalize this to higher rank as follows. 
Start with $SU(2N)_{\frac{1}{2}}$ with $N_A=1$ and $N_F=2N+3$.
The 5-brane web describing this theory is a straightforward generalization of Fig.~\ref{SU(2)+AS+5}a.
Performing the same manipulations on the web as before leads to a web that describes
$SU(2N+2)$ with $N_A=1$ and $N_F=2N+4$, generalizing Fig.~\ref{SU(4)_1+AS+6}c.
Let us now find the dual theory.
The original gauge theory corresponds to the UV fixed point theory $R_{1,2N+1}$, a SCFT with global symmetry
given by $SU(2N+3)\times SU(2)\times U(1)^2$ \cite{BZ}. Comparing with the gauge theory, we see that 
the topological $U(1)$ symmetry is enhanced to $SU(2)$ at the fixed point.
The first step is to gauge the $SU(2)$ and add one flavor, but we cannot do this in the gauge theory since the 
$SU(2)$ cannot be rotated into the flavor symmetry.
However we can use the dual gauge theory for $R_{1,2N+1}$ given by the quiver
$2+SU(3)_{\frac{1}{2}}\times SU(3)_0^{N-3}\times [SU(3)+1]\times SU(2)+3$, the web for which is the 
generalization of Fig.~\ref{SU(2)+AS+5}b. Gauging the $SU(2)$ flavor symmetry and adding an $SU(2)$ flavor then
gives $1+SU(2)\times SU(3)_{\frac{1}{2}}\times SU(3)_0^{N-3}\times [SU(3)+1]\times SU(2)+3$, which corresponds
to the UV fixed point $R_{1,2N+2}$.\footnote{For $N=1$ this reduces to the previous example, 
since $R_{1,4}$ and $R_{0,4}$ are the same theory.}
A dual gauge theory for $R_{1,2N+2}$ is given by the quiver $2+SU(3)_{\frac{1}{2}}\times SU(3)_0^{N-1}+5$ \cite{BZ}.
The final step is to gauge the $SU(2)$ flavor symmetry, leading to $SU(2)\times SU(3)_{\frac{1}{2}}\times SU(3)_0^{N-1} +5$.
%


\subsubsection{Example 3}

\begin{figure}
\center
\includegraphics[width=0.3\textwidth]{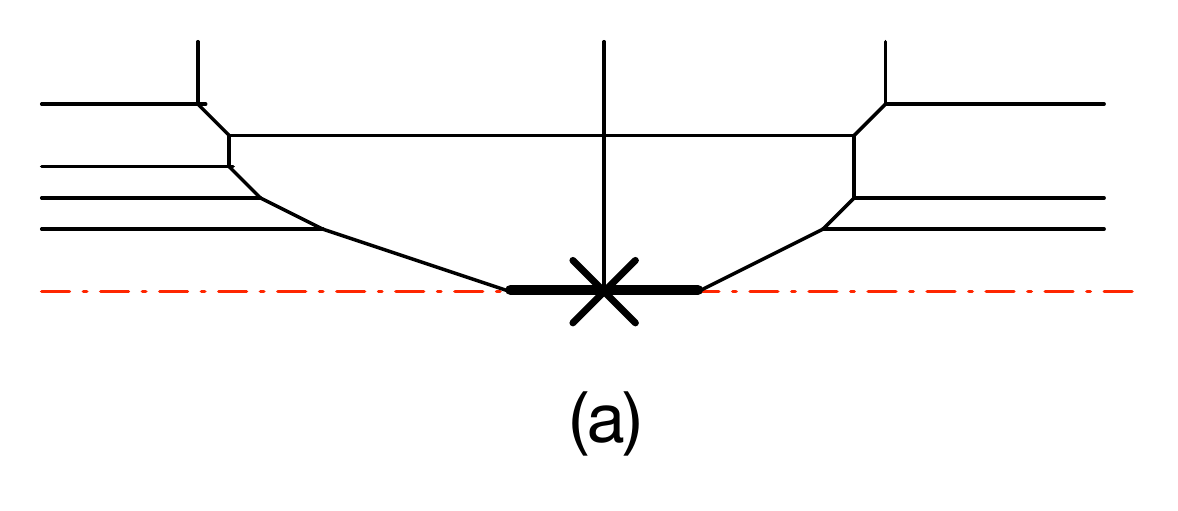} 
\hspace{1cm}
\includegraphics[width=0.3\textwidth]{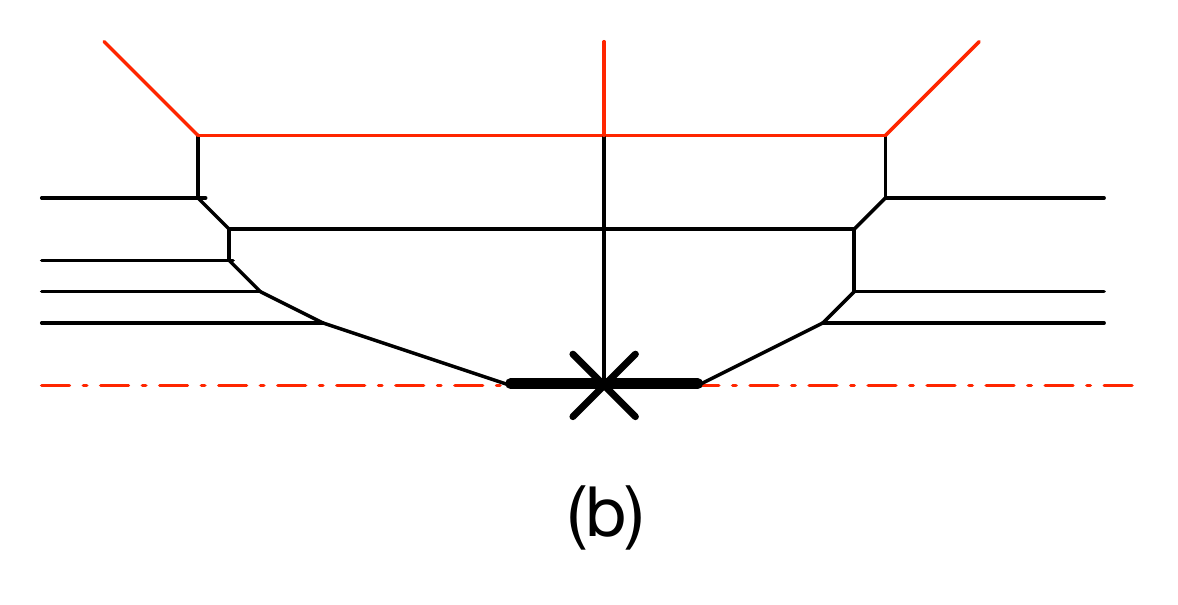} 
\caption{(a) $SU(3)_0$ with $N_A=1$ and $N_F=7$. (b) Gauging an $SU(2)$ and adding a flavor gives
$SU(5)_0$ with $N_A=1, N_F=7$.}
\label{SU(5)+AS+7}
\end{figure}

Next let us give two examples that involve $SU(2N+1)$.
Start with $SU(3)_0$ with $N_A=1$ and $N_F=7$, Fig.~\ref{SU(5)+AS+7}a.
This gauge theory corresponds to a 5d SCFT known as $S_5$, 
which has an enhanced $SU(10)$ global symmetry \cite{BZ}.
Gauging a specific $SU(3)\subset SU(10)$ and adding an $SU(3)$ flavor,
we arrive at $SU(5)_0$ with $N_A=1$ and $N_F=7$, Fig.~\ref{SU(5)+AS+7}b.
We can now use reasoning similar to before to derive a dual for this theory.
The $S_5$ theory has two dual IR gauge theory descriptions.
The first is $SU(3)_0$ with $N_F=8$, which is equivalent to our starting point since
the rank two antisymmetric and fundamental representations are equivalent for $SU(3)$.
The other is the quiver $3+SU(2)\times SU(2)+3$. The global symmetries exhibited by the two gauge theories
are $SU(8)\times U(1)^2$ and $SU(4)^2\times SU(2)\times U(1)^2$, respectively.
Both are enhanced to $SU(10)$ by instantons.
Using the full $SU(10)$ symmetry of the fixed point we can rotate the $SU(3)$ that we gauge into the 
(non-abelian) flavor symmetry of the IR gauge theory, obtaining 
$1+SU(3)\times SU(3)+5$ in the first case, and $1+SU(3)\times SU(2)\times SU(2)+3$ in the second.
We therefore claim that both of these theories are dual to $SU(5)_0$ with $N_A=1, N_F=7$.

Generalizing to higher rank, we find that $SU(2N+1)_0$ with $N_A=1$ and $N_F=2N+3$ corresponds to gauging
an $SU(3)$ in the $S_{2N+1}$ SCFT and adding an $SU(3)$ flavor.
The latter is a UV fixed point with several IR gauge theory manifestations, one of which is $SU(2N-1)_0$ with $N_A=1$ and $N_F=2N+3$ \cite{BZ}.
The full global symmetry at the fixed point (for $N > 2$) is $SU(2N+3)\times SU(3)\times U(1)$, and 
the $SU(3)$ factor is precisely what is gauged. 
As before, we can obtain a dual of our starting theory by embedding $SU(3)$ into one of the other IR gauge theories.
For example, if we take the realization of $S_{2N+1}$ as the quiver theory $3+SU(3)_0^{N-1} +5$ \cite{BZ},
and gauge the $SU(3)$ flavor symmetry and add one flavor, we get $1+SU(3)_0^N +5$.

\subsubsection{Example 4}

\begin{figure}
\center
\includegraphics[width=0.25\textwidth]{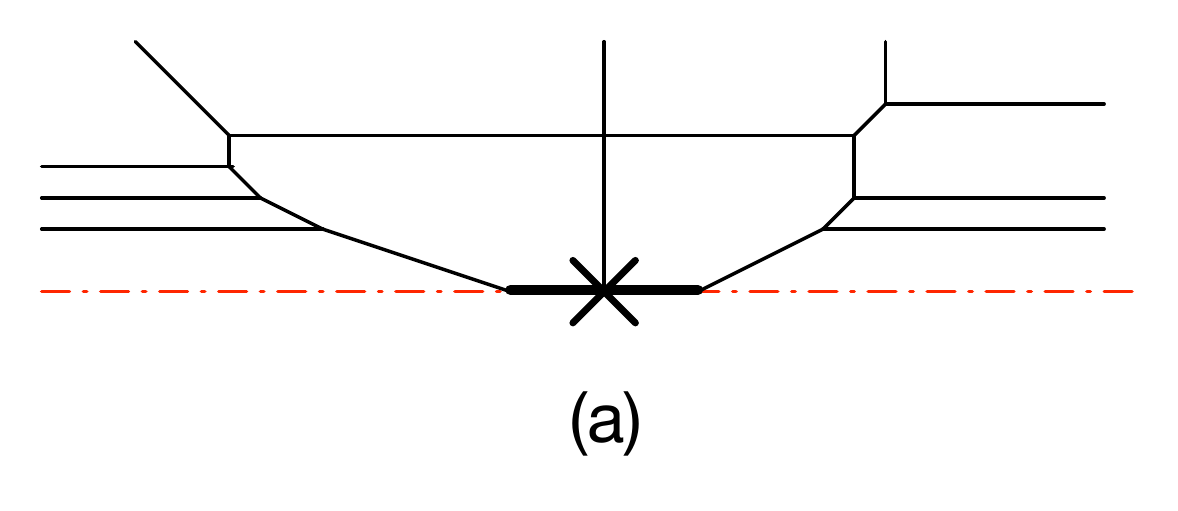} 
\hspace{1cm}
\includegraphics[width=0.25\textwidth]{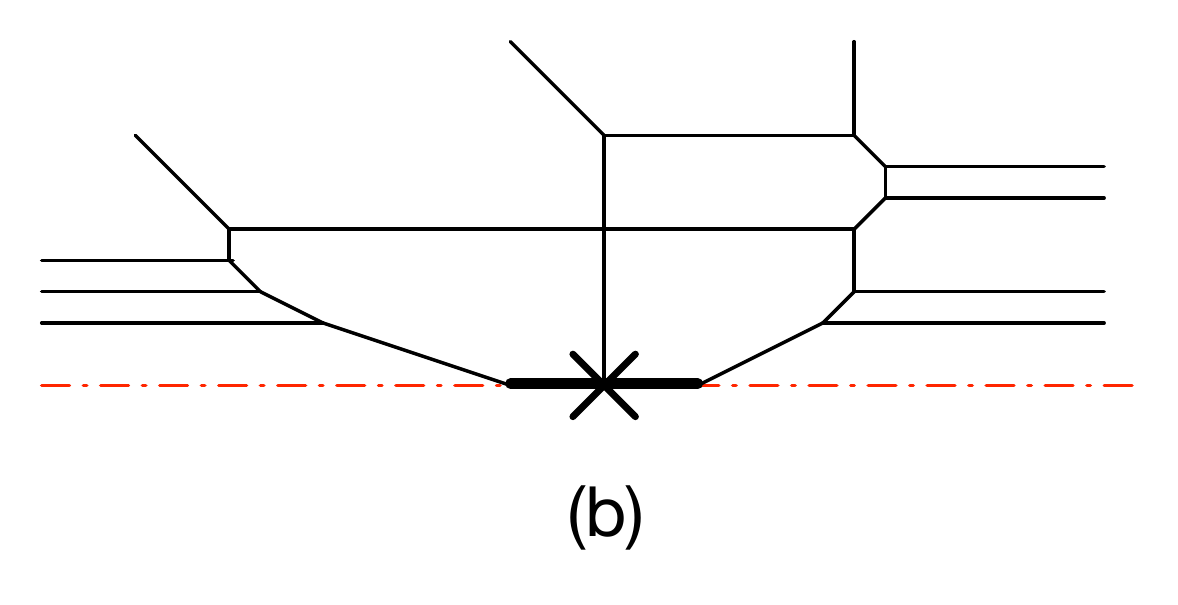} 
\hspace{1cm}
\includegraphics[width=0.25\textwidth]{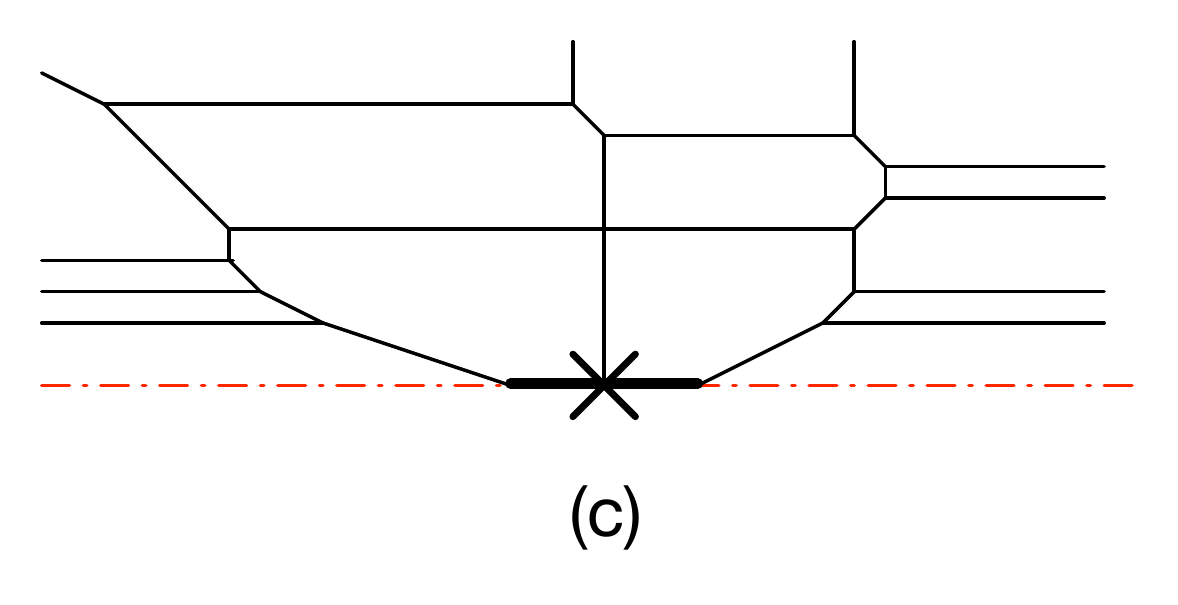} 
\caption{(a) $SU(3)_{\frac{1}{2}}$ with $N_A=1, N_F=6$, (b) gauging $1+SU(2)$, (c) gauging another $SU(2)$ gives $SU(5)_{\pm 1}$
with $N_A=1, N_F=7$.}
\label{SU(5)_1+AS+7}
\end{figure}

Our second $SU(2N+1)$ example is again more intricate, and will involve a two-step gauging of a SCFT.
The starting point is the $R_{1,4}$ theory, which has a gauge theory realization as $SU(3)_{\frac{1}{2}}$ with $N_A=1,N_F=6$ \cite{BZ}.
The global symmetry of the gauge theory is $SU(7)\times U(1)_B\times U(1)_I$, since the antisymmetric and fundamental representations of $SU(3)$ are equivalent.
The full global symmetry at the fixed point is $SU(8)\times SU(2)$.
(Actually $R_{1,4}$ is identical to $R_{0,4}$, but we will use the first description since the more general case below will involve $R_{1,N}$.)
The 5-brane web for the gauge theory is shown in Fig.~\ref{SU(5)_1+AS+7}a.
Now we gauge an $SU(2)  \subset SU(8)$, and add one flavor, resulting in the web shown in Fig.~\ref{SU(5)_1+AS+7}b. 
On the other hand, using $SU(8)$ we can rotate the gauged $SU(2)$ into the flavor $SU(7)$ symmetry of the IR gauge theory, yielding 
$1+SU(2)\times SU(3)_{\frac{1}{2}}+5$ (which we can sort of understand from the S-dual of the web in Fig.~\ref{SU(5)_1+AS+7}b).
We recognize this quiver theory as an IR gauge theory description of the SCFT $R_{1,5}$ \cite{BZ}.

Next, we would like to gauge the $SU(2)$ corresponding to the parallel external $(1,-1)$5-branes in Fig.~\ref{SU(5)_1+AS+7}b,
leading to Fig.~\ref{SU(5)_1+AS+7}c. This corresponds to the gauge theory $SU(5)_{1}$ with $N_A=1, N_F=7$.
As in example 2, this $SU(2)$ is not contained in the flavor symmetry of the quiver gauge theory.
In fact it is a subgroup of the $SU(3)$ part of the full $SU(7)\times SU(3)\times U(1)$ global symmetry of $R_{1,5}$.
On the other hand
$R_{1,5}$ has a dual gauge theory description as $3+SU(3)_{\frac{1}{2}}\times SU(2)+3$, in which this $SU(3)$ corresponds
to a flavor symmetry.
Gauging an $SU(2)\subset SU(3)$ in this description then gives $SU(2)_\pi \times [SU(3)_{\frac{1}{2}}+1]\times SU(2)+3$.

Generalizing to arbitrary rank $N$, we begin with $SU(2N-1)_{\frac{1}{2}}$, $N_A=1, N_F=2N+2$, 
which is an IR gauge theory description of $R_{1,2N}$.
The 5-brane web is a simple generalization of Fig.~\ref{SU(5)_1+AS+7}a.
The full global symmetry of the SCFT (for $N>2$) is $SU(2N+2)\times SU(2)\times U(1)^2$, whereas that of the gauge theory is
$SU(2N+2)\times U(1)^3$. In this case just the topological $U(1)$ symmetry is enhanced to $SU(2)$.
The two-step gauging procedure leads to a web describing $SU(2N+1)_1$ with $N_A=1, N_F=2N+3$.
The dual theory can be determined by considering the dual description of $R_{1,2N}$ as the quiver
$2+SU(3)_{\frac{1}{2}}\times SU(3)_0^{N-2}+5$ \cite{BZ}.
Gauguing the $SU(2)$ flavor symmetry and adding a flavor leads to $1+SU(2)\times SU(3)^{N-1}+5$,
which corresponds to the SCFT $R_{1,2N+1}$. The other gauge theory description of this theory is in turn provided
by $2+SU(3)^{N-2}\times [SU(3)+1] \times SU(2)+3$. Upon gauging the flavor $SU(2)$ symmetry in this description
we finally get $SU(2)_\pi\times SU(3)_{\frac{1}{2}}\times SU(3)_0^{N-3}\times [SU(3)_0+1]\times SU(2)+3$.

\begin{figure}
\center
\includegraphics[width=0.45\textwidth]{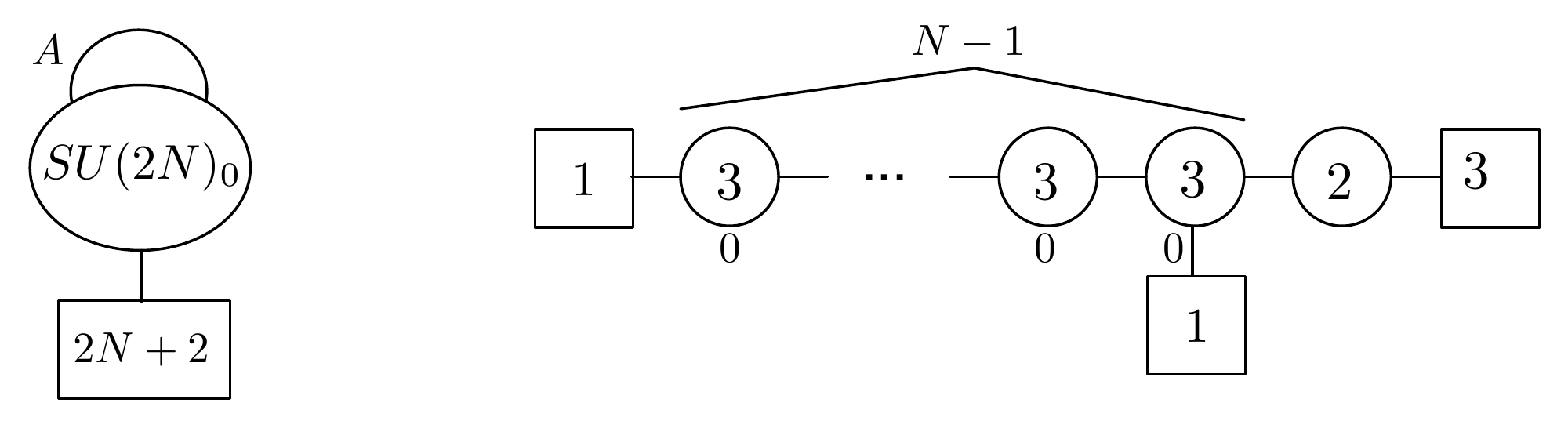} \\
\vspace{0.5cm}
\includegraphics[width=0.45\textwidth]{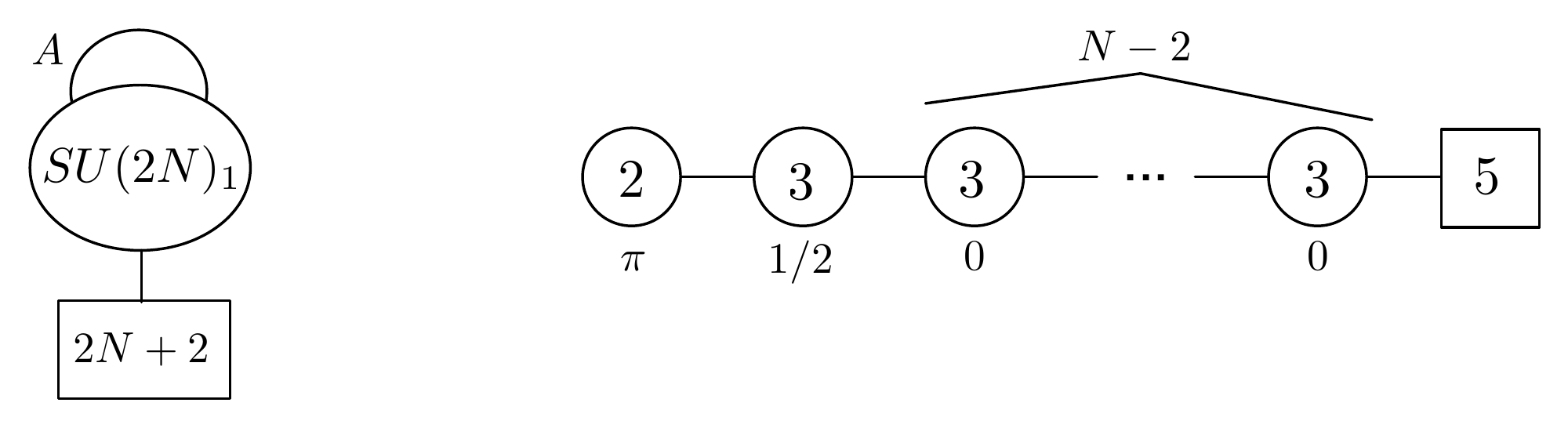} \\
\vspace{0.5cm}
\includegraphics[width=0.45\textwidth]{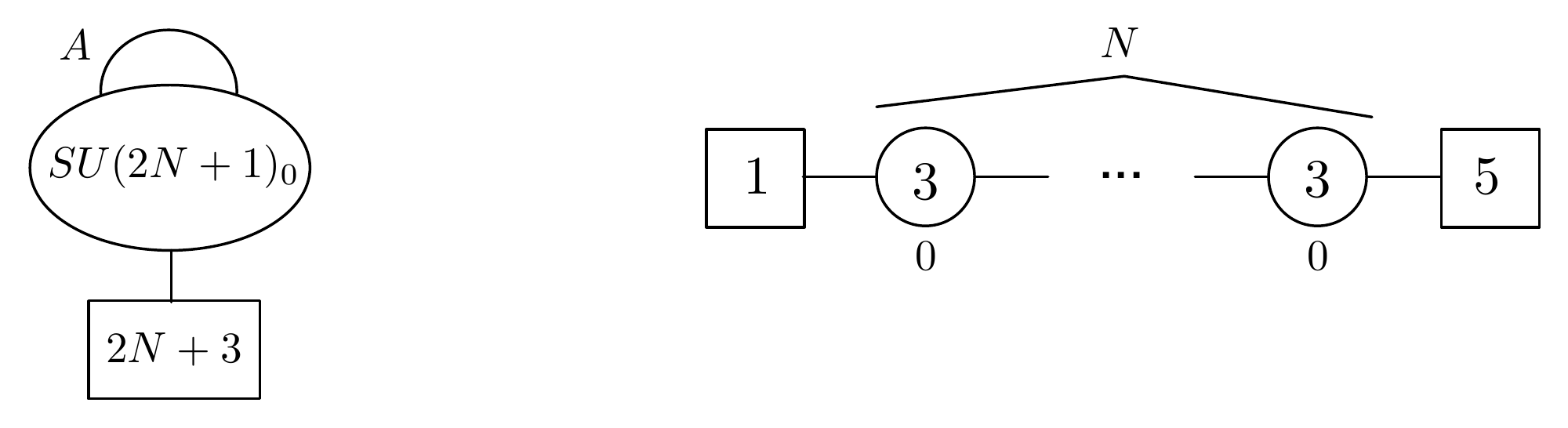}\\
\vspace{0.5cm}
\includegraphics[width=0.45\textwidth]{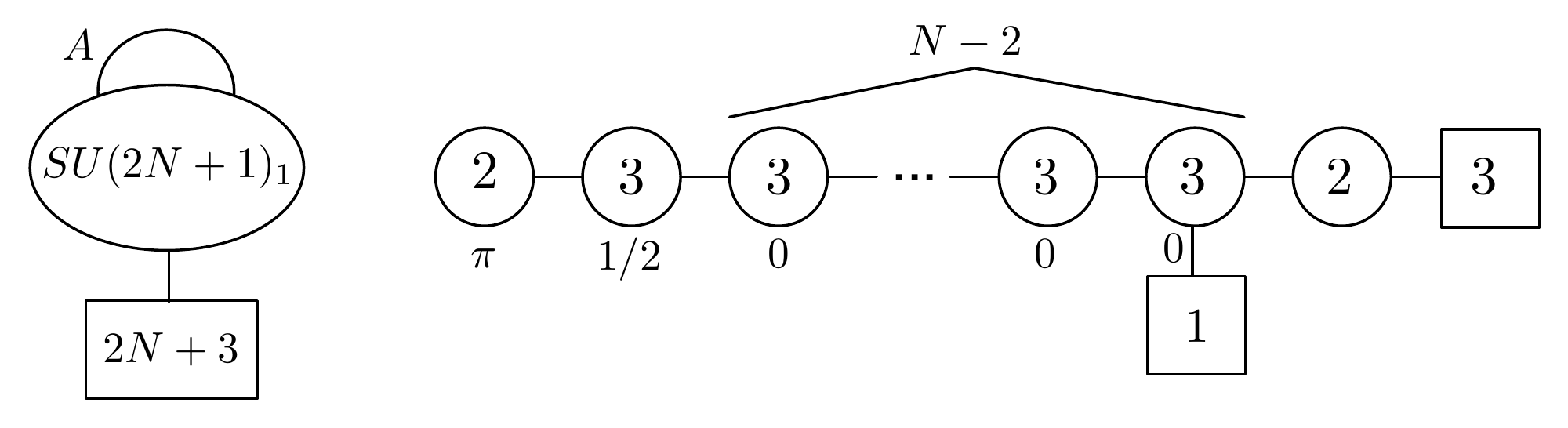}
\caption{A summary of the 5d dualities for $SU(N)$ with $N_A=1$ and $N_F=N+2$. The numbers below the nodes denote the CS level or $\theta$.}
\label{summary}
\end{figure}

A summary of all the general rank dualities discussed in this section is shown in figure \ref{summary}. 
These dualities appear to be related to the 4d duals of $SU(N)$ with $N_A=1$ and $N_F=N+2$ \cite{CD}.
It would be interesting to study this further using, for example, the superconformal index.

\section{$SU(N)$ with a symmetric}

\begin{figure}
\center
\includegraphics[width=0.3\textwidth]{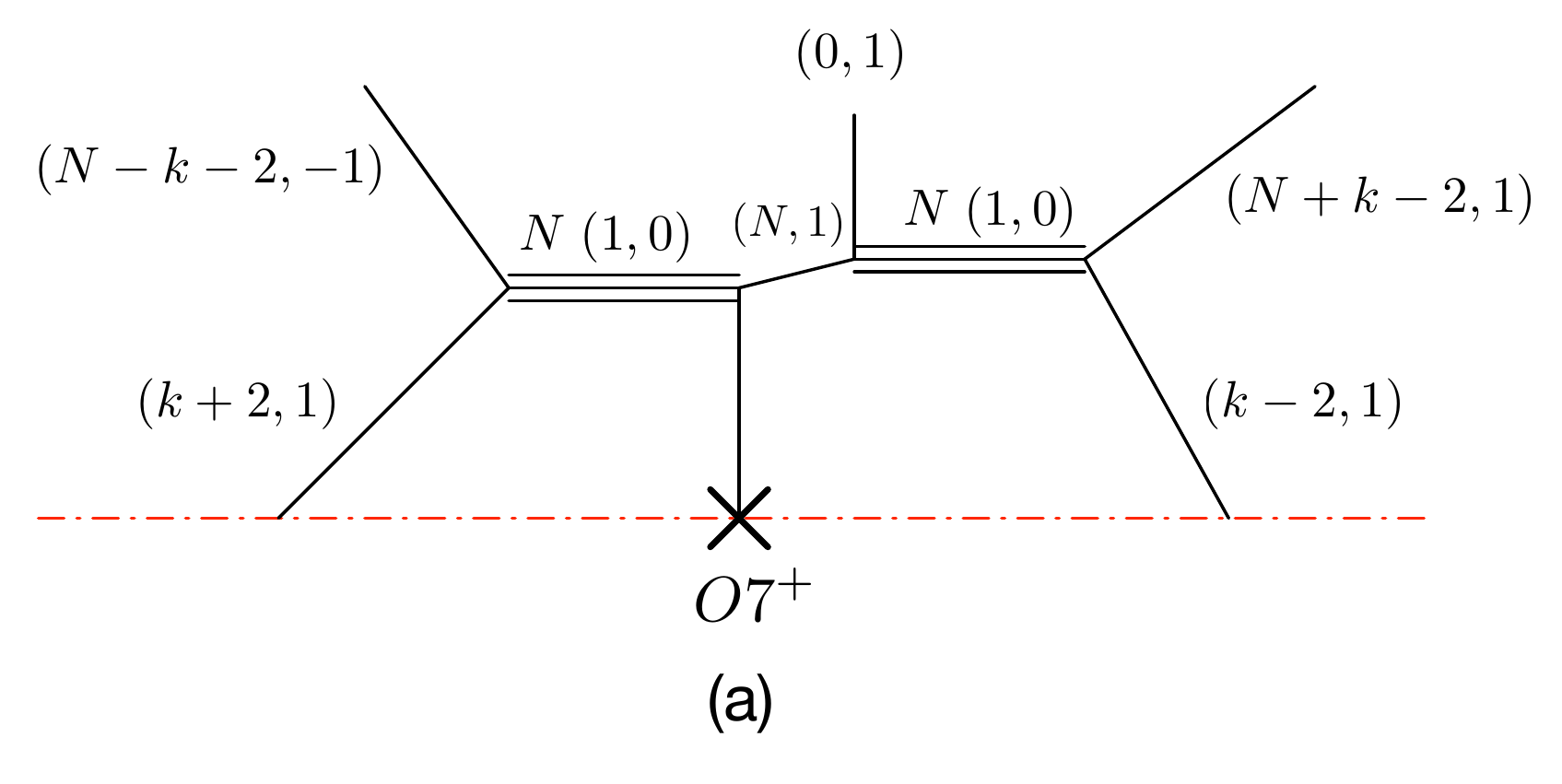} 
\hspace{1cm}
\includegraphics[width=0.3\textwidth]{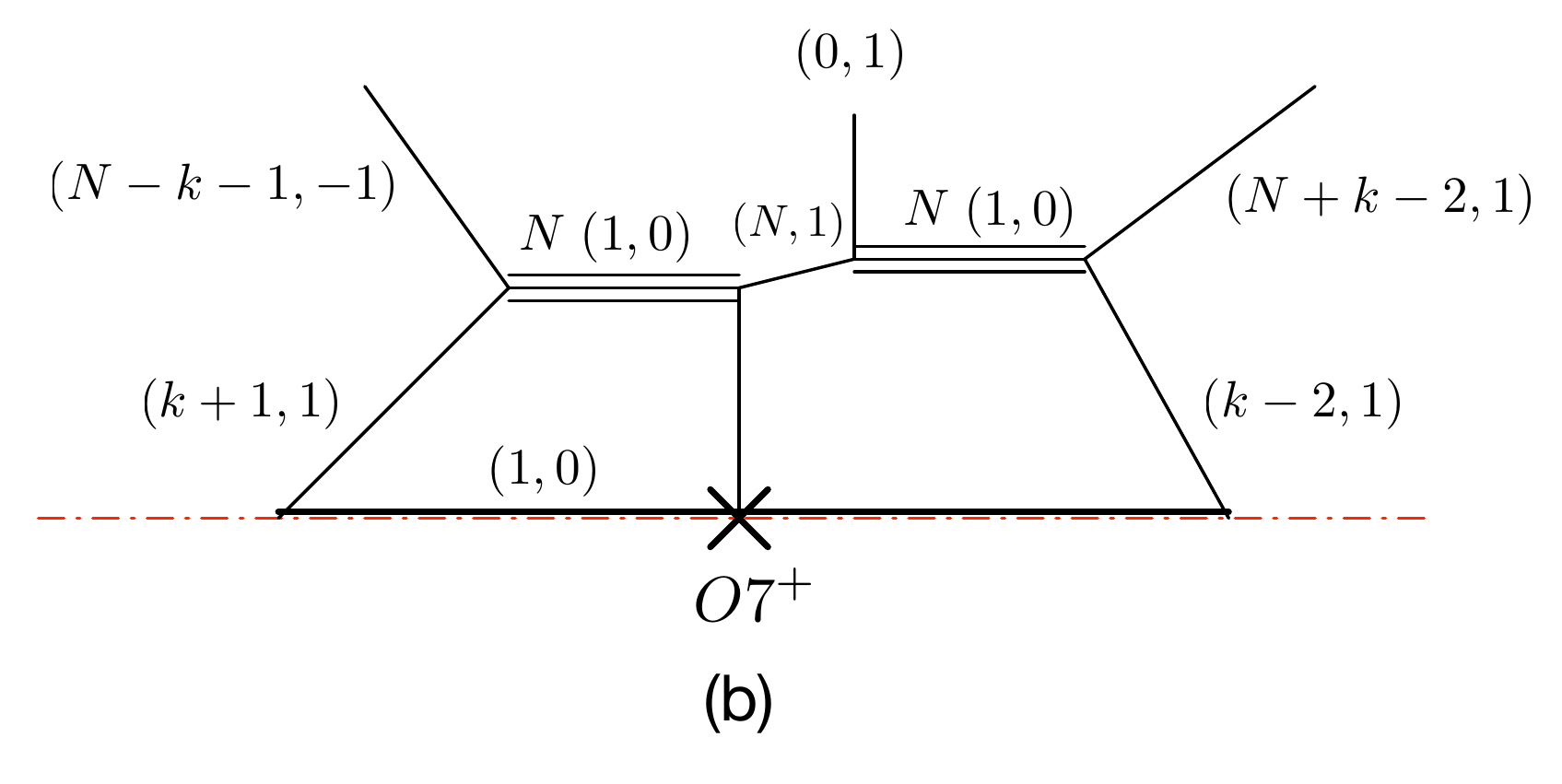} 
\caption{Orientifold 5-brane web for $SU(2N)_k$ with $N_S=1$ (left) and $SU(2N+1)_{k\pm\frac{1}{2}}$ with $N_S=1$ (right).}
\label{SU(N)+S}
\end{figure}

To engineer a symmetric hypermultiplet we replace the $O7^-$ plane with an $O7^+$ plane.
The resulting 5-brane webs for $SU(2N)_k$ and $SU(2N+1)_{k\pm \frac{1}{2}}$ with $N_S=1$ are shown in Fig.~\ref{SU(N)+S}.
In this case the maximal number of flavors that one can add for $SU(M)$ is $N_F=M-3$, which corresponds, as before, to an avoided
intersection of external 5-branes. Also as before, for $N_F=M-3,M-4$ and $M-5$, there are extraneous instanton-charged states
that have to be removed from the partition function.
 
Like in the case with an antisymmetric, the bound we find is very different from the one in \cite{SMI} where these gauge theories were ruled out. However, in the webs we find one can take the limit where all the branes intersect the $O7^+$ plane, corresponding to the fixed point. This strongly suggests that these theories do exist as $5d$ fixed point theories. 

\begin{figure}
\center
\includegraphics[width=0.3\textwidth]{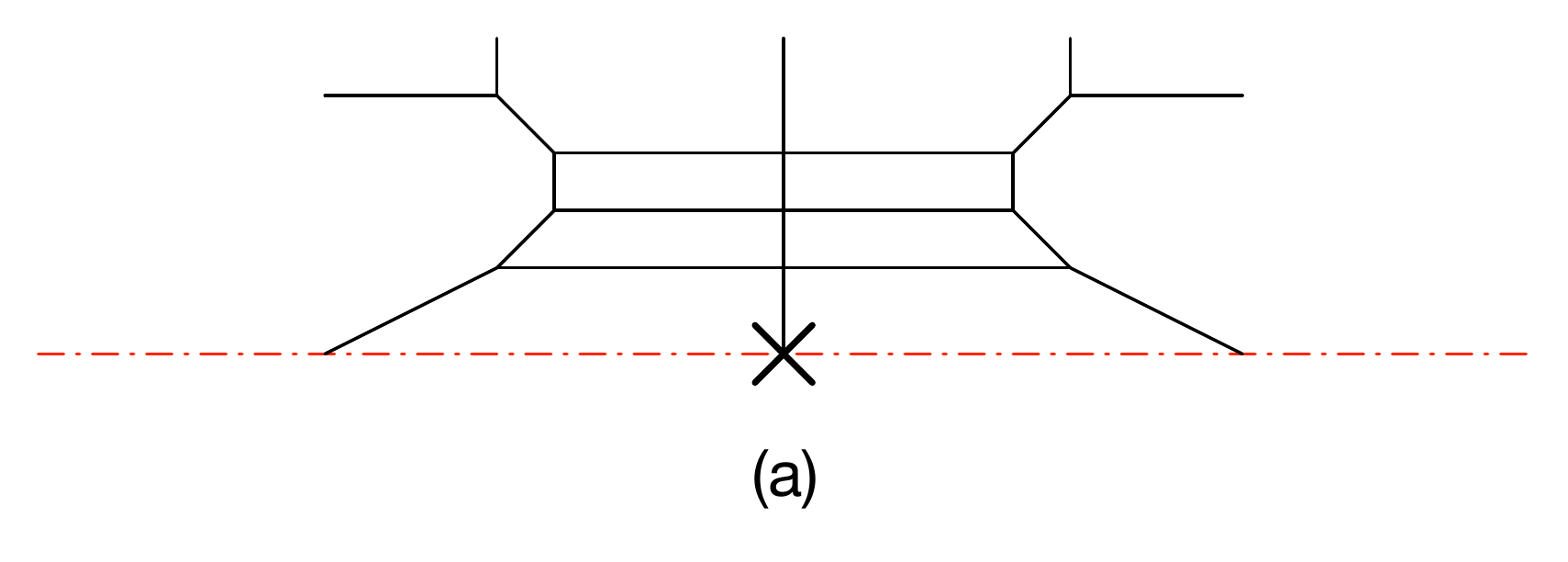} 
\hspace{1cm}
\includegraphics[width=0.3\textwidth]{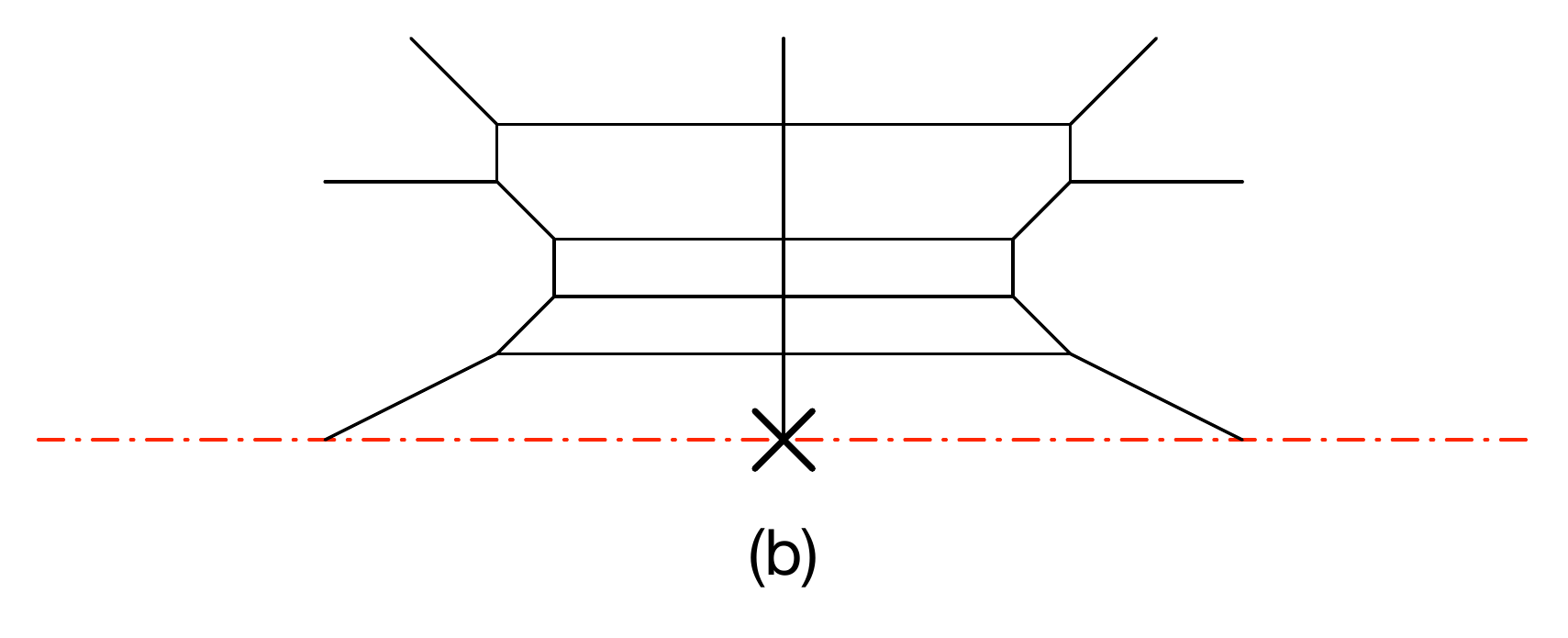} 
\caption{(a) $SU(6)$ with $N_S=1, N_F=2$. (b) $SU(8)$ with $N_S=1, N_F=2$.}
\label{SU(6)+S+2}
\end{figure}

As in the case of the $SO(M)$ theories, the S-dual webs in the present case do not seem to correspond to gauge theories,
but we can use them to relate the original gauge theory to a partial gauging of a lower rank SCFT which
does have a gauge theory description.
For example, the theory with $SU(8)_0$ and $N_S=1, N_F=2$ is obtained by gauging an $SU(3)_0$ subgroup
of the global symmetry of the fixed point corresponding to $SU(6)_0$ with $N_S=1, N_F=2$, and adding one flavor, Fig.~\ref{SU(6)+S+2}.
The generalization to $SU(2N)_0$ with $N_S=1, N_F=2$ leads to a linear quiver with $1+SU(3)_0^{N-3}$, with the last $SU(3)_0$ factor embedded in the
SCFT corresponding to $SU(6)_0$ with $N_S=1, N_F=2$.
Likewise, for $SU(2N+1)_0$ with $N_S=1$ and $N_F=1$ we find a linear quiver $1+SU(3)_0^{N-3}$, with the last $SU(3)_0$ embedded in
the SCFT corresponding to $SU(5)_0$ with $N_S=1, N_F=1$.

\section{Conclusions}

In this paper we explored 5-brane web constructions in the presence of an orientifold $7$-plane. We have shown that these can be used to argue the existence of new fixed points, motivate symmetry enhancement and duality relations, and to assist in index calculations. 
We have concentrated on a simple class of examples, mainly webs with only one full NS$5$-brane. Yet, this method can be used also to construct more complicated webs with an arbitrary number of NS$5$-branes. This then allows us to describe a large class of $SU(N)$ linear quiver theories 
with edge groups $USp(2N)$, $SO(M)$, or $SU(N)$ with a hypermultiplet in the symmetric or antisymmetric representation. 
The studies done here can be straightforwardly generalized to these cases as well. 

Besides adding an $O7$-plane we can also add an $O5$-plane, parallel to the D$5$ branes, without breaking supersymmetry. This also allows a realization of $USp$ and $SO$ gauge theories. Furthermore, this method should also allows realization of $SO$ and $USp$ alternating quiver, which to our knowledge have not been studied in $5d$. Thus, it will be interesting to examine these constructions as well\cite{BZwp}.

The study of the brane realization of $5d$ gauge theories and comparing with field theory data have also lead us to conjecture new results in string theory. Particularly, we expect that there are two varieties of the $O7^-$ plane connected by an $SL(2,Z)$ $T$ transformation. It will be interesting to see if this can also be understood from the string theory perspective. 

\section*{Acknowledgements}

We thank Diego Rodriguez-Gomez for useful conversations.
G.Z. is supported in part by the Israel Science Foundation under grant no.~352/13 and by the German-Israeli Foundation
for Scientific Research and Development under grant no.~1156-124.7/2011.
O.B. is supported in part by the Israel Science Foundation under grant no. 352/13,
the German-Israeli Foundation for Scientific Research and Development under grant no.~1156-124.7/2011,
and the US-Israel Binational Science Foundation under grant no. 2012-041.


\appendix

\section{Instanton partition functions and extraneous states}



The instanton partition function is expressed as a contour integral over the Cartan subalgebra of the dual gauge group, which depends on the gauge group and instanton number. The integrand contains contributions from the gauge group and the matter sector, and is a function of the various fugacities. 
We denote by $s_i$ gauge fugacities for the true gauge group, by $u_a$ gauge fugacities for the dual gauge group, by $f_m$ fundamental flavor fugacities and by $z$ fugacities for other representations. We also use $x, y$ for the superconformal fugacities.
The integral can be evaluated using the residue theorem once supplemented with the appropriate pole prescription which determines which poles should be included. The poles can be classified depending on whether they originate from the contributions of the gauge multiplet, matter hypermultiplets, 
or are poles at zero or infinity. 
The prescription for the poles associated with the gauge multiplets can be found in \cite{KKL,HKKP,BGZ},
and the prescription for those associated with matter mutiplets (in representations other than the fundamental) can be found in
\cite{HKKP}. 

Poles at zero (or infinity) lead to the violation of the $x\rightarrow \frac{1}{x}$ symmetry (required by conformal invariance) in the partition function.
In general, for a single integral ({\em e.g.} for one $SU(N)$ instanton),
\bea
Z_1 [x] - Z_1 \left[\frac{1}{x}\right] &=& Res[u=0] - Res[u \rightarrow \infty] \,.
\eea    
For two integrals ({\em e.g.} two $SU(N)$ instantons) we have
\bea
Z_2 [x] - Z_2 \left[\frac{1}{x}\right]  & = & Res[u_1 \sim x, u_2=0] + Res[u_1 = 0, u_2 \sim x] - Res[u_1 \sim x, u_2\rightarrow\infty] \nonumber \\ & - & Res[u_1 \rightarrow \infty, u_2 \sim x] + Res[u_1 = 0, u_2 = 0] \nonumber \\ & - & Res[u_1 \rightarrow \infty, u_2 \rightarrow \infty] \,.
\eea 
These poles are associated with extraneous states, whose contributions must be removed from the instanton partition function.
The restoration of $x\rightarrow \frac{1}{x}$ symmetry serves as a useful test of this procedure.



In this case we deal with gauge groups $SO(N)$ and $Sp(N)$ and consider only non-complex representations. As a result the residues at zero and infinity differ solely by a sign (this can be seen by the $u\rightarrow \frac{1}{u}$ invariance of the expressions for these groups). Thus, we can simplify to:
%
%
\bea
Z_{1} [x] - Z_{1} \left[\frac{1}{x}\right] &=& 2Res[u=0]
\eea 
\bea
Z_{2} [x] - Z_{2} \left[\frac{1}{x}\right] & = & 2(Res[u_1 \sim x, u_2=0] + Res[u_1 = 0, u_2 \sim x]) \nonumber \\ \label{cvc}
\eea 
and completely ignore poles at infinity.

\subsection{$USp(2N)$}

The $k$-instanton partition function for $USp(2N)$ has two components, $Z_+$ and $Z_-$, corresponding, respectively,
to summing over holonomies of determinant $+1$ and $-1$ in the dual gauge group $O(k)$.
The latter can be viewed as the sector with one gauge QM instanton corresponding to the non-trivial element of $\pi_0(O(k)) = \mathbb{Z}_2$.
For $k=2n+1$ there are $n$ independent holonmies, and therefore $n$ contour integrals, in both cases.
For $k=2n$ there are $n$ independent holonmies of determinant $+1$, but only $n-1$ of determinant $-1$.
The total partition function is given by $Z_+ \pm Z_-$, where the relative sign depends on the value of the discrete $\theta$ parameter.

The expressions for the contributions of the gauge multiplet and fundamental hypermultiplets to each component can be found in the appendix of \cite{BGZ}.
For $k=1$ there is no integral, and both components are invariant under $x\rightarrow \frac{1}{x}$. 

For $k=2$ the determinant $+1$ component $Z_+$ involves a single contour integral, and there are poles
at zero or infinity when $N_F \geq 2N+4$. 
%
%
%
For $N_F=2N+4$ we find that
\bea
Z_2^{USp(2N)+(2N+4)} [x] - Z_2^{USp(2N)+(2N+4)} \left[\frac{1}{x}\right] &=& \frac{(1-x^2)}{(1-x y)(1-\frac{x}{y})} \,.
\eea 
Therefore the combination
\be
\label{twoinstanton}
Z_2^{Sp(N)+(2N+4)} [x] + \frac{x^2}{(1-x y)(1-\frac{x}{y})}
\ee
is $x\rightarrow \frac{1}{x}$ invariant.
This agrees with the 2-instanton correction term that we subtracted in section 2, eq.~(\ref{subtraction}).
For $N_F=2N+5$ we find
\bea
\lefteqn{Z_2^{Sp(N)+(2N+5)} [x] - Z_2^{Sp(N)+(2N+5)} \left[\frac{1}{x}\right] = }\\ \nonumber 
& & \frac{(1-x^2)}{(1-x y)(1-\frac{x}{y})}\left(\chi_{\bf 4N+10}^{SO(4N+10)} - (x+\frac{1}{x})\chi_{\bf 2N}^{USp(2N)} \right) \,,
\eea  
and therefore the combination
\be
Z_2^{USp(2N)+(2N+5)} [x] + \frac{x^2}{(1-x y)(1-\frac{x}{y})}\left( \chi_{\bf 4N+10}^{SO(4N+10)} - x \chi_{\bf 2N}^{USp(2N)} \right) \nonumber
\ee 
is $x\rightarrow \frac{1}{x}$ invariant. This agrees with the 2-instanton correction term, eq.~(\ref{subtraction}), for this case.


For $k=3$ both components of the partition function involve a single contour integral, and again poles at zero appear for $N_F\geq 2N+4$.
Let's concentrate on the case with $N_F=2N+4$, for which we gave a proposal for the full multi-instanton correction factor in eq.~(\ref{multi-instanton}).
For $N_F = 2N+4$ we find
%
%
%
\bea
Z_3^{USp(2N)+(2N+4)} [x] - Z_3^{USp(2N)+(2N+4)} \left[\frac{1}{x}\right] &=& \frac{(1-x^2)}{(1-x y)(1-\frac{x}{y})} 
Z_1^{USp(2N)+(2N+4)}[x] \,,\nonumber \\
\eea  
implying that the combination
\be
\label{threeinstanton}
Z_3^{USp(2N)+(2N+4)} [x] + \frac{x^2}{(1-x y)(1-\frac{x}{y})} Z_1^{USp(2N)+(2N+4)} [x] 
\ee
is $x\rightarrow \frac{1}{x}$ invariant.

For $k=4$, $Z_+$ has two contour integrals and $Z_-$ has one.
Both involve poles at zero for $N_F\geq 2N+4$. We focus again on the case $N_F=2N+4$.
Evaluating the double contour integral for $Z_{4,+}$ requires a little more work.
%
%
We can separate the non-zero poles into mixed ones (where $u_1 \propto u_2$) and non-mixed ones. 
The non-mixed ones come in identical pairs of $(u_1=0, u_2\neq 0)$ and $(u_1\neq0, u_2= 0)$. It is not difficult to see that
\be
Res[u_1=0, u_2] = \frac{1-x^2}{4(1-x y)(1-\frac{x}{y})} Z_{2,+}[u_2] \,,
\ee
where $Z_{2,+}[u_2]$ is the integrand of $Z_{2,+}$.
Combining this with $Res[u_2=0,u_1]$ and with $Z_{4,-}$, and using (\ref{cvc}) we find
\be
\frac{1-x^2}{(1-x y)(1-\frac{x}{y})} Z_2^{USp(2N)+(2N+4)} [x] - \frac{(1-x^2)^2}{2(1-x y)^2(1-\frac{x}{y})^2} \,.\nonumber
\ee
In deriving this we used:
\be
Z_2^{USp(2N)+(2N+4)} [x] = Res[u\sim x] + Res[u=0] , Res[u=0] = \frac{(1-x^2)}{2(1-x y)(1-\frac{x}{y})} \, ,
\ee
where the residues are for the $2$ instanton integrand, $Z_{2,+}+Z_{2,-}$. 
To this we need to add the contributions of the four mixed poles, which give
\bea
& & Res[u_1=x y u_2, u_2=0] + Res[u_1=\frac{x}{u_2 y}, u_2=0] + Res[u_1=\frac{x y}{u_2}, u_2=0] \nonumber \\  & + & Res[u_1=\frac{x u_2}{y}, u_2=0] =  \frac{1-x^4}{4(1-x^2 y^2)(1-\frac{x^2}{y^2})} \,.
\eea
Combining everything we find
\bea
\lefteqn{Z_4^{USp(2N)+(2N+4)} [x] - Z_4^{USp(2N)+(2N+4)} \left[\frac{1}{x}\right] =} \\ \nonumber
 &&  \frac{(1-x^2)}{(1-x y)(1-\frac{x}{y})} Z_2^{USp(2N)+(2N+4)}[x] - \frac{(1-x^2)^2}{2(1-x y)^2(1-\frac{x}{y})^2} + \frac{1-x^4}{2(1-x^2 y^2)(1-\frac{x^2}{y^2})} 
 \\ \nonumber
 && = \frac{(1-x^2)}{(1-x y)(1-\frac{x}{y})} Z_2^{USp(2N)+(2N+4)}[x] + \frac{x(1-x^2)(x(1+x^2)-y-\frac{1}{y})}{(1+x y)(1+\frac{x}{y})(1-x y)^2(1-\frac{x}{y})^2}\,.
\eea  
One can then show that the combination
\be
\label{fourinstanton}
Z_4^{USp(2N)+(2N+4)} [x] + \frac{x^2}{(1-x y)(1-\frac{x}{y})}Z_2^{USp(N)+(2N+4)} [x] + \frac{x^4(1+x^2)}{(1+x y)(1+\frac{x}{y})(1-x y)^2(1-\frac{x}{y})^2} 
\ee
is invariant under $x\rightarrow \frac{1}{x}$.
 
In fact the expressions in (\ref{twoinstanton}), (\ref{threeinstanton}) and (\ref{fourinstanton})
reproduce the multi-instanton correction factor for $N_F=2N+4$, eq.~(\ref{multi-instanton}), to instanton number four.
 

\subsection{$SO(M)$}

The dual gauge group for $k$ instantons of $SO(M)$ is $USp(2k)$. 
The expressions differ slightly between even and odd $M$. We denote $M=2N + \chi$ where $\chi=0$ or 1.
The contributions of the gauge and fundamental (vector) matter multiplets can be lifted from the 4d results of \cite{NS,Shad}.
The gauge multiplet contributes
\bea
\lefteqn{Z^{SO(M)}_{gauge} =  \frac{1}{2^k k!} \frac{(1-x^2)^k}{(1-x y)^k(1-\frac{x}{y})^k} \times} \\ \nonumber 
& & \prod^{k}_{a=0} \frac{(u_a-\frac{1}{u_a})^2(u^2_a+\frac{1}{u^2_a}-x^2-\frac{1}{x^2})}{u_a(x+\frac{1}{x}-u_a-\frac{1}{u_a})^{\chi} \prod^{N}_{i=0}(u_a+\frac{1}{u_a}-x s_i-\frac{1}{x s_i})(u_a+\frac{1}{u_a}-\frac{x}{s_i}-\frac{s_i}{x})} \\ \nonumber   & & \prod^{k}_{a<b} \frac{(u_a+\frac{1}{u_a}-u_b-\frac{1}{u_b})^2(u_a u_b+\frac{1}{u_a u_b}-x^2-\frac{1}{x^2})(\frac{u_a}{ u_b}+\frac{u_b}{u_a}-x^2-\frac{1}{x^2})}{(u_a u_b+\frac{1}{u_a u_b}-x y-\frac{1}{x y})(\frac{u_a}{ u_b}+\frac{u_b}{u_a}-x y-\frac{1}{x y})(u_a u_b+\frac{1}{u_a u_b}-\frac{x}{y}-\frac{y}{x})(\frac{u_a}{ u_b}+\frac{u_b}{u_a}-\frac{x}{y}-\frac{y}{x})} \,,
\eea
and fundamental hypermultiplets contribute
\be
Z^{SO(M)}_{fund.} = \prod^{k}_{a=1} \prod^{N_f}_{m=1} \left(f_m + \frac{1}{f_m} - u_a - \frac{1}{u_a}\right) \,.
\ee

Poles at zero and infinity appear when $N_f \geq M-4$, starting at $k=1$.
For $N_F=N-4$ we find
%
%
\bea
Z_{1}^{SO(M)+(M-4)} [x] - Z_{1}^{SO(M)+(M-4)} \left[\frac{1}{x}\right] &=& \frac{(1-x^2)}{(1-x y)(1-\frac{x}{y})} \,,
\eea 
and the $x\rightarrow \frac{1}{x}$ invariant combination is then given by
\be
Z_{1}^{SO(M)+(M-4)} [x] + \frac{x^2}{(1-x y)(1-\frac{x}{y})} \,. 
\ee
For $N_F=M-3$ we find
\bea
Z_{1}^{SO(M)+(M-3)} [x] - Z_{1}^{SO(M)+(M-3)} \left[\frac{1}{x}\right] = 
 \frac{(1-x^2)}{(1-x y)(1-\frac{x}{y})}\left(\chi_{\bf 2M-6}^{USp(2M-6)} - (x+\frac{1}{x})\chi_{\bf M}^{SO(M)} \right)\,, \nonumber\\
\eea  
and the $x\rightarrow \frac{1}{x}$ invariant combination is then
\be
Z_{1}^{SO(M)+(M-3)} [x] + \frac{x^2}{(1-x y)(1-\frac{x}{y})}\left( \chi_{\bf 2M-6}^{USp(2M-6)}  - x \chi_{\bf M}^{SO(M)} \right) \,.
\ee 

As in the $USp(2N)$ theory, we expect the corrections to the higher instanton contributions for $N_F=M-4$ to sum to a simple form
involving a plethystic exponential. 
The analysis of the 2-instanton contribution is similar to that of the 4-instanton contribution to $Z_+$ in the $USp(2N)$ theory.
There are two contour integrals, and the contributions to the non-invariance can be split into a mixed and non-mixed parts. 
The non-mixed parts give
\be
 \frac{1-x^2}{2(1-x y)(1-\frac{x}{y})} Z_{1}^{SO(M)+(M-4)} [x] - \frac{(1-x^2)^2}{4(1-x y)^2(1-\frac{x}{y})^2} \,, \nonumber
\ee
where we have again used
\be
Z_1^{SO(M)+(M-4)} [x] = Res[u\sim x] + Res[u=0] , Res[u=0] = \frac{(1-x^2)}{2(1-x y)(1-\frac{x}{y})} \,
\ee
residues now referring to the $1$ instanton integrand, $Z_1$.

The mixed parts give
\bea
&&Res[u_1=x y u_2, u_2=0] + Res[u_1=\frac{x}{u_2 y}, u_2=0] + Res[u_1=\frac{x y}{u_2}, u_2=0]\nonumber\\
&&+ Res[u_1=\frac{x u_2}{y}, u_2=0] =  \frac{1-x^4}{4(1-x^2 y^2)(1-\frac{x^2}{y^2})}
\eea
Combining everything we find
\bea
\lefteqn{Z_2^{SO(M)+(M-4)} [x] - Z_2^{SO(M)+(M-4)F} \left[\frac{1}{x}\right] = }\nonumber \\
&& \frac{(1-x^2)}{(1-x y)(1-\frac{x}{y})} Z_1^{SO(M)+(M-4)}[x]  - \frac{(1-x^2)^2}{2(1-x y)^2(1-\frac{x}{y})^2} 
+ \frac{1-x^4}{2(1-x^2 y^2)(1-\frac{x^2}{y^2})} \nonumber \\
&& =
 \frac{(1-x^2)}{(1-x y)(1-\frac{x}{y})} Z_1^{SO(M)+(M-4)}[x]  + \frac{x(1-x^2)(x(1+x^2)-y-\frac{1}{y})}{(1+x y)(1+\frac{x}{y})(1-x y)^2(1-\frac{x}{y})^2} \,.
\eea  
The $x\rightarrow \frac{1}{x}$ invariant combination is given by
\be
Z_2^{SO(M)+(M-4)} [x] + \frac{x^2}{(1-x y)(1-\frac{x}{y})}Z_1^{SO(M)+(M-4)} [x] + \frac{x^4(1+x^2)}{(1+x y)(1+\frac{x}{y})(1-x y)^2(1-\frac{x}{y})^2} .\nonumber
\ee

The one and two-instanton expressions are consistent with a multi-instanton correction factor of the form
\be
PE\left[\frac{x^2 q}{(1-x y)(1-\frac{x}{y})}\right] \mathcal{Z}^{SO(M)+(M-4)F} \nonumber \,.
\ee



\subsection{$SU(N)$}

The contributions from the gauge multiplet as well as from flavor in the fundamental were given in \cite{BGZ}. As we do not us them in this paper, we have not reproduce them. However, we shall write the contribution of matter in the symmetric and antisymmetric representations of $SU(N)$. For the symmetric we can lift from the 4d results of \cite{Shad}, finding:

\bea
& & Z^k_S [u_a]  =  \prod^k_{a=1} (u^2_a z + \frac{1}{u^2_a z} - y - \frac{1}{y}) \prod^N_{i=1} (\sqrt{u_a s_i z}-\frac{1}{\sqrt{u_a s_i z}}) \label{SSU} \\ \nonumber & & \prod^k_{a<b} \frac{(u_a u_b z + \frac{1}{u_a u_b z} - y - \frac{1}{y})}{(u_a u_b z + \frac{1}{u_a u_b z} - x - \frac{1}{x})} 
\eea

The contribution for the antisymmetric of $SU(N)$ was already given in \cite{BGZ}. Unfortunately, there was a mistake there which we take the opportunity to correct. The correct contribution is:

\bea
z_{A}^k[u_a] = \prod^k_{a=1} \frac{\prod^N_{i=1} \left(\sqrt{u_a s_i z} 
- \frac{1}{\sqrt{u_a s_i z}}\right)}{u^2_a z + \frac{1}{u^2_a z} - x - \frac{1}{x}} \prod^k_{a<b} \frac{(u_a u_b z + \frac{1}{u_a u_b z} - y - \frac{1}{y})}{(u_a u_b z + \frac{1}{u_a u_b z} - x - \frac{1}{x})} \label{ASSU}
\eea

Both of these provide additional poles in the integrand. The prescription for dealing with them follows from the results of \cite{HKKP}. Specifically, one defines $p=\frac{1}{z x}$ and $d=\frac{z}{x}$, calculates the integral assuming $x, p, d << 1$, and only at the end return to the original variables. 

Note that as generically with matter contributions, the symmetric and antisymmetric provide negative powers of $u_a$. Specifically, they behaves as $Z^k_S [u_a] \sim u^{-\frac{N \pm 4}{2}}_a$, where the $+$ sign is for the symmetric and the $-$ for the antisymmetric. Since a fundamental goes like $u^{-\frac{1}{2}}_a$, we see that as far as the lowest $u$ power and thus the appearance of zero poles is concerned, a symmetric contributes as $N+4$ fundamentals and an antisymmetric as $N-4$ fundamentals. 

Finally a note regarding signs. When evaluating $SU$ partition function with fundamentals it was noticed that a sign shift of $(-1)^{k(\kappa + \frac{N_f}{2})}$ is also needed. When working with symmetrics or antisymmetrics using (\ref{SSU},\ref{ASSU}) we expect this to generalize to $(-1)^{k(\kappa + \sum_{i} \frac{C_3[R_i]}{2})}$ where the sum is over all the matter field, $R_i$ is the representation of the $i$ matter field under $SU(N)$ and $C_3[R]$ is the cubic Casimir of the representation $R$ (normalized so that $C_3[F]=1$). For antisymmetrics, we can compare with other methods, and we have checked that this is indeed consistent with results obtained by a different formalism. 




\end{document}